\documentclass[
  aps,
  pra,
  superscriptaddress,
  amsmath,amssymb,
  floatfix,
  nofootinbib,
  reprint,   
]{revtex4-2}
\usepackage{cancel}
\usepackage{url}
\usepackage{csquotes}
\usepackage{mathtools}
\usepackage{graphicx}
\usepackage{placeins}
\usepackage{dcolumn}
\usepackage{bm}
\usepackage[colorlinks, linkcolor=red, anchorcolor=green, citecolor=blue]{hyperref}

\usepackage{physics}
\usepackage{dsfont}
\usepackage{tikz,pgfplots}
\usepackage{enumerate}
\usepackage{subfigure}
\usepackage{bbold}
\usepackage{makecell}
\usepackage{array}
\makeatletter
\newcommand{\thickhline}{%
    \noalign {\ifnum 0=`}\fi \hrule height 1pt
    \futurelet \reserved@a \@xhline
}
\newcolumntype{"}{@{\hskip\tabcolsep\vrule width 1pt\hskip\tabcolsep}}
\makeatother
\usepackage{multirow}
\usepackage{tabu}
\usepackage{textcomp,booktabs}
\usepackage{colortbl}
\usepackage[T1]{fontenc}
\definecolor{mygray}{gray}{0.9}
\definecolor{mypink}{rgb}{0.99,0.91,0.95}
\definecolor{mycyan}{cmyk}{0.3,0,0,0}

\newcommand{\id}{\text{id}}

\newcommand{\mB}{\mathcal{B}}
\newcommand{\mT}{\mathcal{T}}
\newcommand{\mE}{\mathcal{E}}
\newcommand{\mF}{\mathcal{F}}

\newcommand{\mD}{\mathcal{D}}
\newcommand{\mI}{\mathcal{I}}
\newcommand{\mJ}{\mathcal{J}}
\newcommand{\mN}{\mathcal{N}}
\newcommand{\mM}{\mathcal{M}}
\newcommand{\mO}{\mathcal{O}}

\newcommand{\mH}{\mathcal{H}}
\newcommand{\mR}{\mathcal{R}}
\newcommand{\mS}{\mathcal{S}}
\newcommand{\mU}{\mathcal{U}}
\newcommand{\mX}{\mathcal{X}}
\newcommand{\T}{\mathbf{T}}

\newcommand{\1}{\mathbb{1}}

\def\ANU{Centre for Quantum Computation and Communication Technologies, Research School of Physics, Australian National University, Acton, ACT 2601, Australia}
\def\CQT{Centre for Quantum Technologies (CQT), National University of Singapore, Singapore 117543, Republic of Singapore}

\makeatletter
\newcommand*{\rom}[1]{\expandafter\@slowromancap\romannumeral #1@}
\makeatother

\newcommand{\orcid}[1]
{
\href{https://orcid.org/#1}{\includegraphics[scale=0.05]{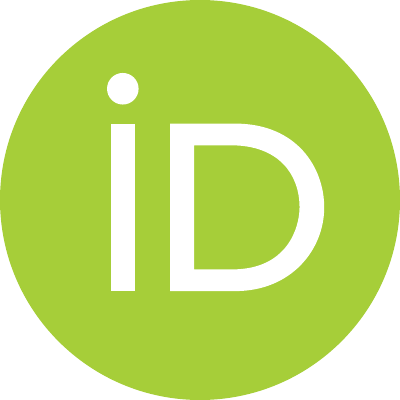}}
}

\usepackage{amsthm}
\newtheorem*{thm*}{Theorem}

\usepackage{soul}
\usepackage{tcolorbox}
\tcbuselibrary{theorems}

\newtcbtheorem[number within=section, use counter*=thm]{mythm}{Theorem}%
{colback=red!5,colframe=red!35!red,fonttitle=\bfseries}{thm}

\usepackage{tcolorbox}
\tcbuselibrary{theorems}

\newtcbtheorem[number within=section, use counter*=thm]{mylem}{Lemma}%
{colback=green!5,colframe=green!35!black,fonttitle=\bfseries}{lem}

\usepackage{tcolorbox}
\tcbuselibrary{theorems}

\newtcbtheorem[number within=section, use counter*=thm]{mydef}{Definition}%
{colback=yellow!5,colframe=yellow!70!black,fonttitle=\bfseries}{def}

\usepackage{tcolorbox}
\tcbuselibrary{theorems}

\newtcbtheorem[number within=section, use counter*=thm]{mycor}{Corollary}%
{colback=blue!5,colframe=blue!70!black,fonttitle=\bfseries}{cor}

\usepackage{tcolorbox}
\tcbuselibrary{theorems}

\newtcbtheorem[number within=section, use counter*=thm]{myprop}{Proposition}%
{colback=pink!5,colframe=pink!70!black,fonttitle=\bfseries}{prop}

\pgfplotsset{compat=1.17}
\begin{document}

\title{\texorpdfstring{$100\pm\Delta t$ Years of Quantum Uncertainty: From Origins to Modern Insights}{$100\pm\Delta t$ Years of Quantum Uncertainty: From Origins to Modern Insights}}


\author{Lorc\'{a}n O. Conlon\orcid{0000-0002-0921-5003}}
\affiliation{Quantum Innovation Centre (Q.InC), Agency for Science Technology and Research (A*STAR), 2 Fusionopolis Way, Innovis \#08-03, Singapore 138634, Republic of Singapore}
\affiliation{\CQT}
\affiliation{Joint Center for Quantum Information and Computer Science, NIST/University of Maryland, College Park, MD, 20742, USA}

\author{Biveen Shajilal\orcid{0000-0002-1374-8566}}
\affiliation{Quantum Innovation Centre (Q.InC), Agency for Science Technology and Research (A*STAR), 2 Fusionopolis Way, Innovis \#08-03, Singapore 138634, Republic of Singapore}

\author{Jie Zhao\orcid{0000-0002-7382-1964}}
\affiliation{\ANU}

\author{Tim C. Ralph\orcid{0000-0003-0692-8427}}
\affiliation{Centre for Quantum Computation and Communication Technology, School of Mathematics and Physics, University of Queensland, St Lucia, Queensland 4072, Australia}

\author{Gerd Leuchs\orcid{0000-0003-1967-2766}}
\affiliation{Physik Department, Naturwissenschaftliche Fakult\"at, Friedrich-Alexander-Universit\"at Erlangen-N\"urnberg, 91058 Erlangen, Germany}
\affiliation{Max-Planck-Institut f\"ur die Physik des Lichts, 91058 Erlangen, Germany}
\affiliation{Department of Physics, Faculty of Science, University of Ottawa, Ottawa ON K1N7N9, Canada}

\author{Ulrik L. Andersen\orcid{0000-0002-1990-7687}}
\affiliation{Center for Macroscopic Quantum States (bigQ), Department of Physics, Technical University of Denmark, Fysikvej, Kongens Lyngby, Denmark}

\author{Syed M. Assad\orcid{0000-0002-5416-7098}}
\affiliation{Quantum Innovation Centre (Q.InC), Agency for Science Technology and Research (A*STAR), 2 Fusionopolis Way, Innovis \#08-03, Singapore 138634, Republic of Singapore}

\author{Ping Koy Lam\orcid{0000-0002-4421-601X}}
\email{pingkoylam@gmail.com}
\affiliation{Quantum Innovation Centre (Q.InC), Agency for Science Technology and Research (A*STAR), 2 Fusionopolis Way, Innovis \#08-03, Singapore 138634, Republic of Singapore}
\affiliation{\CQT}

\author{Yunlong Xiao\orcid{0000-0001-5414-7946}}
\affiliation{Quantum Innovation Centre (Q.InC), Agency for Science Technology and Research (A*STAR), 2 Fusionopolis Way, Innovis \#08-03, Singapore 138634, Republic of Singapore}
\affiliation{Institute of High Performance Computing (IHPC), Agency for Science, Technology and Research (A*STAR), 1 Fusionopolis Way, \#16-16 Connexis, Singapore 138632, Republic of Singapore}


\date{\today}
             

\begin{abstract} 
Heisenberg's uncertainty principle is a cornerstone of quantum mechanics, marking a decisive departure from classical physics. 
Conceived almost a century ago through a thought experiment showing that measuring an electron's position inevitably disturbs its momentum, it began as a deceptively simple idea that sparked countless studies and grew into the rich research field it is today. 
This review traces its development into a spectrum of mathematical formulations -- known as uncertainty relations -- and explores their interconnections and wide-ranging applications. 
We highlight its central role in quantum metrology, where it underpins strategies for extracting information from quantum systems with ever-increasing precision, and its links to multiparameter estimation and squeezed states. 
This review, dedicated to the centenary of the uncertainty principle, reflects on how it has deepened our understanding of quantum theory and driven practical advances, and looks ahead to a century poised for further surprising and transformative discoveries.

\end{abstract}


\maketitle

\tableofcontents


\section{Introduction}

Uncertainty is a defining feature of our physical world, present in both classical and quantum systems. 
In the macroscopic realm, for example, when we roll a die, each face has a one-in-six chance of appearing; when we check the weather, forecasts speak in probabilities. These examples illustrate uncertainty in our knowledge of a system. A more mathematical form of classical uncertainty appears in wave mechanics: via the Fourier transform, a signal cannot be perfectly localized in both time and frequency -- the so-called time-bandwidth product. 
In quantum mechanics, however, this limitation takes on a fundamentally different character.
Here, the act of measurement itself -- inherently probabilistic, as dictated by the Born rule~\cite{born1926quantenmechanik} -- imposes intrinsic limits, giving rise to Heisenberg's uncertainty principle. 
This subtle difference reflects the fundamental departure of quantum mechanics from the classical world.

Quantum uncertainty was first explored in Heisenberg's famous $\gamma$-ray microscope thought experiment~\cite{heisenberg1927illustrativeTranslation}. 
To determine an electron's position, one must illuminate it with light; yet the very photon that scatters to reveal the electron's location imparts an uncontrollable ``kick'' to its momentum through the {\it Compton effect}. 
Thus, the act of pinpointing position inevitably alters momentum in a discontinuous and unpredictable way. 
Unlike in classical physics: the corresponding observables do not commute, and measuring one inevitably disturbs the other. 
This heuristic insight gave rise to what we now call {\it Heisenberg's uncertainty principle} -- one of the cornerstones of quantum theory -- written in every textbook on quantum mechanics.

Heisenberg's original work offered no precise mathematical expression for the trade-off between the uncertainties of an electron's position and momentum, leaving it open to multiple interpretations~\cite{heisenberg1927illustrativeTranslation}. 
Heisenberg, in his own words, discusses the \textquote{mean error} in the determination of position and the corresponding \textquote{discontinuous change of momentum}. 
In this sense, it is common to interpret Heisenberg's uncertainty principle through a trade-off between error and disturbance caused by measurements.  
In the same work, Heisenberg also considered a wavefunction with a Gaussian spread in both position and momentum, highlighting the uncertainty inherent in state preparation. 

The century that followed was marked by sustained efforts to cast the uncertainty principle into a precise mathematical framework.
As the subject evolved, new conceptual viewpoints continually reshaped its interpretation, giving rise to a rich landscape of formulations now collectively known as uncertainty relations. 
In a sense, {\it there are a thousand uncertainty principles in a thousand physicists' eyes}.
Yet beneath this diversity, nearly all developments can ultimately be traced back to two fundamental guiding questions:
\begin{itemize}
  \item How should we quantify the disturbance caused by a prior measurement on the system?
  \item Is it possible to prepare a state that is simultaneously an eigenstate of the observables of interest?
\end{itemize}
These questions frame two complementary perspectives on the uncertainty principle. 
The first leads to the {\it measurement uncertainty relation}, which captures the error-disturbance trade-off inherent in probing a single quantum system. 
The second leads to the {\it preparation uncertainty relation}, which considers large ensembles of identically prepared states, partitioned into subsets, each measured with a different observable. 
Together, they reveal the uncertainty principle not as a single statement, but as a rich landscape of limits on what can be known and prepared in the quantum world.

The uncertainty principle has played an essential role in building the foundations of modern quantum mechanics, most notably in entanglement theory. 
The uncertainty principle served as the fulcrum for the Einstein, Podolsky and Rosen (EPR) argument~\cite{PhysRev.47.777}, which in turn sparked Schr\"{o}dinger’s identification of entanglement~\cite{schrodinger1935discussion}. 
Today, uncertainty relations remain central in the detection and quantification of entanglement~\cite{duan2000inseparability,simon2000peres}. 
However, understanding the uncertainty principle at a fundamental level is only part of its story. 
Just as compelling is how its many forms have driven practical advances. 
One of the first -- and perhaps most influential -- was in gravitational-wave detection. 
In the 1970s, physicists debated whether quantum mechanics could push beyond classical sensitivity limits~\cite{maddox1988beating}. 
The question was resolved in the affirmative, and quantum principles are now integral to detecting these faint ripples in spacetime. 
Since then, the toolkit of sensing has expanded to include the use of quantum states, such as highly entangled states~\cite{bollinger1996optimal,doi:10.1126/science.1104149,giovannetti2011advances,marciniak2022optimal} and squeezed light~\cite{caves1981quantum}, exotic interactions, such as interactions without definite causal order~\cite{zhao2020quantum,yin2023experimental} and non-linear interactions~\cite{roy2008exponentially,boixo2007generalized}, and collective or entangling measurements~\cite{conlon2023approaching,hou2018deterministic}.
Together, these advances have shaped quantum sensing into one of the clearest demonstrations that the uncertainty principle is not merely a constraint, but a powerful tool.

Quantum key distribution (QKD) is another area that builds directly on the uncertainty principle. 
The original protocol, proposed by Bennett and Brassard in 1984, exploits the fact that conjugate variables of a quantum state cannot be measured simultaneously -- an embodiment of uncertainty at the heart of its security~\cite{bennett2014quantum}. 
In recent years, rigorous security proofs for quantum cryptography have been developed using memory assisted entropic uncertainty relations~\cite{Berta2010}. The uncertainty principle also has relevance for quantum computing. 
In optical quantum computing, for example, Gottesman-Kitaev-Preskill (GKP) qubits encode information in states whose quality can be quantified by how tightly quantum uncertainty is squeezed. 
As quantum technologies progress, further unexpected and powerful applications are certain to follow.
Even outside the laboratory, the uncertainty principle continues to capture the imagination, appearing throughout popular culture as a symbol of the strange and profound limits -- and possibilities -- imposed by the laws of physics~\cite{endgame,endgame2}.

For almost a century, the uncertainty principle has stood as a cornerstone of quantum theory -- defining the limits of what can be known while continually opening new frontiers of inquiry. 
From Heisenberg's original insight, it has evolved into a rich tapestry of uncertainty relations, each exposing a different facet of nature's fundamental constraints. 
These advances have reshaped our understanding of quantum mechanics and found many applications.
Yet, there remains no single account that weaves together its many formulations, explores the limitations of each, and surveys its wide-ranging applications. 
To mark the centenary of Heisenberg's uncertainty principle, we present a comprehensive review that interlaces these diverse threads into a coherent whole -- charting its conceptual evolution, surveying its many applications, and looking ahead to the challenges and opportunities that will define its second century.


\subsection{Outline of the Review}
\label{sec:Outline}

This review does not attempt the impossible task of covering every development, formulation, and application of the uncertainty principle. 
Instead, our goal is to highlight the most significant advances in understanding the principle and to focus in depth on one application of broad relevance to experimental physics: quantum sensing. 
The review is organized into two main parts. 
The first part (Sec.~\ref{sec:Uncertainty_Relations}) surveys the landscape of uncertainty relations. 
Secs.~\ref{sec:VUR}-\ref{sec:ET} summarize and compare the principal formulations, highlight the links between them, and review the key applications of each.
Specifically, Sec.~\ref{sec:VUR} reviews the conventional variance-based uncertainty relations, followed in Sec.~\ref{sec:EUR} by the information-theoretic formulations expressed in terms of entropies, and in Sec.~\ref{sec:MUR} by more general approaches based on majorization. 
Sec.~\ref{sec:DUR} extends the discussion from states to channels and, more broadly, to non-Markovian quantum dynamics. 
These sections focus on {\it preparation uncertainty relations}. 
Sec.~\ref{sec:EDUR} turns to developments in measurement uncertainty relations, Sec.~\ref{sec:TEUR} examines the various formulations of time-energy uncertainty relations, and Sec.~\ref{sec:ET} discusses other notable variants.
This part concludes with Sec.~\ref{sec:experimentaltestsUC}, which reviews key experimental tests of these relations. 
The second part (Sec.~\ref{sec:Metrology}) examines how the uncertainty principle shapes our interaction with the physical world, with a particular emphasis on sensing. Sec.~\ref{sec:SQLHS} discusses the connections between the uncertainty principle and quantum metrology, showing how surpassing classical sensitivity limits (achieving Heisenberg scaling) can be viewed as an expression of the uncertainty principle itself. 
Sec.~\ref{subsec:multiparameter} explores its role in the simultaneous estimation of multiple parameters, and Sec.~\ref{sec:ET2} summarises other important aspects of multiparameter estimation. Sec.~\ref{section:squeezedlight} considers squeezed states -- their connection to the uncertainty principle, methods of realization, and applications. 
We conclude in Sec.~\ref{conclusion} with an outlook that synthesizes the key challenges and opportunities that lie ahead.
Additional material clarifying the details of the review is provided in the Appendix.


\section{
Part 1 -- 
From Heisenberg to Modern Frameworks: 
The Many Faces of the Uncertainty Principle
}
\label{sec:Uncertainty_Relations}

In the century since Heisenberg's seminal insight, the uncertainty principle has blossomed into a rich and rigorous theoretical framework, giving rise to a diverse landscape of formulations. 
These broadly fall into two principal categories, each illuminating a distinct facet of quantum uncertainty. 
{\it Measurement uncertainty relations} articulate the fundamental trade-off between the precision of a measurement and the disturbance it induces on incompatible observables, reflecting the operational constraints intrinsic to a single measurement process (see Sec.~\ref{sec:EDUR}). 
In contrast, {\it preparation uncertainty relations}, more commonly presented in textbooks, bound the statistical spread of measurement outcomes across independent and identically distributed (i.i.d.) ensembles (see Sec.~\ref{sec:VUR}, \ref{sec:EUR}, \ref{sec:MUR}, \ref{sec:DUR}, \ref{sec:TEUR}, and~\ref{sec:ET}). 
Despite omitting direct quantification of measurement-induced effects, these relations have become indispensable tools in quantum information and computation, underpinning applications from entanglement detection to quantum cryptography. 
In what follows, we survey key developments across both paradigms, highlighting how uncertainty has transformed from a limit on knowledge into a versatile resource for quantum science and technologies.


\subsection{Variance-Based Uncertainty Relations}
\label{sec:VUR}

Variance and standard deviation were the earliest rigorous tools used to quantify uncertainty in quantum theory. 
Defined through the statistical spread of outcomes from repeated measurements on identically prepared systems, they reflect a natural extension of the classical fluctuation and measurement imprecision, and hence form the foundation for the earliest formulations of the uncertainty principle. 
Secs.~\ref{sec:VUR_PBF} and~\ref{sec:VUR_SBF} review these variance-based uncertainty relations in detail, highlighting both the product and sum formulations.


\subsubsection{Product-Based Formulation}\label{sec:VUR_PBF}

The earliest attempt to cast Heisenberg's thought experiment~\cite{heisenberg1927illustrativeTranslation} into a rigorous mathematical form was independently undertaken by Kennard~\cite{kennard1927quantenmechanik} and Weyl~\cite{weyl1928gruppentheorie}. 
Their formulations led to the first precise expression of the uncertainty principle, now known as the position-momentum uncertainty relation, which reads
\begin{align}\label{eq:VUR_Kennard}
    \Delta x\Delta p\geqslant\frac{\hbar}{2},
\end{align}
where $\Delta X:=(\langle X^2\rangle-\langle X\rangle^2)^{1/2}$ is the standard deviation, $x$ and $p$ denote the position and momentum observables, respectively.
This relation is widely taken to embody Heisenberg's original insight, but in truth, it reflects a different aspect of quantum uncertainty. 
Rather than addressing what can be known from a single system, it captures the spread in outcomes when position and momentum are measured across many i.i.d. states. 
In modern terms, this is a preparation uncertainty relation. 
It does not speak to the disturbance caused by measuring one observable on a system and then trying to measure another -- in other words, the measurement uncertainty that was central to Heisenberg's argument. 
We return to this distinction in Sec.~\ref{sec:EDUR}.

Robertson later generalized the position-momentum uncertainty relation to any pair of observables $A$ and $B$, demonstrating that~\cite{PhysRev.34.163}
\begin{align}\label{eq:VUR_Robertson}
    \Delta A\Delta B\geqslant\frac{1}{2}|\langle[A,B]\rangle|.
\end{align}
Here, the lower bound is set by the commutator $[A,B]:=AB-BA$, with all expectation values and variances taken in the chosen quantum state. Upon specializing to the canonically conjugate pair $x$ and $p$, which satisfy $[x,p]=i\hbar$, one immediately recovers Eq.~\eqref{eq:VUR_Kennard}. For the treatment of angle and angular momentum uncertainty relations, see Refs.~\cite{PhysRevLett.18.182,Franke-Arnold_2004,PhysRevLett.97.243601}.

The proof of Robertson's uncertainty relation in Eq.~\eqref{eq:VUR_Robertson} is a direct consequence of the Cauchy–Schwarz inequality. To this end, define the operator $\bar{A}:=A-\langle A\rangle\1$, so that in the state $\psi$, $\Delta A=\|\bar{A}\ket{\psi}\|$. Rewriting both $\Delta A$ and $\Delta B$ in norms, Cauchy–Schwarz yields
\begin{align}
    \Delta A\Delta B
    =\|\bar{A}\ket{\psi}\|\cdot\|\bar{B}\ket{\psi}\|
    \geqslant
    |\langle\bar{A}\bar{B}\rangle|
\end{align}
By splitting the operator product $\bar{A}\bar{B}$ into its Hermitian and anti-Hermitian parts, one finds
\begin{align}\label{eq:VUR_comm-anticomm}
    |\langle\bar{A}\bar{B}\rangle|
    =
    \sqrt{\left(\frac{1}{2i}\langle[\bar{A},\bar{B}]\rangle\right)^2
    +\left(\frac{1}{2}\langle\{\bar{A},\bar{B}\}\rangle\right)^2}.
\end{align}
Here, $\{\bar{A},\bar{B}\}:=\bar{A}\bar{B}+\bar{B}\bar{A}$ denotes the anti-commutator of the operators $\bar{A}$ and $\bar{B}$. 
Omitting the anti-commutator term in Eq.~\eqref{eq:VUR_comm-anticomm} recovers Robertson uncertainty relation in Eq.~\eqref{eq:VUR_Robertson}, whereas its inclusion leads to Schr\"odinger's uncertainty relation~\cite{schrodinger1930heisenbergschen}
\begin{align}\label{eq:VUR_Schrodinger}
    \Delta A^2\Delta B^2
    \geqslant
    \left|\frac{1}{2}\langle[A,B]\rangle\right|^2+
    \left|\frac{1}{2}\langle\{A,B\}\rangle-\langle A\rangle\langle B\rangle\right|^2.
\end{align}

In many quantum mechanics textbooks, Eq.~\eqref{eq:VUR_Robertson} is often introduced as a mathematical extension of Eq.~\eqref{eq:VUR_Kennard}, yet these two embody fundamentally different physics. 
Kennard's relation in Eq.\eqref{eq:VUR_Kennard} imposes a state-independent lower bound that captures the intrinsic incompatibility of position and momentum. 
By contrast, the Robertson bound in Eq.~\eqref{eq:VUR_Robertson} depends explicitly on the state: if the system is prepared in an eigenstate of either $A$ or $B$, then the bound vanishes -- even though $[A,B]\neq0$ -- and the relation becomes trivial, thus failing to signal any genuine incompatibility.

For an uncertainty relation to possess operational content, its bound should be derivable with strictly less information than that required to compute the actual uncertainty itself. 
In Robertson's formulation, computing the lower bound requires full knowledge of the quantum state and both observables -- the very inputs needed to calculate the standard deviations -- thereby nullifying its practical value: one might just as well evaluate the true uncertainty outright. 
More generally, whenever a lower bound $\mB(A, B, \psi)$, e.g. the right‐hand side of Eqs.~\eqref{eq:VUR_Robertson} and \eqref{eq:VUR_Schrodinger}, and the joint uncertainty $\mJ(A, B, \psi)$, e.g. $\Delta A\Delta B$, depend on the identical data, namely, the classical descriptions of $A$, $B$, and $\psi$, the bound carries no operational content, as for any $p\in[0,1]$, the convex combination $p\mJ(A, B, \psi)+(1-p)\mB(A, B, \psi)$ provides a tighter bound without requiring any additional information
\begin{align}\label{eq:VUR_trivial}
    \mJ\geqslant p\mJ+(1-p)\mB\geqslant\mB.
\end{align}
 Therefore, only when obtaining the uncertainty bound requires strictly less data, experimental constraints, or computational complexity does the resulting inequality carry genuine operational significance.

Additional challenges concerning state‐dependent bounds emerge when one extends uncertainty relations to three or more observables. 
For example, consider the canonical triple $x$, $p$, and $r:=-x-p$, each pair of which satisfies the same commutation relation as position and momentum. 
In this setting, one obtains the following natural generalization of Kennard's uncertainty relation~\cite{PhysRevA.90.062118}
\begin{align}\label{eq:VUR_Kechrimparis_Weigert}
    \Delta x\Delta p\Delta r\geqslant\left(\frac{\tau\hbar}{2}\right)^{\frac{3}{2}}.
\end{align}
This bound is genuinely nontrivial: it cannot be recovered by merely combining pairwise uncertainty relations for canonically conjugate pair, but instead relies on the triple constant $\tau:=2/\sqrt{3}$. 
Moreover, it is tight -- there exists a generalized squeezed state that attains equality -- emphasizing the fundamental physical significance of $\tau$. 

In contrast, attempts to extend Robertson's form to three observables introduces new difficulties. 
For instance, Ref.~\cite{PhysRevLett.118.180402} proposes an uncertainty relation for angular momentum components $S_x$, $S_y$, and $S_z$ in a spin-$1/2$ system
\begin{align}\label{eq:VUR_Ma}
    \Delta S_x\Delta S_y\Delta S_z\geqslant\left|\frac{\tau^3}{8}\langle S_x\rangle\langle S_y\rangle\langle S_z\rangle\right|^{\frac{1}{2}}.
\end{align}
Although the parameter $\tau$ reappears in this setting, it inherits the same limitations already identified for Robertson's uncertainty relation.

The above results admit several natural extensions. First, uncertainty relations can be generalized from pairs (or triplets) to any finite collection of $N$ observables~\cite{PhysRev.46.794}. 
Second, one can also formulate analogous uncertainty relations for unitary operators~\cite{PhysRevLett.100.190401,PhysRevA.93.052118,PhysRevA.94.042104,PhysRevLett.120.230402,xiao2022near,PhysRevA.100.022116,Xiao2017Experimental,Qu2021Experimental}, and even certain non-Hermitian~\cite{PhysRevD.86.064038,PhysRevA.92.052120,PhysRevA.99.022108,zhao2022uncertainty,PhysRevA.107.042201,PhysRevLett.132.070203,PhysRevA.111.012221}, operators. 
In all such cases, however, any state-dependent lower bound remains fundamentally constrained by the same conceptual and operational limitations that underlie Robertson's uncertainty relation.


\subsubsection{Sum-Based Formulation}
\label{sec:VUR_SBF}

One systematic approach to overcome the limitations of product‐form uncertainty relations is to reformulate the trade-off as a sum of variances or standard deviations. Rather than bounding
$\Delta A\Delta B$, we consider the combined quantity $\Delta A^2+\Delta B^2$. In this framework, the uncertainty vanishes if and only if the state $\psi$ is a common eigenstate of both $A$ and $B$. It then follows immediately that 
\begin{align}\label{eq:VUR_Sum}
    \Delta A^2+\Delta B^2
    \geqslant
    2\Delta A\Delta B
    \geqslant
    |\langle[A,B]\rangle|.
\end{align}
The above variance sum relation still suffers from the usual drawbacks of state‐dependent bounds (see Sec.~\ref{sec:VUR_PBF}): both the left‐ and right‐hand sides rely on the same information about observables $A$, $B$, and quantum state $\psi$, so the lower bound can be tightened trivially and carries little operational meaning. By contrast, reformulating uncertainty in a sum‐of‐variances form enables genuinely state‐independent bounds. For example, in a $d$-dimensional system, the spin‐$s$ operators $S_x$, $S_y$, and $S_z$, with $d=2s+1$, satisfy~\cite{PhysRevA.68.032103} 
\begin{align}
    \Delta S_x^2 + \Delta S_y^2 + \Delta S_z^2\geqslant s=\frac{d-1}{2}.
\end{align}
where the bound is state‐independent and tight. Here the eigenvalues of a spin‐$s$ operator range over $\{-s,-s+1,\cdots,s-1,s\}$. The corresponding bound for two spin components is considerably more intricate, and studies show that it grows asymptotically as $s^{2/3}$~\cite{PhysRevA.84.022107,Dammeier_2015}.

In particular, when $s=1/2$, each $S_i$ ($i=x,y,z$) takes the values $\pm1/2$. Rescaling by $1/s$ then yields the Pauli matrices $\sigma_i$~\cite{Pauli1927}, whose variances satisfy the corresponding uncertainty relation
\begin{align}
    \Delta \sigma_x^2 + \Delta \sigma_y^2 + \Delta \sigma_z^2\geqslant2.
\end{align}
Moreover, since each Pauli operator $\sigma_i$ satisfies 
$\Delta \sigma_i^2\leqslant1$, it immediately follows that for any two distinct Pauli matrices, we have $\Delta \sigma_i^2 + \Delta\sigma_j^2\geqslant1$.
This class of uncertainty relations admits a natural extension to complete sets of local orthogonal observables (LOOs)~\cite{PhysRevLett.95.150504}. 
For a $d$-dimensional quantum system, one may introduce a collection of $d^2$ orthogonal observables $\{\mO_i\}$, forming an operator basis with respect to the Hilbert-Schmidt inner product, from which the following inequality follows~\cite{PhysRevA.74.010301}
\begin{align}
    \sum_{i=1}^{d^2}\Delta \mO_i^2\geqslant d-1.
\end{align}
In the qubit setting, this construction reduces to the familiar Pauli operators, where the LOOs are given by $\{\1/\sqrt{2}, \sigma_x/\sqrt{2}, \sigma_y/\sqrt{2}, \sigma_z/\sqrt{2}\}$.

Employing Bloch‐sphere notation, the spectral projections of an observable $A$ can be written as $(\1\pm\vec{a}\cdot\vec{\sigma})/2$, where $\vec{\sigma}:=(\sigma_x, \sigma_y, \sigma_z)^{\T}$ are the Pauli matrices and $\vec{a}\in\mathbb{R}^3$ is a unit vector. A general qubit state can be expressed as $\rho=(\1+\vec{r}\cdot\vec{\sigma})/2$ with Bloch vector $|r|\leqslant1$, from which the standard deviation of $A$ follows as $\Delta A_{\rho}:=(1-(\vec{r}\cdot\vec{a})^2)^{1/2}$. Busch, Lahti, and Werner demonstrate in~\cite{PhysRevA.89.012129} that for any two dichotomic ($\pm1$-valued) qubit observables $A$ and $B$, the uncertainties satisfy two complementary, tight bounds. Firstly, the sum of standard deviations obeys the state-independent relation
\begin{align}\label{eq:VUR_Sum_qubit_SD}
    \Delta A_{\rho}+\Delta B_{\rho}
    \geqslant
    \|\vec{a}\times\vec{b}\|,
\end{align}
where $\times$ denotes the cross product between vectors. Eq.~\eqref{eq:VUR_Sum_qubit_SD} is tight -- saturated precisely when the Bloch vector of state aligns or anti‐aligns with one of the measurement axes, i.e., $\vec{r}=\pm\vec{a}$ or $\vec{r}=\pm\vec{b}$. Secondly, the corresponding sum-of-variances bound reads
\begin{align}\label{eq:VUR_Sum_qubit_two_V}
    \Delta A_{\rho}^2+\Delta B_{\rho}^2\geqslant1-|\vec{a}\cdot\vec{b}|,
\end{align}
with the tight lower bound set by the inner product between vectors $\vec{a}$ and $\vec{b}$. Since adding any multiple of the identity leaves the variance invariant, we may, without loss of generality, restrict to traceless observables of the form $A_i:=\vec{n_i}\cdot\vec{\sigma}$, where $\vec{n_i}$ is not assumed to be normalized. In Ref.~\cite{PhysRevA.100.032118}, Xiao {\it et al.} solve the general uncertainty problem for an arbitrary finite set of $N$ such observables, showing that the optimal sum‐of‐variances bound is
\begin{align}\label{eq:VUR_Sum_qubit_N_V}
    \sum_{i=1}^{N}\left(\Delta A_i\right)_{\rho}^2
    \geqslant
    \sum_i\|\vec{n_i}\|^2+\lambda_{\max}\left(\sum_i\ketbra{\vec{n_i}}{\vec{n_i}}\right).
\end{align}
Here, $\lambda_{\max}$ denotes the largest eigenvalue of the Hermitian matrix $\sum_i\ketbra{\vec{n_i}}{\vec{n_i}}$ acting on $\mathbb{R}^3$. Equality holds precisely when the state is chosen as the eigenvector associated with this maximal eigenvalue.

Despite extensive investigations into uncertainty relations, a truly unified framework for bounding the joint uncertainties of arbitrary observables in terms of their standard deviations, and for identifying the optimal such bound, remains elusive. 
Consequently, two complementary strategies have been developed. 
The first imposes specific constraints on the observables -- such as limiting their eigenvalue spectrum or enforcing algebraic relations -- to derive tighter, often optimal, bounds. 
For instance, by restricting to dichotomic operators $\{A_i\}_i$ with $A_i^2=\1$, and examining their anti-commutativity graph $G$, whose vertices represent $A_i$, joining two vertices exactly when the corresponding operators anti-commute (i.e.\ $\{A_i,A_j\}=0$), de Gois, Hansenne, and G\"uhne use the Lov\'asz theta number $\vartheta(G)$~\cite{1055985} of the graph to establish nontrivial lower bounds on the sum of variances~\cite{PhysRevA.107.062211}. In particular, the joint uncertainty obeys
\begin{align}\label{eq:VUR_Sum_GHG}
    \sum_{i=1}^{N}\left(\Delta A_i\right)_{\rho}^2
    =
    N-\sum_{i=1}^{N}\langle A_i\rangle_{\rho}^2
    \geqslant
    N-\vartheta(G).
\end{align}
This approach works because the Lov\'asz theta number provides an upper bound on the quadratic form $\sum_{i}\langle A_i\rangle_{\rho}^2$.

Building on this insight, Bermejo Mor\'an and Huber employ state-polynomial optimization~\cite{doi:10.1137/090760155,PhysRevLett.123.140503,Klep2024} to define a semidefinite programming (SDP) hierarchy that systematically tightens the $\vartheta(G)$-based bound~\cite{PhysRevLett.132.200202}. 
Specifically, letting $\beta(G)$ denote the maximal value of $\sum_{i}\langle A_i\rangle_{\rho}^2$ over all quantum states, one obtains the following chain of inequalities
\begin{align}
    \alpha(G)\leqslant\beta(G)\leqslant\cdots\leqslant
    \vartheta_2(G)\leqslant\vartheta_1(G)=
    \vartheta(G),
\end{align}
where each $\vartheta_i(G)$ admits an SDP characterization and is therefore efficiently computable~\cite{PhysRevLett.132.200202}, while $\alpha(G)$ is the independence number of the graph $G$.
At its first level, the hierarchy reproduces the Lov\'asz theta number as $\vartheta_1(G)=\vartheta(G)$; each subsequent level enforces stronger moment constraints and thus yields improved variance bounds. 
Moreover, if at any hierarchy level the SDP upper bound coincides with the graph's independence number $\alpha(G)$, the resulting uncertainty bound is optimal. 
As an illustrative example, consider the wheel graph $W_6$, namely
\begin{equation}
W_6 = \text{
\begin{tikzpicture}[baseline=(center.base), scale=0.6]
    \tikzset{
        node/.style={circle, fill=black, inner sep=1.5pt},
        edge/.style={thick}
    }
    \node[node] (center) at (0,0) {};
    \foreach \i in {1,...,5} {
        \node[node] (v\i) at ({90 + 72 * (\i - 1)}:1.2cm) {};
    }
    \foreach \i in {1,...,5} {
        \pgfmathtruncatemacro{\next}{1 + mod(\i, 5)}
        \draw[edge] (center) -- (v\i);
        \draw[edge] (v\i) -- (v\next);
    }
\end{tikzpicture}
}
,
\end{equation}
which gives rise to the hierarchy of inequalities
\begin{align}
    2=\alpha(W_6)=\vartheta_2(W_6)<\sqrt{5}=
    \vartheta_1(W_6)=\vartheta(W_6).
\end{align}
From this chain, one finds that the optimal value of $\beta(W_6)$ is 2, leading directly to the optimal uncertainty relation
\begin{align}\label{eq:Uncertainty_W_6}
    \sum_{i=1}^{6}\left(\Delta A_i\right)_{\rho}^2\geqslant4.
\end{align}
It is important to emphasize that uncertainty relations derived from graph-theoretic approaches are intrinsically restricted to dichotomic observables whose incompatibility structure is represented by the associated anticommutativity graph.

While this framework presumes specific algebraic relations among the observables, it promises wide applicability: given the central role of the Lov\'asz theta number in characterizing Bell nonlocality~\cite{PhysRevLett.112.040401}, contextuality~\cite{Acin2015}, and zero‐error communication~\cite{6319408}, the graph‐theoretic uncertainty relations may deepen the conceptual links between uncertainty relations and these foundational quantum phenomena.
Graph-theoretical methods constitute a powerful framework for formulating uncertainty relations, but they are not the sole avenue. 
Approaches grounded in Lie algebraic structures offer complementary perspectives~\cite{PhysRevA.98.042121}, particularly in systems with underlying symmetries.

The second strategy departs from tightening variance-based bounds and instead lowers uncertainty by employing alternative metrics from information and statistical theory, such as the Wigner-Yanase skew information and the quantum Fisher information. In their information-theoretic analysis of conservation laws and measurement theory, Wigner and Yanase defined the skew information as~\cite{wigner1963information}
\begin{align}\label{eq:skew_information}
    I(\rho,A):=-\frac{1}{2}\Tr[[\sqrt{\rho},A]^2],
\end{align}
where the inner square brackets denote the commutator. Building on this, Luo derived the following skew-information-based uncertainty relation~\cite{PhysRevLett.91.180403}
\begin{align}\label{eq:SI_Luo}
    I(\rho,A)I(\rho,B)\geqslant\frac{1}{4}|\Tr[\rho[A,B]]|^2,
\end{align}
which links the skew information across a set of observables to their collective non-commutativity in state. Because variance never falls below skew information $\Delta A_{\rho}^2\geqslant I(\rho,A)$, Eq.~\eqref{eq:SI_Luo} therefore delivers a stronger form than Robertson's relation, and in the pure‐state limit, where $\Delta A_{\psi}^2= I(\psi,A)$, it reduces exactly to Robertson's uncertainty relation in Eq.~\eqref{eq:VUR_Robertson}. 

This idea carries over to sum‐of‐uncertainty relations as well. Before presenting related results, we recall the requisite notation and key concepts. 
In particular, T\'oth and Petz showed that for any mixed state, the variance admits a concave‐roof representation~\cite{PhysRevA.87.032324}
\begin{align}\label{eq:Variance_concave_roof}
    \left(\Delta A\right)_{\rho}^2
    =
    \sup_{\{p_k,\psi_k\}}\sum_{k}p_k
    \left(\Delta A\right)_{\psi_k}^2,
\end{align}
where the supremum is taken over all pure‐state decompositions $\rho=\sum_k p_k\psi_k$. Their conjecture that the quantum Fisher information $J_\text{S}(\rho,A)$ equals four times the convex roof of the variance was later confirmed by Yu in Ref.~\cite{yu2013quantumfisherinformationconvex}, who proved
\begin{align}\label{eq:QFI_convex_roof}
    \frac{1}{4}J_\text{S}(\rho,A)
    =
    \inf_{\{p_k,\psi_k\}}\sum_{k}p_k
    \left(\Delta A\right)_{\psi_k}^2.
\end{align}
It then follows that $\left(\Delta A\right)_{\rho}^2\geqslant J_\text{S}(\rho,A)/4$. Both the quantum Fisher information $J_\text{S}(\rho,A)$ and the Wigner–Yanase skew information $I(\rho,A)$ (up to a scalar) are special cases of the generalized quantum Fisher information~\cite{PhysRevA.87.032324}, with $J_\text{S}(\rho,A)$ attaining the maximal value.
As it will be used extensively in Sec.~\ref{sec:Metrology}, we present the formal definition of the quantum Fisher information in Eq.~\eqref{eq:SLDQFIBG} of that section.

By exploiting the convexity of the quantum Fisher information and the concavity of the variance, Chiew and Gessner showed that any sum‐of‐variances uncertainty relation can be tightened by replacing a single variance term with one‐quarter of the corresponding quantum Fisher information~\cite{PhysRevResearch.4.013076}. For instance, integrating the quantum Fisher information substitution into the graph‐theoretic uncertainty framework yields a stronger uncertainty relation. Accordingly, Eq.~\eqref{eq:VUR_Sum_GHG} admits the following refinement
\begin{align}\label{eq:VUR_Sum_GHG_QFI}
    \frac{1}{4}J_\text{S}(\rho,A_1)
    +
    \sum_{i=2}^{N}\left(\Delta A_i\right)_{\rho}^2
    \geqslant
    N-\vartheta(G).
\end{align}
An analogous procedure can be used to strengthen any sum-of-variances uncertainty relation. 

Thus far, we have reviewed several sum‐form of uncertainty relations, whose common feature is that their bounds are state-independent -- precisely why we formulate them in terms of summations. 
However, state-dependent results also exist. For example, Maccone and Pati derive two strong forms~\cite{PhysRevLett.113.260401}: The first reads
\begin{align}\label{eq:VUR_Sum_MP_1}
    \Delta A^2+\Delta B^2\geqslant\pm i
    \langle[A,B]\rangle
    +
    |\bra{\psi}A\pm i B\ket{\psi^{\perp}}|^2
\end{align}
where the inequality holds for any state $\ket{\psi^{\perp}}$ orthogonal to the state $\ket{\psi}$, with the sign chosen so that the first term on the right-hand side is positive. The second is
\begin{align}\label{eq:VUR_Sum_MP_2}
    \Delta A^2+\Delta B^2\geqslant
    \frac{1}{2}
    |\bra{\psi^{\perp}_{A+B}}A + B\ket{\psi}|^2.
\end{align}
Here, $\ket{\psi^{\perp}_{A+B}}\propto\overline{(A+B)}\ket{\psi}$ denotes a state orthogonal to $\ket{\psi}$. Ref.~\cite{Xiao2016} further extends these to weighted forms. 
By introducing state orthogonal to the system of interest, Maccone and Pati strengthen Robertson's original bound in Eq.~\eqref{eq:VUR_Robertson} as well
\begin{align}\label{eq:VUR_Product_MP}
    \Delta A\Delta B\geqslant\pm\frac{i}{2}\langle[A,B]\rangle
    /
    (1-\frac{1}{2}|\bra{\psi}
    \frac{A}{\Delta A}\pm i \frac{B}{\Delta B}
    \ket{\psi^{\perp}}|^2).
\end{align}
Nevertheless, like all state-dependent bounds, they inherit the limitations discussed in Sec.~\ref{sec:VUR_PBF}.

Before concluding our review of variance‐based uncertainty relations, we make three remarks. First, these inequalities have found numerous applications in characterizing quantum correlations: they can witness entanglement~\cite{PhysRevLett.84.2722,PhysRevA.69.052327,PhysRevLett.92.117903,PhysRevLett.96.110402,PhysRevA.76.014305,PhysRevA.78.052317,PhysRevLett.101.130402,PhysRevA.80.052335,PhysRevA.81.062128,PhysRevA.81.012324,PhysRevA.84.052325,PhysRevA.86.042124,PhysRevLett.108.230503,PhysRevA.93.042310,xiao2017framework,PhysRevLett.119.170404,PhysRevA.97.042333}, EPR steering~\cite{PhysRevA.40.913,PhysRevA.92.062130,PhysRevA.93.012108,PhysRevA.96.052326,PhysRevA.103.062224}, and Bell nonlocality~\cite{PhysRevA.99.012105,PhysRevA.100.012123}. In particular, Reid's uncertainty‐based EPR criterion~\cite{PhysRevA.40.913} -- formulated well before the modern notion of steering~\cite{PhysRevLett.98.140402} -- remains a seminal tool.
Second, variance-based uncertainty relations extend naturally beyond projective measurements to the more general framework of positive operator-valued measures (POVMs). 
Further discussion can be found in Ref.~\cite{PhysRevA.76.042114} and the references therein. 
Third, although we have considered both product‐form and sum‐form uncertainty relations, they are all formulated in terms of variances. A natural question is whether such variance-based relations connect to other frameworks -- most notably entropic uncertainty relations. The answer is affirmative~\cite{PhysRevA.86.024101}, and further discussion of entropic uncertainty relations will be provided in Sec.~\ref{sec:EUR}.


\subsection{Entropic Uncertainty Relations}
\label{sec:EUR}

Shannon entropy quantifies the uncertainty associated with a random variable~\cite{Shannon1948}: a higher entropy signals greater unpredictability in its outcomes, whereas zero entropy corresponds to complete determinism. This foundational concept naturally extends to characterizing the uncertainty of measurement outcomes when observing a quantum system, leading to entropic uncertainty relations. In this section, we trace the development of entropic uncertainty relations, highlight the advantages of entropic formulations over traditional variance-based approaches in expressing the uncertainty principle, and examine their far-reaching implications in quantum information science. Specifically, Sec.~\ref{subsec:QU_without_QM} presents the standard entropic uncertainty relations without memory; Sec.~\ref{subsec:Guessing_Games} discusses the modern guessing game framework; and Secs.~\ref{subsec:QU_with_CM} and~\ref{subsec:QU_with_QM} review extensions incorporating classical and quantum memory, respectively. For completeness, Appendix~\ref{appendix:Entropies} provides a brief overview of the entropic measures used throughout this review.


\subsubsection{Uncertainty Without Memory}\label{subsec:QU_without_QM}

Historically, Everett was the first to consider whether the uncertainty principle could be formulated in terms of entropies~\cite{RevModPhys.29.454}. 
Hirschman answered this question affirmatively by introducing the first entropic uncertainty relation for position and momentum in his notes on entropy~\cite{Hirschman1957}. 
This pioneering work was later refined and strengthened by Beckner~\cite{Beckner1975}, and Bia\l ynicki-Birula and Mycielski~\cite{Bialynicki-Birula1975}, whose formulation is given by
\begin{align}\label{eq:EUR_diff_position_momentum}
    h(x)+h(p)\geqslant\log(e\pi\hbar),
\end{align}
where $h$ refers to the differential entropy, as defined in Eq.~\eqref{eq:diff_ent}. 
Similar to Kennard's uncertainty relation in terms of standard deviations, the left-hand side of Eq.~\eqref{eq:EUR_diff_position_momentum} represents the joint uncertainty associated with position and momentum, which depends on both the observables and the state. 
In contrast, the right-hand side of Eq.~\eqref{eq:EUR_diff_position_momentum} is state-independent, making it applicable to all quantum states. 
The entropic uncertainty relation provides more insight compared to the standard deviation form, as it: 
(\romannumeral1) implies Kennard's form (Eq.~\eqref{eq:VUR_Kennard}), and (\romannumeral2) indicates that the uncertainty relation becomes a strict inequality when the probability distribution is non-Gaussian.

In the seminal work of Ref.~\cite{PhysRevLett.50.631}, Deutsch introduced a general entropic uncertainty relation for bounded operators, offering a perspective distinct from the traditional variance-based formulation. 
He identified a fundamental limitation of Robertson's uncertainty relation -- its dependence on the quantum state. 
In continuously evolving quantum systems, such a state-dependent bound is problematic, as it fails to account for the broader dynamical nature of quantum uncertainty.
Moreover, from a modern perspective, no quantum system is truly isolated; interactions with the environment inevitably induce decoherence, further altering the system and the state of interest. 
In contrast, Heisenberg's original uncertainty relation for position and momentum is independent of the specific quantum state, suggesting that a fundamental formulation of quantum uncertainty should be expressed in terms of
\begin{align}\label{eq:EUR_Core}
    \mJ(M,N,\psi)\geqslant\mB(M,N).
\end{align}
In Eq.~\eqref{eq:EUR_Core}, the left-hand side $\mJ$ quantifies the joint uncertainty of measuring the state $\ket{\psi}$ using $M:=\{\ket{u_j}\}_j$ and $N:=\{\ket{v_k}\}_k$. 
The right-hand side $\mB$ provides a lower bound that depends solely on the measurements themselves, encapsulating their inherent incompatibility. 
Deutsch further emphasized that relabeling the eigenvalues of a discrete observable carries no physical significance, necessitating that any well-defined uncertainty measure remain invariant under such transformations. 
Invariance under permutation of measurement outcomes is therefore a fundamental property, making Shannon entropy a natural candidate, which leads to the following entropic uncertainty relation
\begin{align}\label{eq:EUR_Deutsch}
    H(M)+H(N)\geqslant2\log \frac{2}{1+\sqrt{c_1}}.
\end{align}
Here, $H(M)$ represents the Shannon entropy of the probability distribution resulting from measuring the state $\ket{\psi}$ of the system in the eigenbasis $\ket{u_j}$. 
It is given by $H(M) = -\sum_j p_j \log p_j$, where $p_j=|\braket{\psi}{u_j}|^2$ is the probability of observing the system in the eigenbasis $\ket{u_j}$. 
A similar notation applies to $H(N)$. 
Let
\begin{align}\label{eq:c_jk}
    c_{jk}:=|\braket{u_j}{v_k}|^2,
\end{align}
then, the quantity $c_1$, i.e.,
\begin{align}\label{eq:c_1}
    c_1:=\max_{jk}c_{jk},
\end{align}
denotes the maximal overlap between the measurements $M$ and $N$. 
In this case, the joint uncertainty of a quantum state measured in $M$ and $N$ is given by the sum of the Shannon entropies $H(M)+H(N)$, with the bound determined by $c_1$.
Since the Shannon entropy $H(\cdot)$ is concave in $\ket{\psi}$, Eq.~\eqref{eq:EUR_Deutsch} extends naturally to the case of mixed states. 

While Deutsch's relation captures uncertainty for finite-dimensional measurements, it does not directly extend to continuous observables such as position-momentum or angle-angular momentum. 
Partovi resolved this issue by incorporating the resolution of the measuring device~\cite{PhysRevLett.50.1883}, ensuring a consistent entropic uncertainty framework across both discrete and continuous cases. 
However, for continuous observables, Partovi's Shannon entropic uncertainty relations are not tight. 
The optimal bound for the angle-angular momentum pair was later derived by Bia\l ynicki-Birula, who also provided an improved entropic uncertainty relation for the position-momentum pair~\cite{BIALYNICKIBIRULA1984253}. 
Extensions to more general angular distributions are discussed in~\cite{BIALYNICKIBIRULA1985384}. 
Additionally, Bia\l ynicki-Birula extended the framework to R\'enyi entropies (see Eq.~\eqref{eq:Renyi_Entropy}), presenting the corresponding R\'enyi entropic uncertainty relations for position and momentum~\cite{PhysRevA.74.052101}, given by
\begin{align}\label{eq:REUR_PM}
    H_{\alpha}(x)+H_{\beta}(p)\geqslant-\frac{1}{2}\left(\frac{\log\alpha}{1-\alpha}+\frac{\log\beta}{1-\beta}\right)-\log\frac{\delta x\delta p}{\pi\hbar}.
\end{align}
The lower bound incorporates the area of the phase
space $\delta x\delta p$, which is determined by the resolution of the measuring devices. 
Here, the parameters $\alpha$ and $\beta$ satisfy the H\"older conjugacy constraint 
\begin{align}\label{eq:alpha_beta}
    \frac{1}{\alpha}+\frac{1}{\beta}=2.
\end{align}
In the limit $\alpha \to $1 and $\beta \to 1$, Eq.~\eqref{eq:REUR_PM} reduces to the results established in Ref.~\cite{BIALYNICKIBIRULA1984253}. 
Extensions of the R\'enyi entropic uncertainty relation beyond the H\"older conjugacy constraint of Eq.~\eqref{eq:alpha_beta} have been explored in Refs.~\cite{ZOZOR20084800} and~\cite{6471825}. 
In the qubit case, when $\alpha=\beta=1/2$, Ref.~\cite{PhysRevA.85.012108} establishes tighter bounds, while for $\alpha=\beta=2$, improved results are provided in Refs.~\cite{1027784,Bosyk_2013}. 
For arbitrary $\alpha$ and $\beta$, Ref.~\cite{Zozor_2013} derives the uncertainty relation based on the conjugacy curve in the $\alpha-\beta$ plane and identifies the states that minimize uncertainty. 
The optimal bounds are ultimately presented in Ref.~\cite{doi:10.1142/S0219749915500458}.  

\begin{figure}[t]
    \centering   
    \includegraphics[width=0.48\textwidth]{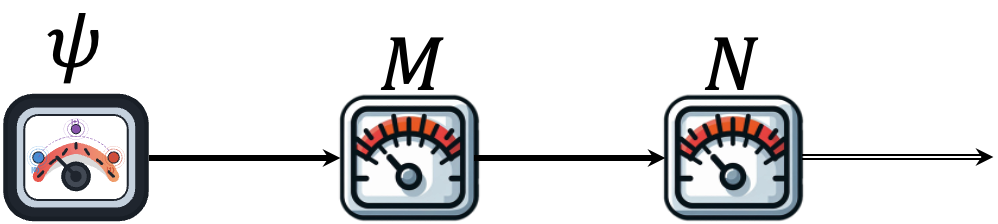}
    \caption{(Color online) \textbf{Successive Measurements}. 
    Once the initial state $\psi$ is prepared, the measurements $M$ and $N$ are performed sequentially. Let $\sigma$ (see Eq.~\eqref{eq:EUR_successive_output}) denote the post-measurement state after applying $M$. The entropic uncertainty relation then characterizes the uncertainty associated with measuring $\psi$ using $M$ and subsequently measuring $\sigma$ using $N$, as given in Eq.~\eqref{eq:EUR_successive}.
   }
    \label{fig:Successive_Measurements}
\end{figure}

The entropic uncertainty relations discussed so far apply to scenarios where i.i.d. copies of quantum state are measured separately using $M$ and $N$. 
However, these relations do not account for the fundamental limitations imposed by sequential measurements on the same system (Fig.~\ref{fig:Successive_Measurements}). 
To address this, Srinivas introduced an entropic uncertainty relation for successive measurements~\cite{Srinivas1985}. 
Without loss of generality, we assume that the measurement $M=\{\ket{u_j}\}_j$ is performed first on the initial state $\psi$, resulting in the following post-measurement state
\begin{align}\label{eq:EUR_successive_output}
    \sigma:=\sum_j|\braket{\psi}{u_j}|^2\ketbra{u_j}{u_j}.
\end{align}
Subsequently, performing the second measurement $N$ yields the joint uncertainty, which is expressed as
\begin{align}\label{eq:EUR_successive}
    H(M)_{\psi}+H(N)_{\sigma},
\end{align}
where $\sigma$ is defined in Eq.~\eqref{eq:EUR_successive_output}. 
A key observation that links successive measurements to the standard setting is that the uncertainty associated with measuring $M$ on both $\psi$ and $\sigma$ is identical, i.e., $H(M)_{\psi}=H(M)_{\sigma}$. 
This leads to the following equation
\begin{align}
    H(M)_{\psi}+H(N)_{\sigma}=H(M)_{\sigma}+H(N)_{\sigma}.
\end{align}
Consequently, the lower bound for standard entropic uncertainty relations remains valid for successive measurements. 
Therefore, in the following discussion of this review, we focus exclusively on the standard case where measurements are performed separately.

In the follow-up study of canonically conjugate observables -- now known as mutually unbiased bases (MUBs)~\cite{DURT2010}, namely $|\braket{u_j}{v_k}|^2=1/d$ holds for any $j$ and $k$ -- Kraus conjectured a stronger form of the entropic uncertainty relation~\cite{PhysRevD.35.3070}. 
He proposed that for a $d$-dimensional quantum system, the lower bound of the entropic uncertainty relation for two MUBs should be $\log d$, a result he confirmed for $d = 2, 3, 4$. 
Just a year later, Maassen and Uffink rigorously proved this conjecture, establishing the following trade-off~\cite{maassen1988generalized}
\begin{align}\label{eq:EUR_Maassen_Uffink_Renyi}
    H_{\alpha}(M)+H_{\beta}(N)\geqslant-\log c_1.
\end{align}
The R\'enyi entropies $H_{\alpha}$ and $H_{\beta}$ of orders $\alpha$ and $\beta$ are evaluated subject to the constraint given in Eq.~\eqref{eq:alpha_beta}.
Their proof directly stems from Riesz-Thorin theorem in functional analysis~\cite{Riesz1926}, and the result is valid for arbitrary measurements $M$ and $N$. 
In the limit, when $\alpha\to1$ and $\beta\to1$, we arrive at the following Shannon entropic uncertainty relation
\begin{align}
\label{eq:EUR_Maassen_Uffink}
    H(M)+H(N)\geqslant-\log c_1,
\end{align}
which is one of the most fundamental and widely recognized entropic uncertainty relations in the literature. This relation confirms Kraus' conjecture~\cite{PhysRevD.35.3070}. 
In particular, when measurements are mutually unbiased, $c_1$ (see Eq.~\eqref{eq:c_1}) simplifies to $1/d$.

The next question that arises naturally is whether $-\log c_1$ is a tight bound? 
If the measurements are a pair of MUBs, then choosing an eigenstate from one of the bases leads to the equation in Eqs.~\eqref{eq:EUR_Maassen_Uffink_Renyi} and~\eqref{eq:EUR_Maassen_Uffink}, making them optimal in the MUBs case. 
However, this does not hold in general. 
Even in the Shannon entropic uncertainty relation given by Eq.~\eqref{eq:EUR_Maassen_Uffink}, the bound is not necessarily optimal, as it depends solely on the maximal overlap between measurements. 
In contrast, measurement incompatibility arises from contributions across all overlaps, not just the largest one. 
Garrett and Gull first explored  Eq.~\eqref{eq:EUR_Maassen_Uffink} in the qubit case, considering the scenario where the transformation matrices between measurements are real (i.e., all $\braket{u_j}{v_k}$ are real), employing numerical methods~\cite{GARRETT1990453}. 
S\'anches-Ruiz later extended the results of~\cite{GARRETT1990453} to the complex case and derived an analytical bound~\cite{SANCHESRUIZ1998189}. 
Finally, Ghirardi, Marinatto, and Romano derived the optimal Shannon entropic uncertainty relation for an arbitrary pair of observables in a two-dimensional Hilbert space, as detailed in Ref.~\cite{GHIRARDI200332}.

Various refinements of the Maassen-Uffink uncertainty relations have been proposed, introducing different forms and tighter bounds. 
However, for dimensions higher than two, improvements across the full range of $c_1$ (see Eq.~\eqref{eq:c_1}) remained elusive until 2014, when the quantum data processing inequality from quantum information theory was utilized to enhance the result~\cite{PhysRevA.89.022112}.
Note that Ref.~\cite{PhysRevA.89.022112} focuses primarily on strengthening entropic uncertainty relations in the presence of quantum memory, and we therefore postpone its discussion to Sec.~\ref{subsec:QU_with_QM}.

Beyond rank-$1$ projective measurements, experiments can also employ general POVMs to probe the system of interest, such as $\mM = \{M_j\}_{j}$ and $\mN = \{N_k\}_{k}$. 
In this case, the following quantity can be used to quantify their incompatibility, 
\begin{align}\label{eq:c_1_POVM}
    c_{1}^{\text{POVM}}:=\max_{jk}\|\sqrt{M_j}\sqrt{N_{k}}\|^2_{\infty}.
\end{align}
where $\|\cdot\|_{\infty}$ denotes the spectral norm, i.e., the largest singular value of the operator~\cite{Horn_Johnson_2012}. 
With this, the entropic uncertainty relation for POVMs takes the following form~\cite{4e6149f1-76a0-31c8-9387-cc0cb9ad7b07}:
\begin{align}
\label{eq:EUR_Maassen_Uffink_POVM}
    H(M)+H(N)\geqslant-\log c_1^{\text{POVM}}.
\end{align}

To close, we have focused on entropic uncertainty relations formulated for a single quantum system -- the measured system. While these results do not involve any ancillary systems, they already offer powerful tools for entanglement detection. In particular, applying such uncertainty relations locally can lead to separability conditions for the composite system, as illustrated by~\cite{PhysRevA.70.022316}. The same line of reasoning has also been adapted to develop steering criteria~\cite{PhysRevA.98.050104,PhysRevA.101.022324,PhysRevA.104.052425}, highlighting the broader reach of entropic uncertainty in detecting quantum correlations~\cite{PhysRevA.70.012102,PhysRevA.87.062103,PhysRevA.91.042103}. For formulations involving other types of measurements, such as weak measurements, see Refs.~\cite{Halpern2019,PhysRevLett.126.100403,PhysRevA.108.032206}.


\subsubsection{Guessing Games}
\label{subsec:Guessing_Games}

One of the most significant advances in the exploration of the uncertainty principle is its extension to cases involving quantum memory, a development that reveals a fundamental connection with quantum cryptography. To gain a clear conceptual understanding of this extension and its relationship to the conventional uncertainty principle, which does not account for quantum memory, physicists often employ the ``guessing game'' framework to formulate the principle.

\begin{figure}[t]
    \centering   
    \includegraphics[width=0.48\textwidth]{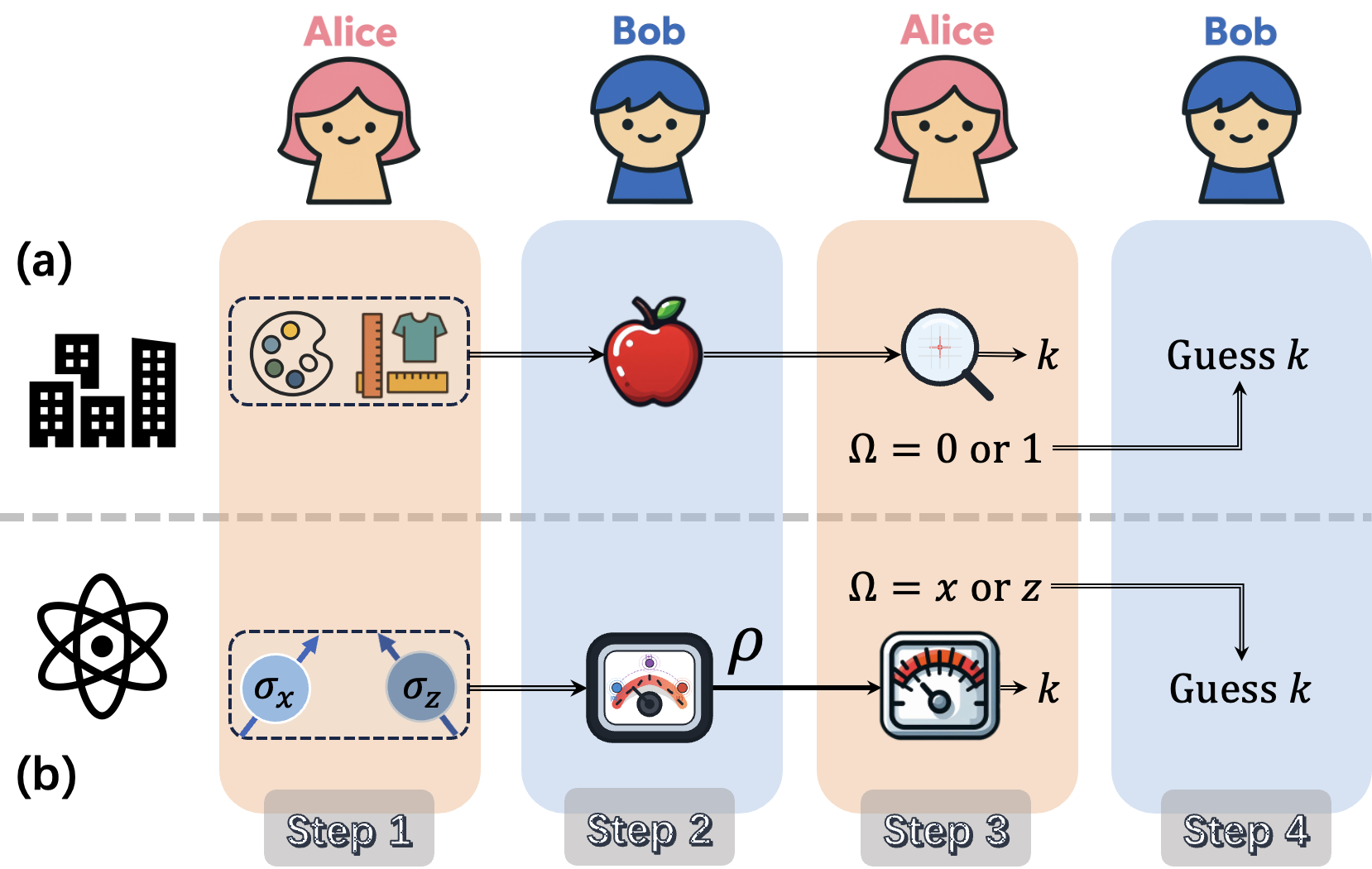}
    \caption{(Color online) \textbf{Guessing Games}. 
    (a) Classical Game: Alice tells Bob she's interested in either the color or shape of an object. Bob prepares a red apple and sends it to her. Alice sets a switch $\Omega$: if $\Omega = 0$, she observes the color; if $\Omega = 1$, the shape. After the observation, she reveals which property she chose, and Bob is asked to guess the outcome. Since Bob prepared the object, his guess is always correct -- there's no uncertainty in the classical case. (b) Quantum Game: Alice measures either $\sigma_x$ or $\sigma_z$, determined by a switch $\Omega$. Bob can prepare any quantum state but chooses $\ket{0}$, which minimizes uncertainty between the two observables. If $\Omega = x$, Alice measures $\sigma_x$; if $\Omega = z$, she measures $\sigma_z$, obtaining outcome $k \in \{+1, -1\}$. After the measurement, she informs Bob of her measurement choice, and he must guess the outcome. Unlike the classical case, quantum uncertainty principle prevents Bob from guessing with certainty.
   }
    \label{fig:Guessing_Game}
\end{figure}

Let us first consider a guessing game in the classical setting, as depicted in Fig.~\ref{fig:Guessing_Game}(a). 
In this game, Alice tries to observe either the color or the shape of an object prepared by Bob. 
A switch, labeled $\Omega$, determines which property she measures: when $\Omega=0$, she measures the color; 
when $\Omega=1$, she measures the shape. 
The game proceeds as follows: Alice begins by informing Bob of the two properties -- color and shape -- that she may measure. 
Bob then prepares an object, such as a red apple, and sends it to Alice. 
Upon receiving the object, Alice randomly selects one of the two properties to measure and communicates her choice (but not the measurement outcome) to Bob. 
Bob's task is to guess the result of her measurement. 
In the classical world, where both properties coexist and can be simultaneously accessed without disturbance, Bob can always predict the correct outcome with certainty. 

However, the incompatibility of certain measurements ensures that this predictability does not carry over to the quantum realm. Consider a qubit system, where Alice is interested in measuring the spin along the 
$x$- and $z$-axes, as illustrated in Fig.~\ref{fig:Guessing_Game}(b). She will provide Bob with all the information about her measurement. Upon receiving the classical description of the measurement, Bob may choose to prepare the state $\ket{0}$ and send it to Alice. She then performs a measurement of the spin along either the $x$-axis (if $\Omega = x$) or the $z$-axis (if $\Omega = z$). After obtaining the measurement outcome, Alice informs Bob of her chosen axis, and Bob's task is to predict the measurement result.

If Alice announces that she measured along the $z$-axis, Bob can be certain that the outcome $k$ must be $+1$, since the state $\ket{0}$ he prepared corresponds to the $+1$ eigenvalue of the spin operator along the $z$-axis. In contrast, if Alice announces that she measured along the $x$-axis, the outcome $k$ will be $+1$ with a $50\%$ probability and $-1$ with a $50\%$ probability, as $p(\pm1,x)=\frac{1}{2}|\braket{\pm}{0}|^2$. This serves as a building block for evaluating the joint Shannon entropy $H(K, \Omega)$ and the conditional Shannon entropy $H(K | \Omega)$ associated with the measurement outcomes.
\begin{align}
    H(K | \Omega)
    =
    &H(K, \Omega)-H(\Omega)\notag\\
    =
    &\left(1+\frac{1}{2}\left(H(X)+H(Z)\right)\right)-1\notag\\
    =
    &\frac{1}{2}\left(H(X)+H(Z)\right),
\end{align}
where $H(X)$ and $H(Z)$ denote the Shannon entropies of the outcome distributions resulting from spin measurements along the $x$- and $z$-axes, respectively. 

The measurements $X$ and $Z$ form a pair of mutually unbiased bases (MUBs), for which the entropic uncertainty bound in Eq.~\eqref{eq:EUR_Maassen_Uffink} evaluates to $1$ in a qubit system. When Bob prepares the state $\ket{0}$, the resulting joint uncertainty exactly saturates this bound. Therefore, preparing and sending the state $\ket{0}$ is one of Bob's optimal strategies for minimizing his uncertainty about Alice's measurement outcome. Although Bob cannot deterministically predict the result of a spin-$x$ measurement, he possesses complete knowledge of its probability distribution. 

While our discussion of the guessing game focuses on a simple scenario, it naturally extends to the more general setting involving multiple measurements, each chosen according to a specified probability distribution. Unlike the conventional formulation discussed in Ref.~\cite{RevModPhys.89.015002}, we assume that Bob is fully informed of the possible measurements before preparing the quantum state, as outlined in Fig.~\ref{fig:Guessing_Game}. This prior knowledge allows Bob to adopt a more active strategy by tailoring the state preparation to minimize the resulting measurement uncertainty. Nonetheless, even classically, joint uncertainty can arise if Bob is unaware of which property will be tested, as illustrated in Fig.~\ref{fig:Guessing_Game}(a). The crucial distinction in the quantum case is that uncertainty remains unavoidable even when Bob knows the measurement choices in advance, reflecting an intrinsic incompatibility of observables rather than mere ignorance.

Moving beyond this setting, we consider what if Bob can prepare a bipartite state and store one subsystem as quantum memory, often referred to as quantum side information in the literature? This leads to the memory-assisted quantum uncertainty principle, which we will examine in Secs.~\ref{subsec:QU_with_CM} and~\ref{subsec:QU_with_QM}. 


\subsubsection{Uncertainty with Classical Memory}
\label{subsec:QU_with_CM}

So far, we have primarily focused on scenarios where all measurements are performed on a single quantum system -- referred to as the measured system. Accordingly, the uncertainty relations developed thus far apply to this single-system setting. But what happens if there exists another system that is correlated with the measured system? Such a correlated system provides side information and is often termed a memory system. For notational consistency, we will henceforth refer to this auxiliary system simply as the memory. In what follows, we begin by examining the case where the memory is classical. Specifically, let us consider an ensemble $\{p_i, \rho_i\}$, and assume the existence of a reference system $R$ that stores the classical information indicating which state $\rho_i$ is prepared in each round. This gives rise to the following quantum-classical state
\begin{align}\label{eq:QC_State}
    \rho_{AR}:=\sum_{i}p_i\rho_i\otimes\ketbra{i}{i},
\end{align}
where system $A$ contains the quantum states and system $R$ carries the classical memory. Applying measurement $M=\{\ket{u_j}\}_j$ to system $A$ maps the state $\rho_{AR}$ to the post-measurement state $\rho_{MR}$, given by
\begin{align}
    \rho_{MR}
    =
    &\sum_j\ketbra{u_j}{u_j}\left(\sum_{i}p_i\rho_i\otimes\ketbra{i}{i}\right) \ketbra{u_j}{u_j}\label{eq:rho_MR_1}\\
    =
    &\sum_ip_i\sum_j\bra{u_j}\rho_i\ket{u_j}\ketbra{u_j}{u_j}\otimes\ketbra{i}{i}.\label{eq:rho_MR_2}
\end{align}
The conditional entropy of the measurement $M$ given access to the classical memory $R$ is then defined as
\begin{align}
    H(M|R):=&H(MR)-H(R)\\
    =&
    \sum_i p_i H(M|R=i)\\
    =&
    \sum_i p_i H(M)_{\rho_i}.
\end{align}
The mutual information $I(M;R)$, defined as
\begin{align}
    I(M;R):= &H(M) - H(M|R)\\
    =&
    H(M)_{\rho_A}-\sum_i p_i H(M)_{\rho_i}
\end{align}
measures the amount of information that the classical memory $R$ contains about the measurement outcome $M$. It quantifies the reduction in uncertainty about $M$ due to the knowledge of $R$. Here, $\rho_A$ denotes the reduced state of $\rho_{AR}$, i.e., $\rho_A:=\Tr_{R}[\rho_{AR}]$.
Similarly, we can define the mutual information $I(N;R)$ with respect to the measurement $N=\{\ket{v_k}\}_k$. The information gain trade-off between $M$ and $N$ can then be formulated as
\begin{align}\label{eq:IER_Hall}
    I(M;R)+I(N;R)\leqslant\log d^2c_1,
\end{align}
which is known as the {\it information exclusion relation}~\cite{PhysRevLett.74.3307}. 
The upper bound arises directly from the fact that the Shannon entropies $H(M)_{\rho_A}$ and $H(N)_{\rho_A}$ are bounded above by $\log d$ in a $d$-dimensional system, in conjunction with the Maassen-Uffink uncertainty relation given in Eq.~\eqref{eq:EUR_Maassen_Uffink}. Hall introduced the information exclusion principle and, more generally, provided a framework for deriving the corresponding upper bound based on the state-dependent upper bound of joint uncertainty and entropic uncertainty relations. He further applied this information-theoretic trade-off to orthogonal spin components, position and momentum, and number and phase observables of a harmonic oscillator. The information exclusion principle thus establishes the first information-theoretic trade-off for incompatible measurements with classical memory. Moreover, the information exclusion relation for measurements on two distinct systems can be derived from Eq.~\eqref{eq:IER_Hall}, and has subsequently been utilized to analyze the security of quantum key distribution protocols involving qudits~\cite{PhysRevLett.88.127902}.

The action of measurement $M=\{\ket{u_j}\}_j$ on a quantum system can be represented as a quantum channel, defined by $\mM(\cdot):=\sum_j\ketbra{u_j}{u_j}\cdot\ketbra{u_j}{u_j}$. Using this channel representation, Eq.~\eqref{eq:rho_MR_1} can be equivalently expressed as $\rho_{AR}=\sum_{i}p_i\mM(\rho_i)\otimes\ketbra{i}{i}$. 
More generally, one can consider a scenario where the measurement channel $\mM$ is replaced by a quantum channel $\mE$, yielding the generalized quantum-classical state $\sum_i p_i \mE(\rho_i) \otimes \ketbra{i}{i}$.
This sets the stage for a channel-based extension of the information exclusion principle.

An alternative formulation of the information exclusion relation involves applying a single measurement to two quantum states related by a basis transformation, which connects the two measurement settings. For example, consider the maximally correlated state, $\mu_{AR}$, as defined below, 
\begin{align}
    \mu_{RA}=\sum_{i}\frac{1}{d}\ketbra{i}{i}\otimes\ketbra{i}{i},
\end{align}
and a quantum channel $\mE:A\to A'$. By exploring the information exclusion relation between these states, particularly when $\mU(\cdot):=U\cdot U^{\dagger}$ is the discrete Fourier transform (DFT), Christandl and Winter introduced the following channel-based information exclusion relation~\cite{1499048}
\begin{align}\label{eq:IER_Christandl_Winter}
    I(A';R)_{(\mE\circ\mU)\otimes\mI(\mu)}+I(A';R)_{\mE\otimes\mI(\mu)}\leqslant
    I(A;A')_{\mI\otimes\mE(\phi^{+}_d)},
\end{align}
where, $\mI$ denotes the identity channel, and $\phi^{+}_d$ is the maximally entangled state on a 
$d$-dimensional Hilbert space. The expression $\mI\otimes\mE(\phi^{+}_d)$ then corresponds to the Choi state of the quantum channel $\mE$. 
Building on Eq.~\eqref{eq:IER_Christandl_Winter}, Christandl and Winter demonstrated that the squashed entanglement~\cite{10.1063/1.1643788} of the so-called flower states~\cite{PhysRevLett.94.200501} can be reduced by an arbitrary amount through the loss of a single qubit from a local subsystem -- a counterintuitive phenomenon known as locking~\cite{PhysRevLett.92.067902}. This result highlights the fundamental role of classical-memory-assisted entropic uncertainty relations, showcasing their power in advancing our understanding of quantum information.


\subsubsection{Uncertainty with Quantum Memory}
\label{subsec:QU_with_QM}

While classical memory offers a window into the interplay between quantum information theory and uncertainty, a far richer structure emerges when the memory itself is quantum~\cite{PhysRevA.68.042301}. In this setting, uncertainty is no longer framed by conventional information exclusion principles, but instead unfolds within the correlations of a bipartite quantum state -- one part encoding the system of interest, the other functioning as a quantum memory. This shift marks more than a mere generalization; it redefines the conceptual landscape of quantum uncertainty. Notably, the idea of conditioning uncertainty on quantum memory predates even the modern discourse on classical memory. As far back as the famed EPR argument~\cite{PhysRev.47.777}, quantum entanglement was already being employed to question the completeness of quantum mechanics, implicitly invoking memory-assisted uncertainty to scrutinize the physical reality of position and momentum. Thus, quantum memory not only enriches our understanding of uncertainty -- it lies at the heart of quantum theory.

The first entropic uncertainty relation incorporating quantum memory was introduced by Renes and Boileau in Ref.~\cite{PhysRevLett.103.020402}. They conjectured that if the measurements $M$ and $N$ are modeled as quantum channels $\mM$ and $\mN$ (see Sec.~\ref{subsec:QU_with_CM}), respectively, then for any tripartite state $\rho_{ABC}$, the following inequality holds
\begin{align}\label{eq:EUR_QM_Renes_Boileau}
    H(M|B)+H(N|C)\geqslant-\log c_1,
\end{align}
where $H(M|B)$ and $H(N|C)$ are the conditional entropies associated with subsystems $B$ and $C$, respectively. Renes and Boileau proved this relation in the special case where $M$ and $N$ are linked by the Fourier transform. The conjecture was later rigorously proven by Berta {\it et al.}~\cite{Berta2010}, who derived the following trade-off for a bipartite quantum state $\rho_{AB}$ in the presence of a single quantum memory system $B$~\cite{Winter2010}
\begin{align}\label{eq:EUR_QM_Berta}
    H(M|B)+H(N|B)\geqslant-\log c_1+H(A|B).
\end{align}
Here, the additional quantity $H(A|B)$ quantifies the entanglement between the measured system $A$ and the quantum memory $B$. When they are maximally entangled, $H(A|B)$ takes on a negative value, i.e., $-\log d$, thereby reducing the joint uncertainty. Experimental confirmations of this relation have been demonstrated in Ref.~\cite{Prevedel2011,Li2011}.  

While Eq.~\eqref{eq:EUR_QM_Renes_Boileau} involves two additional systems, Eq.~\eqref{eq:EUR_QM_Berta} pertains to a single memory system. Despite the apparent mathematical differences between these inequalities, they are, in fact, equivalent. For any bipartite state $\rho_{AB}$ that satisfies Eq.~\eqref{eq:EUR_QM_Berta}, we can consider its purification $\psi_{ABC}$, which acts on the extended systems $ABC$. This leads to 
\begin{align}
    &H(M|B)_{\rho}+H(N|B)_{\rho}-H(A|B)_{\rho}\\
    =
    &H(MB)_{\rho}+H(NB)_{\rho}-H(C)_{\psi}-H(B)_{\rho}\label{eq:EUR_23eq_1}\\
    =
    &H(MB)_{\rho}+H(NC)_{\psi}-H(C)_{\psi}-H(B)_{\rho}\label{eq:EUR_23eq_2}\\
    =
    &H(M|B)_{\psi}+H(N|C)_{\psi},
\end{align}
where the subscripts of $\rho_{AB}$ and $\psi_{ABC}$ are omitted to simplify the notation.
In Eq.~\eqref{eq:EUR_23eq_1}, we use the identity $H(C)_{\psi} = H(AB)_{\rho}$, and in Eq.~\eqref{eq:EUR_23eq_2}, we apply $H(NB)_{\rho} = H(NC)_{\psi}$, both of which follow from the assumption that $\psi$ is a pure state. 

\begin{figure}[t]
    \centering   
    \includegraphics[width=0.48\textwidth]{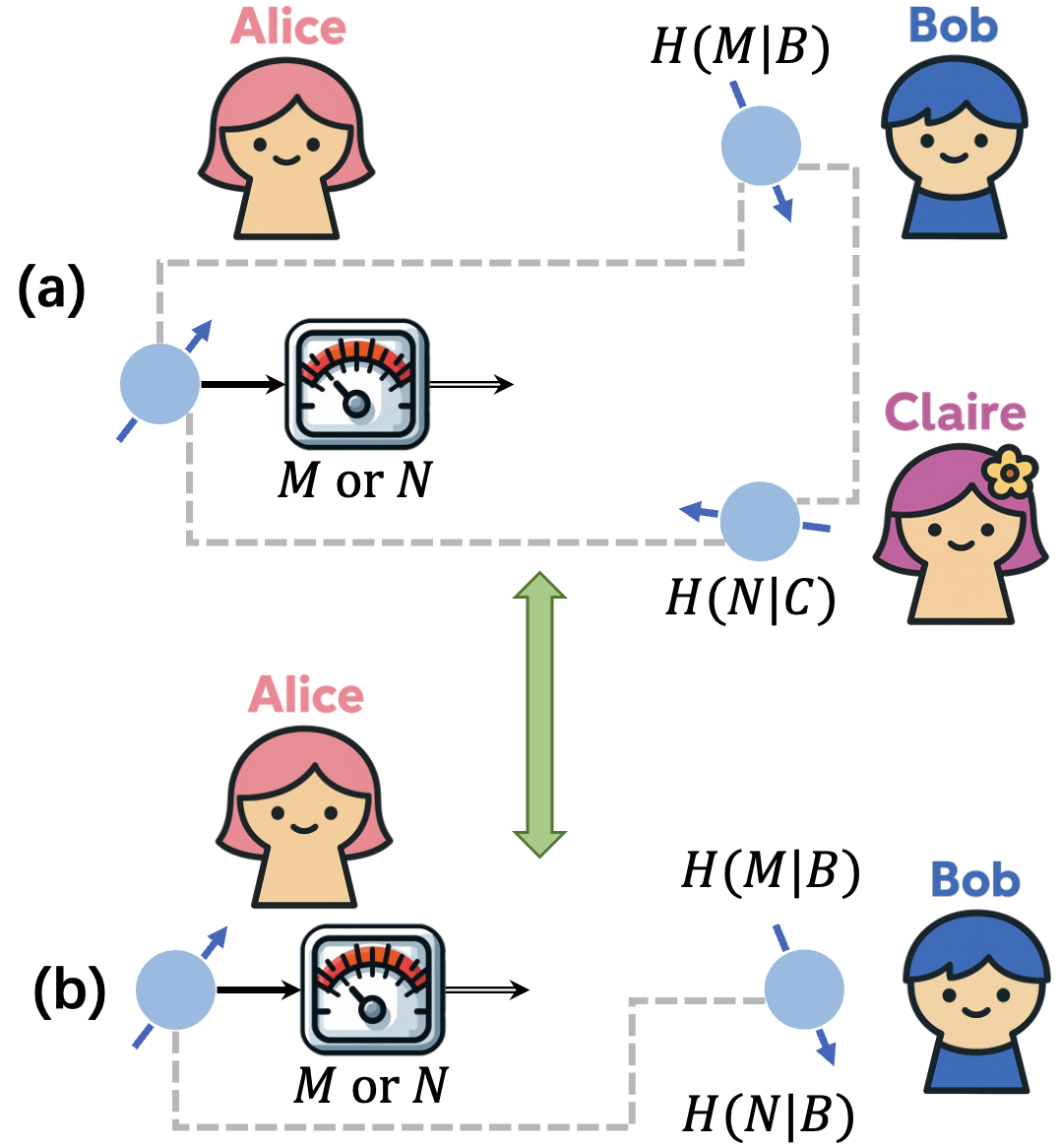}
    \caption{(Color online) \textbf{Uncertainty under Memory Effects}. 
    (a) Consider a setting where Alice, Bob, and Claire share a tripartite quantum state with respective subsystems $A$, $B$, and $C$. Two alternative measurements, $M$ and $N$, can be performed on subsystem $A$. When $M$ is measured, Bob uses his memory system $B$ to guess the outcome; when $N$ is measured, Claire uses her memory $C$ for the same purpose. This tripartite scenario captures uncertainty trade-offs in the presence of distributed quantum memory, as formalized in Eq.~\eqref{eq:EUR_QM_Renes_Boileau}. (b) In the bipartite case, only Alice and Bob are involved. Bob uses his memory $B$ to infer the outcome of either $M$ or $N$ measured by Alice on $A$, with the corresponding uncertainty relation given by Eq.~\eqref{eq:EUR_QM_Berta}. Mathematically, the tripartite uncertainty relation is equivalent to the bipartite one.
   }
    \label{fig:Bipartite_Tripartite}
\end{figure}

The equivalence between the bipartite uncertainty relation in Eq.~\eqref{eq:EUR_QM_Berta} and the tripartite relation in Eq.~\eqref{eq:EUR_QM_Renes_Boileau} helps resolve a fundamental question: does the memory-assisted entropic uncertainty relation still conform to the basic requirement set by Deutsch in Eq.~\eqref{eq:EUR_Core}? At first glance, it may appear not to, since the right-hand side of Eq.~\eqref{eq:EUR_QM_Berta} depends on the quantum state. However, this is not the case. Eq.~\eqref{eq:EUR_QM_Berta} can be reformulated into Eq.~\eqref{eq:EUR_QM_Renes_Boileau}, where the left-hand side reflects the dependence on both the quantum state and the choice of incompatible measurements, while the right-hand side -- the uncertainty lower bound -- is state-independent. Therefore, it indeed satisfies the criterion established in Eq.~\eqref{eq:EUR_Core}.

The original proof of Eq.~\eqref{eq:EUR_QM_Berta} relies on constructing the following uncertainty relations for smooth entropies (see Appendix~\ref{appendix:Entropies})
\begin{align}\label{eq:EUR_QM_Berta_Original}
    H^{5\sqrt{\varepsilon}}_{\min}(M|B)+H^{\varepsilon}_{\max}(NB)
    \geqslant
    \log \frac{\varepsilon^2}{c_1} + H^{\varepsilon}_{\min}(AB).
\end{align}
By further applying the quantum asymptotic equipartition property for min- and max-entropies, we obtain Eq.~\eqref{eq:EUR_QM_Berta}. Later, Frank and Lieb provided an alternative, more direct proof, demonstrating that the result could be derived using the strong subadditivity property of von Neumann entropy~\cite{Lieb2002}, offering a simplification~\cite{Frank2013}.

Shortly thereafter, Coles {\it et al.}~\cite{PhysRevA.83.062338} investigated the information-theoretic trade-offs in tripartite quantum systems, addressing the central question, ``How much of what kind of information about what is located where?'' They extended the concept of ``type of information''~\cite{PhysRevA.76.062320} to encompass arbitrary POVMs -- a POVM is a set of positive linear
operators, $\{\Pi_{k}\}_k$, that sum to the identity $\sum_{k}\Pi_{k}=\1_d$. This can be viewed as an extension of the concept of physical reality~\cite{PhysRev.48.696}, originally discussed in Ref.~\cite{PhysRev.47.777}. Additionally, they derived entropic uncertainty relations and information exclusion relations incorporating quantum memory for general POVMs. In particular, for two POVMs, $M = \{M_j\}_{j}$ and $N = \{N_k\}_{k}$, the entropic uncertainty relation takes the form
\begin{align}\label{eq:EUR_QM_Coles11}
    H(M|B)+H(N|C)\geqslant
    -\log c_{1}^{\text{POVM}},
\end{align}
and the uncertainty bound for a single POVM $M$ has also been introduced
\begin{align}
    H(M|B)\geqslant-\log
    \max_{jk}\|\sqrt{M_j}\sqrt{M_{k}}\|_{\infty}.
\end{align}
Information exclusion relations involving channels and POVMs are also derived in Ref.~\cite{PhysRevA.83.062338}. 
Aside from conditional entropy, Ref.~\cite{PhysRevLett.106.110506} presents a generalized framework for uncertainty relations based on smooth entropies, which applies to POVMs
\begin{align}\label{eq:EUR_QM_Tomamichel_Renner}
    H^{\varepsilon}_{\min}(M|B)+H^{\varepsilon}_{\min}(N|C)\geqslant
    -\log c_{1}^{\text{POVM}}.
\end{align}

Generalizing this direction of study, Coles {\it et al.} developed an abstract framework for entropic uncertainty relations~\cite{PhysRevLett.108.210405}. Through an axiomatic definition of relative entropy, they constructed conditional entropies and their duals, proving that these duals adhere to uncertainty relations akin to Eqs.~\eqref{eq:EUR_QM_Coles11} and~\eqref{eq:EUR_QM_Tomamichel_Renner}. This abstract approach resulted in a R\'enyi entropic uncertainty relation incorporating quantum side information
\begin{align}\label{eq:EUR_QM_Coles12}
    H_{\alpha}(M|B)+H_{\beta}(N|C)\geqslant
    -\log c_{1}^{\text{POVM}}.
\end{align}
Here, the R\'enyi entropies $H_{\alpha}$ and $H_{\beta}$, with orders $\alpha$ and $\beta$, are evaluated under the constraint in Eq.\eqref{eq:alpha_beta}, which provides a natural extension of the Maassen–Uffink uncertainty relation in Eq.~\eqref{eq:EUR_Maassen_Uffink_Renyi}.
In Eqs.~\eqref{eq:EUR_QM_Coles11},~\eqref{eq:EUR_QM_Tomamichel_Renner}, and~\eqref{eq:EUR_QM_Coles12}, $c_{1}^{\text{POVM}}$ denotes the maximal overlap between the POVMs, as previously defined in Eq.~\eqref{eq:c_1_POVM}. 

Until now, it has appeared that the maximal overlap between incompatible measurements determines the entropic uncertainty relations. This intuition holds well for qubit systems, where the overall transformation matrix between measurement bases is indeed governed by this maximal overlap. However, in general, this is not the case -- the full structure of the transformation matrix, including all its elements, should influence the formulation of the uncertainty lower bound. But how exactly do these contributions manifest? This question was addressed by Coles and Piani in Ref.~\cite{PhysRevA.89.022112}, where they established a stronger entropic uncertainty relation along with corresponding information exclusion relations. Their results apply to both rank-$1$ projective measurements and general POVMs. For simplicity, we focus here on the case of rank-$1$ projective measurements. Given two such measurements, $M=\{\ket{u_j}\}_j$ and $N=\{\ket{v_k}\}_k$, we define the following two quantities, which serve as more fine-grained refinements of the standard maximal overlap $c_1$:
\begin{align}
    d_j:=&\max_{k}c_{jk},\label{eq:dj}\\
    e_k:=&\max_{j}c_{jk},\label{eq:ek}
\end{align}
where $c_{jk}$ is defined in Eq.~\eqref{eq:c_jk}, and introduce the operator $G(\lambda)$ with $0\leqslant\lambda\leqslant1$ as follows
\begin{align}\label{eq:G_p}
    G(\lambda):=-\lambda\sum_j\log d_j\ketbra{u_j}{u_j}
    -(1-\lambda)\sum_k\log e_k\ketbra{v_k}{v_k}.
\end{align}
To further refine the uncertainty bound, we define the optimal solution to the following SDP~\cite{doi:10.1137/1038003} as $g$
\begin{align}
    g:=\max_{0\leqslant \lambda\leqslant1}
    \min\quad&\Tr[\rho_A\cdot G(\lambda)]\label{eq:EUR_CP_SDP_0}
    \\
    \text{s.t.}\quad&\,\,\rho_A\geqslant0,\,\,\,\Tr[\rho_A]=1.\label{eq:EUR_CP_SDP_1}
\end{align}
When additional information is available -- such as the marginal state $\rho_A$ of the bipartite system $\rho_{AB}$ -- a tighter bound, denoted by $g^{\ast}$, can be derived as
\begin{align}
    g^{\ast}:=\max_{0\leqslant \lambda\leqslant1}
    \Tr[\rho_A\cdot G(\lambda)].
\end{align}
These values establish a natural hierarchy, with each level incorporating progressively more information to yield sharper estimates.
\begin{align}\label{eq:EUR_Hierarchy}
    g^{\ast}\geqslant g
    \geqslant
    -\log c_1+\frac{1-\sqrt{c_1}}{2}\log\frac{c_1}{c_2},
\end{align}
where $c_2$ denotes the second-largest overlap between the measurement bases.
Each quantity in this hierarchy can replace the conventional $-\log c_1$ term in Eq.~\eqref{eq:EUR_QM_Berta}, thereby leading to a strengthened entropic uncertainty relation. Notably, the final expression in Eq.~\eqref{eq:EUR_Hierarchy} yields an analytical, state-independent lower bound. When the auxiliary system $B$ is trivial, this bound strengthens the Maassen–Uffink uncertainty relation (see Eq.~\eqref{eq:EUR_Maassen_Uffink}) 
\begin{align}
\label{eq:EUR_Coles_Piani}
    H(M)+H(N)\geqslant-\log c_1+\frac{1-\sqrt{c_1}}{2}\log\frac{c_1}{c_2},
\end{align}
where the second term on the right-hand side is always non-negative, since $c_1\geqslant c_2$ by definition (see Eq.~\eqref{eq:c_1}). This ensures that the bound is valid across the full range of $c_1$, and guarantees a strict improvement whenever $c_1>c_2$. Moreover, this refined bound explicitly highlights the role of the second-largest overlap $c_2$, thereby capturing more features of measurement incompatibility beyond the leading term.

Eq.~\eqref{eq:EUR_Coles_Piani} addresses the problem outlined in Sec.~\ref{subsec:QU_without_QM}, while also suggesting several promising avenues for future exploration. First, it raises the question of whether similar results could extend to smooth entropies, which play a fundamental role in non-asymptotic quantum Shannon theory, or to other dual entropies. Additionally, the fact that only the first and second largest overlaps feature in the analytical bound for the entropic uncertainty relation invites further investigation. Could this result be strengthened by incorporating additional non-negative terms into Eq.~\eqref{eq:EUR_Coles_Piani}, potentially involving the third, fourth, and more overlaps? We explore these questions in detail in Sec.~\ref{sec:MUR}.

Memory-assisted entropic uncertainty relations represent a profound refinement of the uncertainty principle, revealing fundamental links between quantum uncertainty, entanglement, and the role of quantum correlations in shaping physical reality. 
Their significance extends beyond foundational insights: they have become indispensable tools in quantum certification, verification~\cite{Zhu2021}, and cryptography, particularly in the security proof of quantum key distribution protocols (see Ref.~\cite{RevModPhys.89.015002} for an in-depth review).
Although initially formulated for discrete-variable systems, these relations are not confined to this setting. 
Extensions to continuous-variable regimes have been successfully developed, with the position-momentum uncertainty relation in the presence of quantum memory standing out as an example~\cite{furrer2014position}. 
Together, these advances highlight the versatility and far-reaching implications of incorporating memory into the entropic formulation of quantum uncertainty~\cite{PhysRevA.86.032338,PhysRevA.86.042315,PhysRevA.86.062334,PhysRevLett.110.020402,PhysRevA.87.022314,PhysRevA.89.010302,Coles2014,PhysRevA.91.012115,PhysRevA.93.062111,PhysRevA.94.052323,PhysRevA.97.052307,PhysRevLett.126.190503,PhysRevA.103.052412,PhysRevE.106.054107,PhysRevE.109.064103,PhysRevA.111.032418}.


\subsection{Majorization Uncertainty Relations}
\label{sec:MUR}

The probability distribution resulting from a quantum measurement encodes the fundamental uncertainty inherent in the system. 
While entropic functions map this distribution to non-negative real values -- providing scalar summaries of uncertainty -- each entropy highlights only certain aspects of the distribution. 
Consequently, different entropic uncertainty relations have found utility in different operational and theoretical settings. 
Nonetheless, no single entropic measure can fully capture the structure or extent of the uncertainty encoded in the probability distribution. 
A more comprehensive and fine-grained approach is to analyze the full probability vector obtained from the measurement. 
This naturally leads to the framework of majorization uncertainty relations, also known in the literature as universal uncertainty relations, which offer a more complete characterization of quantum uncertainty. 
In the following, we offer an overview of the direct-product and direct-sum formulations of majorization uncertainty relations, which are explored in Secs.~\ref{subsec:Direct_Product} and~\ref{subsec:Direct_Sum}, respectively. 
To support these discussions, Appendix~\ref{appendix:Majorization} describes the mathematical framework of majorization, which plays a central role in both approaches. 
Much like their entropic counterparts (see Sec.~\ref{subsec:Guessing_Games}), majorization-based uncertainty relations can also be interpreted through the lens of a guessing game -- this perspective is outlined in Appendix~\ref{appendix:Lottery_Games}.


\subsubsection{Direct-Product Formulation}
\label{subsec:Direct_Product}

Partovi pioneered the study of quantum uncertainty by formulating the joint uncertainty of incompatible measurements through majorization~\cite{PhysRevA.84.052117} (see Appendix~\ref{appendix:Majorization}). 
He proposed a general form of the majorization uncertainty relation based on the direct product of probability distributions and derived optimal bounds for two and three mutually unbiased bases (MUBs) in two-dimensional systems. 
He also calculated the leading term of the bound for the position and momentum majorization uncertainty relation, where the deviation from unity highlights the presence of fundamental quantum uncertainty. 
Notably, in his seminal work, Partovi observed that constructing uncertainty bounds by maximizing the largest partial sums -- such as the largest single element, the largest sum of two elements, and so on -- of the joint probability distribution does not, in general, result in a non-increasing probability vector, and thus may not yield the optimal bound. 
To resolve this issue, he emphasized the necessity of implementing a flatness process $\mF$, which is detailed in Appendix~\ref{appendix:Majorization}.

Continuing along this line of inquiry, Friedland, Gheorghiu, and Gour~\cite{PhysRevLett.111.230401} employed majorization theory~\cite{marshall2010inequalities} to address a fundamental question: Given two probability vectors $\vec{x}$ and $\vec{y}$, how can we determine which one is more uncertain? 
Their answer is grounded in the idea of random relabeling of measurement outcomes: if $\vec{x}$ can be obtained from $\vec{y}$ by randomly permuting the outcome labels -- without retaining knowledge of which permutation was applied -- then $\vec{x}$ should be regarded as more uncertain than $\vec{y}$. 
Formally, such a random relabeling process corresponds to applying a convex combination of permutation matrices, which defines a doubly stochastic matrix $D$. 
In this setting, the relation between $\vec{x}$ and $\vec{y}$ is precisely captured by 
\begin{align}\label{eq:x=Dy}
    \vec{x}=D\vec{y}.
\end{align}
Any Schur-concave function, which reverses the majorization order, thus defines a valid uncertainty measure. Specifically, Eq.~\eqref{eq:x=Dy} ensures that for any such function $f$, we have $f(\vec{x})\geqslant f(\vec{y})$, yielding an infinite family of inequalities. 

It is worth noting that in the original work~\cite{PhysRevLett.111.230401}, two conditions were introduced to characterize uncertainty measures and to compare the uncertainty between distributions: (\romannumeral1) invariance under permutations, and (\romannumeral2) monotonicity under random relabeling. However, the first condition is redundant. Since any permutation is a deterministic and invertible instance of a random relabeling, invariance under permutations is already implied by the second condition.

For real vectors in $d$-dimensional space $\mathbb{R}^{d}$, the majorization relation~~\cite{marshall2010inequalities} defines a pre-order: it is reflexive and transitive, but not necessarily antisymmetric. However, when all vectors are arranged in non-increasing order, the relation becomes a partial order. Throughout this appendix, unless stated otherwise, we assume all vectors are sorted in non-increasing order and denote the set of such vectors by $\mathbb{R}^{d,\downarrow}$. For any vector $\vec{x} \in \mathbb{R}^d$, we denote its non-increasing rearrangement by $\vec{x}^{\downarrow}$. For example, if $\vec{x}=(1,2,3)^{\T}$, then $\vec{x}^{\downarrow}=(3,2,1)^{\T}$.

Given two vectors $\vec{x}, \vec{y} \in \mathbb{R}^{d,\downarrow}$, we say that $\vec{y}$ weakly majorizes $\vec{x}$, denoted $\vec{x}\prec_{\text{w}}\vec{y}$, if the following condition holds
\begin{align}\label{eq:week_majorization}
    \sum_{i=1}^{k}x_i
    \leqslant
    \sum_{i=1}^{k}y_i,
    \quad\forall\,\,\,\, 1\leqslant k\leqslant d.
\end{align}
If, in addition, the total sums of the vectors $\vec{x}$ and $\vec{y}$ are equal; that is, 
\begin{align}\label{eq:majorization}
    \sum_{i=1}^{d}x_i
    =
    \sum_{i=1}^{d}y_i,
\end{align}
then $\vec{y}$ is said to majorize $\vec{x}$, denoted $\vec{x}\prec\vec{y}$. 

A concrete example further illustrates this concept. Consider two probability vectors in $d$ dimensions: the uniform distribution $(1/d, \ldots, 1/d)$ and the deterministic distribution $(1, 0, \ldots, 0)$. The uniform distribution represents a state of maximal uncertainty, where all outcomes are equally likely, while the deterministic distribution reflects complete certainty, with a single outcome occurring with probability one. This intuition is reinforced by applying Shannon entropy, a canonical example of a Schur-concave function, to majorization inequalities. 
For any $d$-dimensional probability vector $\vec{x}$, we have $(1/d, \ldots, 1/d)\prec\vec{x}\prec(1, 0, \ldots, 0)$, and the following inequality holds 
\begin{align}
    \log d\geqslant H(\vec{x})\geqslant0.
\end{align}

\begin{figure}[t]
    \centering   
    \includegraphics[width=0.35\textwidth]{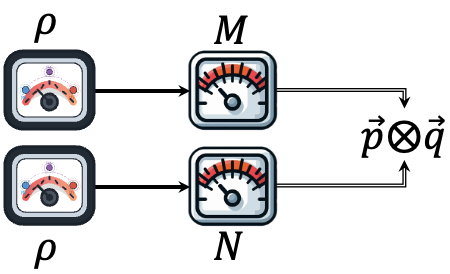}
    \caption{(Color online) \textbf{Majorization Uncertainty Relations in Direct-Product Form}.
    When independent and identically distributed (i.i.d.) copies of the state $\rho$ are prepared and measured using $M$ and $N$, the resulting joint probability distribution, expressed in the direct-product form $\vec{p}\otimes\vec{q}$, satisfies the majorization uncertainty relation given in Eq.~\eqref{eq:Maj_DP}
    .
    }
    \label{fig:Direct_Product}
\end{figure}

Within this framework, Friedland, Gheorghiu, and Gour study the joint uncertainty encoded in the product distribution $\vec{p}\otimes\vec{q}$, where $\vec{p}$ and $\vec{q}$ arise from measurements $M$ and $N$ on a quantum state (see Fig.~\ref{fig:Direct_Product}). Denoting the set of all such distributions by $S_{\otimes}$, the goal is to characterize its least upper bound (LUB) (see Appendix~\ref{appendix:Majorization}), i.e., $\vee S_{\otimes}$, leading to the formulation of the corresponding direct-product majorization uncertainty relations
\begin{align}\label{eq:Maj_DP}
    \vec{p}\otimes\vec{q}\prec\vee S_{\otimes},
\end{align}
where the right-hand side is both state-independent and optimal. This formulation is known as the universal uncertainty relation~\cite{PhysRevLett.111.230401}, as it encapsulates a broad class of uncertainty measures -- all Schur-concave functions -- within a single unified framework.

In general, computing $\vee S_{\otimes}$ is highly nontrivial, as it involves an optimization over the space of all product distributions. However, in certain simple scenarios, such as when both $M=\{\ket{u_j}\}_j$ and $N=\{\ket{v_k}\}_k$ are rank-$1$ projective measurements, explicit expressions for the leading components $\varpi_{\otimes,1}$ and $\varpi_{\otimes,2}$ (see Eq.~\eqref{eq:varpi}) of the ordered vector $\vec{p}\otimes\vec{q}$ in $\vee S_{\otimes}^{\downarrow}$ can be obtained. Here, the subscript $\otimes$ indicates that these values pertain specifically to the direct product setting in Eq.~\eqref{eq:Maj_DP}. 
The explicit construction of the majorization bound, built upon the quantity $\varpi_{\otimes,n}$, is presented in Appendix~\ref{appendix:Majorization}.
In particular, the largest component $\varpi_{\otimes,1}$ was previously solved by Deutsch in Ref.~\cite{PhysRevLett.50.631}, and is given by
\begin{align}\label{eq:Maj_DP_1}
    \varpi_{\otimes,1}=\left(\frac{1+\sqrt{c_1}}{2}\right)^2,
\end{align}
where the maximal overlap $c_1$ is defined in Eq.~\eqref{eq:c_1}. The maximal sum of the two largest components, corresponding to 
$\varpi_{\otimes,2}$, is determined by
\begin{align}\label{eq:Maj_DP_2}
    \varpi_{\otimes,2}=\left(\frac{1+\max\sqrt{c_{jk}+c_{j'k'}}}{2}\right)^2,
\end{align}
where the maximum is taken over all distinct index pairs, subject to either $j=j'$ with $k\neq k'$, or $k= k'$ with $j\neq j'$. For general $n$, evaluating $\varpi_{\otimes,n}$ exactly becomes intractable. Nevertheless, a tractable upper bound can be obtained through the following relaxation
\begin{align}
    \varpi_{\otimes,n}
    &=
    \max_{|S|=n}
    \max_{\rho}
    \sum_{(j,k)\in S}p_jq_k\label{eq:Maj_DP_DS_1}\\
    &\leqslant
    \max_{
    \substack{
    |S_1|=n_1\\ 
    |S_2|=n_2\\
    n_1+n_2=n+1
    }
    }
    \max_{\rho}
    \left(\sum_{j\in S_1}p_j\right)
    \left(\sum_{k\in S_2}q_k\right)\label{eq:Maj_DP_DS_2}\\
    &\leqslant
    \max_{
    \substack{
    |S_1|=n_1\\ 
    |S_2|=n_2\\
    n_1+n_2=n+1
    }
    }
    \max_{\rho}
    \left(\frac{\sum_{j\in S_1}p_j+\sum_{k\in S_2}q_k}{2}\right)^2\label{eq:Maj_DP_DS_3},
\end{align}
where the last inequality follows from the arithmetic mean-geometric mean inequality.

We will momentarily set aside the construction of the upper bound for the direct-product form and will revisit it in the context of direct-sum majorization uncertainty relations. There, semidefinite programming (SDP)~\cite{doi:10.1137/1038003} -- originally introduced for this purpose in~\cite{PhysRevResearch.3.023077} -- offers a unified framework for deriving both types of bounds. Related results were also established in~\cite{Puchala_2013}, where the authors compared the bound obtained by applying the Shannon entropy to Eq.~\eqref{eq:Maj_DP} with the well-known Maassen-Uffink bound in Eq.~\eqref{eq:EUR_Maassen_Uffink}. Their analysis shows that, for randomly chosen unitary transformation matrices between rank-$1$ projective measurements in $5$-dimensional systems, the majorization-based approach yields a tighter bound with $98\%$ probability. Additional discussions and numerical comparisons can be found in Refs.~\cite{PhysRevLett.111.230401,Puchala_2013,PhysRevA.89.052115,RevModPhys.89.015002}.

Beyond formulating direct-product majorization uncertainty relations, the authors of~\cite{Puchala_2013} also introduced a classical analog of the Maassen-Uffink uncertainty relation (see Eq.~\eqref{eq:EUR_Maassen_Uffink}). Specifically, for a probability distribution $\vec{p}$, if a stochastic matrix $T := (t_{jk})$, where $t_{jk} \geqslant 0$  and $\sum_j t_{jk} = 1$, is applied to $\vec{p}$, the resulting output probability vector $\vec{q}$ satisfies the following entropic uncertainty relation, 
\begin{align}\label{EUR_Classical}
    H(\vec{p})+H(\vec{q})
    \geqslant
    -\log t_1,
\end{align}
where $t_1$ denotes the maximal $t_{jk}$. This can be viewed as a classical analogue of entropic uncertainty relations, applied directly to probability distributions.


\subsubsection{Direct-Sum Formulation}
\label{subsec:Direct_Sum}

\begin{figure}[t]
    \centering   
    \includegraphics[width=0.48\textwidth]{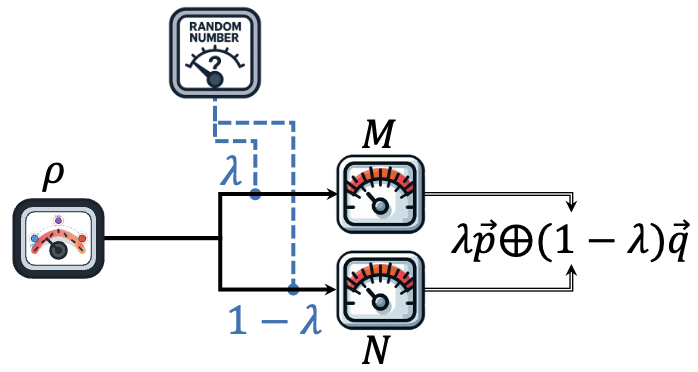}
    \caption{(Color online) \textbf{Majorization Uncertainty Relations in Direct-Sum Form}.
    A random number generator outputs $0$ with probability $\lambda$, triggering measurement $M$, and $1$ with probability $1-\lambda$, leading to measurement $N$. The resulting joint uncertainty, expressed in the direct-sum form $\lambda\vec{p}\oplus(1-\lambda)\vec{q}$, is governed by the majorization uncertainty relation, as outlined in Eq.~\eqref{eq:Maj_DS}
    .
    }
    \label{fig:Direct_Sum}
\end{figure}

We now proceed to the direct-sum majorization uncertainty relation. Consider a system in which a random number generator outputs $0$ with probability $\lambda$, prompting the observer to perform measurement $M$; with probability $1-\lambda$, it outputs $1$, leading the observer to perform measurement $N$ instead. This setup is depicted in Fig.~\ref{fig:Direct_Sum}. Under this probabilistic measurement protocol, the joint uncertainty is described by the following form~\cite{Yuan2023}
\begin{align}\label{eq:Maj_DS}
    \lambda\vec{p}\oplus(1-\lambda)\vec{q}\prec\vee S_{\oplus}(\lambda).
\end{align}

The first direct-sum majorization uncertainty relation was introduced in~\cite{PhysRevA.89.052115} for rank-$1$ projective measurements, taking the form $\vec{p}\oplus\vec{q}$. In that case, the leading component of the majorization upper bound is always $1$. Eq.~\eqref{eq:Maj_DS} above can be viewed as a weighted direct-sum generalization of this original form. For each value of $\lambda$, the corresponding probability distribution defines a set $S_{\oplus}(\lambda)$, which is bounded above by its LUB $\vee S_{\oplus}(\lambda)$ under majorization. In the following, we consider the general case where $M$ and $N$ are POVMs, and we describe a systematic method to construct $\vee S_{\oplus}(\lambda)$. Merge the effects of measurements $M$ and $N$ into a single set, denoted $T(\lambda) := \lambda M \cup (1-\lambda) N$. We then define $\varpi_{\oplus,n}$ as the maximum sum of $n$ distinct elements of the probability distribution within the set $\vee S_{\oplus}(\lambda)$, given explicitly by
\begin{align}
    \varpi_{\oplus,n}(\lambda)
    =
    \max_{|S|=n}
    \max_{\rho}
    \quad
    &
    \Tr[\left(\sum_{j\in S}T_j(\lambda)\right)\cdot\rho]
    \label{eq:Maj_DP_DS_01}\\
    \text{s.t.}
    \quad
    &
    \rho\geqslant0, \,\,\, \Tr[\rho]=1,
    \label{eq:Maj_DP_DS_02}
\end{align}
where $T_j(\lambda)$ are the elements of the set $T(\lambda)$. 
This approach allows us to compute $\varpi_{\otimes,n}$ to establish the bound for $\vec{p}\otimes\vec{q}$, as described in Eq.~\eqref{eq:Maj_DP_DS_01}. 
In fact, the final inequality in Eq.~\eqref{eq:Maj_DP_DS_3} can now be rewritten as
\begin{align}
    \varpi_{\otimes,n}\leqslant
    \left(\frac{2\varpi_{\oplus,n+1}(1/2)}{2}\right)^2.
\end{align}
From the optimal value $\varpi_{\oplus,n}(\lambda)$ obtained in Eq.~\eqref{eq:Maj_DP_DS_01}, we construct the vector $\vec{\omega}_{S_{\oplus}(\lambda)}$ as defined in Eq.~\eqref{eq:Maj_omega}. 
Upon applying the flatness process $\mF$ (see Appendix~\ref{appendix:Majorization}), we obtain the LUB $\vee S_{\oplus}(\lambda)=\mF(\vec{\omega}_{S_{\oplus}(\lambda)})$, which serves as the optimal bound for the direct-sum majorization uncertainty relation.

The weighted majorization uncertainty relation in Eq.~\eqref{eq:Maj_DS} was first proposed and experimentally demonstrated in a high-dimensional photonic system in~\cite{Yuan2023}, which also realized the direct-product form proposed in Refs.~\cite{PhysRevLett.111.230401,Puchala_2013}. Additionally, Ref.~\cite{Yuan2023} provides concrete examples showing how the flatness process can be employed to tighten the majorization-based uncertainty bound. For qubit systems, unweighted majorization uncertainty relations have also been experimentally investigated using coherent light in Refs.~\cite{Wang2019,WANG2022128105}.

As highlighted by Partovi in~\cite{PhysRevA.84.052117}, the optimal bound for majorization uncertainty relations -- whether in the direct-product or direct-sum form -- does not, in general, belong to the set of probability distributions that can be realized by quantum states. In other words, this bound may not be achievable by any physical quantum state. Therefore, within the context of majorization uncertainty relations, a minimal uncertainty state does not necessarily exist.

Up to this point, we have described two main forms of majorization uncertainty relations. It appears that, despite their generality, the majorization approach cannot outperform the Maassen-Uffink bound across the entire range of $c_1$ (see Eq.~\eqref{eq:c_1}), as demonstrated in Refs.~\cite{PhysRevLett.111.230401,Puchala_2013,PhysRevA.89.052115,RevModPhys.89.015002}. This limitation holds when Shannon entropy is applied directly to the majorization form, whether in the direct-product or direct-sum case. However, by integrating majorization techniques with the data-processing inequality, Ref.~\cite{PhysRevA.89.052115} establishes results that are tighter than the Maassen-Uffink bound (see Eq.~\eqref{eq:EUR_Maassen_Uffink}).

To streamline the analysis, we focus again on rank-$1$ projective measurements, where the data-processing inequality leads to
\begin{align}
    H(M)-S(\rho)
    =
    &
    D(\rho\|\mM(\rho))\label{eq:Hyper_EUR_Maj_M_0}\\
    \geqslant
    &
    D(\mN(\rho)\|\mN\circ\mM(\rho))\label{eq:Hyper_EUR_Maj_M_1}\\
    =
    &
    H(N)-\sum_k q_k\log\sum_j p_j c_{jk},
    \label{eq:Hyper_EUR_Maj_M_2}
\end{align}
where $D$ denotes the relative entropy, $\mM$ and $\mN$ represent the quantum channels associated with the measurements $M$ and $N$, respectively. The quantities $p_j$ and $q_k$ represent the probabilities of obtaining outcomes associated with $\ketbra{u_j}{u_j}$ and $\ketbra{v_k}{v_k}$, while $c_{jk}$ is defined in Eq.~\eqref{eq:c_jk}. This technique follows directly from the method introduced by Coles and Piani in Ref.~\cite{PhysRevA.89.022112} for deriving their Eq.~\eqref{eq:EUR_Coles_Piani}. By further rearranging the inequalities above and invoking the concavity of the logarithm, we obtain
\begin{align}\label{eq:EUR_RPZ_0}
    H(M)+H(N)\geqslant-\log\sum_{jk} p_j q_k c_{jk}+S(\rho).
\end{align}
We now vectorized the overlap matrix $C:=(c_{jk})$ into a vector $\vec{c}$, using a suitable ordering of the index pairs $(j,k)$. With this representation, the first term on the right-hand side of Eq.~\eqref{eq:EUR_RPZ_0} becomes $-\log (\vec{p}\otimes\vec{q})\cdot\vec{c}$. Substituting $\vec{p}\otimes\vec{q}$ with the deterministic distribution recovers the Maassen-Uffink uncertainty relation in Eq.~\eqref{eq:EUR_Maassen_Uffink}. Alternatively, choosing $\vee S_{\otimes}$ in Eq.~\eqref{eq:Maj_DP} can yield a tighter bound
\begin{align}
    H(M)+H(N)\geqslant-\log (\vee S_{\otimes}\cdot\vec{c})+S(\rho).
\end{align}
Starting from $H(N) - S(\rho)$, we can derive an inequality in the spirit of Eq.~\eqref{eq:Hyper_EUR_Maj_M_2} 
\begin{align}\label{eq:eq:Hyper_EUR_Maj_N}
    H(M)+H(N)\geqslant
    -\sum_i p_i\log\sum_l q_l c_{il}
    +S(\rho).
\end{align}
Taking the average of Eqs.~\eqref{eq:Hyper_EUR_Maj_M_2} and~\eqref{eq:eq:Hyper_EUR_Maj_N} subsequently leads to the following entropic uncertainty relation.
\begin{align}
    H(M)+H(N)\geqslant-\frac{1}{2}
    \sum_{ik}p_iq_k\log\sum_{jl}
    p_jq_lc_{jk}c_{il}+S(\rho).
\end{align}
For each index pair $(i, k)$, we vectorize the product terms $c_{jk} c_{il}$ into a vector $\vec{c}_{(i,k)}$, and define a vector $\vec{f}$ with entries $\log (\vee S_{\otimes} \cdot \vec{c}_{(i,k)})$. This construction enables us to derive the following entropic uncertainty relation via the majorization technique
\begin{align}
    H(M)+H(N)\geqslant-\frac{1}{2}
    \vee S_{\otimes} \cdot \vec{f}
    +S(\rho).
\end{align}

Armed with the majorization-based uncertainty framework -- particularly its direct-sum formulation -- we now revisit the question raised at the end of Sec.~\ref{subsec:QU_with_QM}: Can Eq.~\eqref{eq:EUR_Coles_Piani} be sharpened by drawing on more than just the two largest entries of the overlap matrix? Xiao et al. answered this in the affirmative~\cite{Xiao_2016}. By using the quantity $2\varpi_{\oplus,n}(1/2)$, they derived a tighter entropic uncertainty relation that incorporates finer structure from the overlap matrix
\begin{align}
\label{eq:EUR_Xiao2016JPA}
    &H(M)+H(N)\notag\\
    \geqslant
    &-\log c_1
    +
    \sum_{i=1}^{\lceil(d+1)/2\rceil}\frac{2-2\varpi_{\oplus,2i}(1/2)}{2}
    \log\frac{c_i}{c_{i+1}}.
\end{align}
Here, $c_i$ denotes the $i$-th largest element of the overlap matrix. 
By definition, the ratio $c_i/c_{i+1}$ is always greater than or equal to 1, ensuring that all terms in the bound remain non-negative. 
When $i = 1$, the expression simplifies to $2 - 2\varpi_{\oplus,2}(1/2) = 1 - \sqrt{c_1}$, recovering Eq.~\eqref{eq:EUR_Coles_Piani} as a special case. 
Without loss of generality, we assume that each measurement yields $d$ possible outcomes. 
Remark that, this refined bound continues to hold even in the presence of quantum memory.

The use of majorization uncertainty relations to refine entropic uncertainty bounds illustrates the deep structural insights that majorization offers in quantum theory. 
This perspective not only strengthens the entropic framework but also highlights a unifying thread across distinct formulations of quantum uncertainty. 
Owing to the close relationship between majorization and entropy, it is natural to extend majorization uncertainty relations to the detection of quantum correlations~\cite{PhysRevA.82.012335,PhysRevA.86.022309,PhysRevA.96.032122}.
Beyond the standard forms reviewed here, majorization uncertainty relations also admit alternative formulations~\cite{PhysRevA.91.032123,Narasimhachar_2016} -- for instance, in scenarios involving memory, conditional versions can be established~\cite{PhysRevA.97.042130,PhysRevA.108.L050202}.
Finally, in parallel with the guessing game formulation discussed in Sec.~\ref{subsec:Guessing_Games}, majorization uncertainty relations can likewise be recast as interactive protocols, offering an operational interpretation. 
A detailed construction of this protocol is provided in Appendix~\ref{appendix:Lottery_Games}.


\subsection{Dynamical Uncertainty Relations}
\label{sec:DUR}

To date, investigations into uncertainty -- or information gain -- trade-offs have largely focused on static quantum states. Yet in realistic settings, quantum systems are never truly stationary; every state undergoes evolution under some quantum dynamics. In fact, both state preparation and POVMs can be viewed as particular instances of such processes, and classical information may even be embedded directly into quantum channels -- as exemplified by Pauli rotations in superdense coding~\cite{PhysRevLett.69.2881,PhysRevLett.76.4656}. These observations motivate a natural and pressing extension: What is the most general framework for probing an arbitrary quantum process? How should we rigorously define and implement ``measurements'' on dynamics? And if certain process measurements prove mutually incompatible, do they obey uncertainty relations analogous to those known for static states? This section explores the emerging unified theory of dynamical uncertainty, which addresses these questions by characterizing the fundamental limits of information extraction from arbitrary quantum evolutions. In particular, Sec.~\ref{subsec:Uncertainty_QChannel} describes uncertainty relations for quantum channels, while Sec.~\ref{subsec:Uncertainty_QCF} extends this framework to general non-Markovian dynamics and discusses applications in quantum causal inference.


\subsubsection{Uncertainty of Quantum Channels}
\label{subsec:Uncertainty_QChannel}

\begin{figure}[t]
    \centering   
    \includegraphics[width=0.48\textwidth]{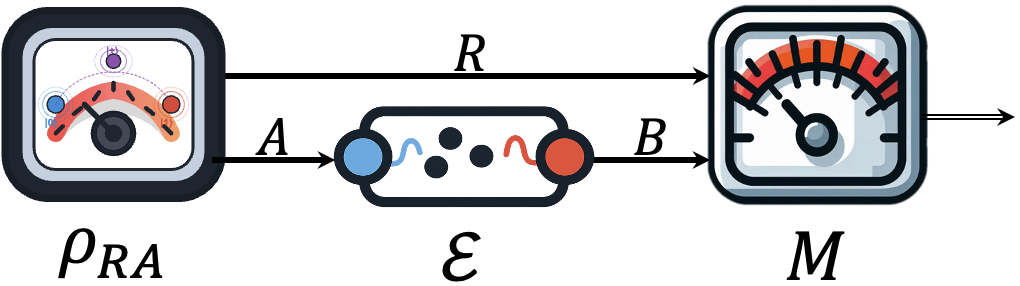}
    \caption{(Color online) \textbf{Process POVM (PPOVM)}. The measurement of a quantum channel consists of preparing a bipartite state on the channel's input and an auxiliary reference system, and then performing a measurement on the reference together with the channel's output.
   }
    \label{fig:PPOVM}
\end{figure}

Many fundamental tasks in quantum theory  -- such as preparing a state or performing a measurement -- can be understood as specific instances of quantum operations. These operations are unified under the formalism of quantum channels, which describe how a quantum system evolves between two points in time~\cite{nielsen_chuang_2010,wilde_2013,watrous_2018}. Mathematically, a quantum channel $\mE:A\to B$ is a completely positive, trace-preserving (CPTP) linear map. To extract information about such a process, one typically prepares an initial bipartite state $\rho_{RA}$, where $R$ is a reference system, and lets the channel act on subsystem $A$. This results in an output state $\id\otimes\mE(\rho_{RA})$, on which a measurement $M = \{M_j\}_{j}$ is performed to infer properties of the channel. The pair $\mT_1:=\{\rho_{RA}, M\}$ constitutes what is known as a process POVM (PPOVM)~\cite{PhysRevA.77.062112}, a generalization of conventional measurements to the setting of quantum dynamics, as illustrated in Fig.~\ref{fig:PPOVM}.

By Born's rule, the probability of observing outcome $j$ under PPOVM $\mT_1$ is given by $\Tr[M_j\cdot\id\otimes\mE(\rho)]$. 
For two PPOVMs $\mT_1$ and $\mT_2$, let $\vec{p}(\mE,\mT_1)$ and $\vec{q}(\mE,\mT_2)$ denote the respective outcome distributions when probing channel $\mE$. 
These distributions obey majorization uncertainty relations in both the direct‐product and direct‐sum formulations~\cite{PhysRevResearch.3.023077}. 
Applying any Schur‐concave function -- such as Shannon or R\'enyi entropies -- to these relations immediately yields entropic uncertainty relations for quantum channels, with the tightest direct-sum bound determined by the conditional min‐entropies (see Eq.~\eqref{eq:Entropy_Condi_Min}). 
In addition, one can further obtain channel‐analogues of the Maassen-Uffink relation (see Eq.~\eqref{eq:EUR_Maassen_Uffink_Renyi}), establishing complementary trade‐offs between the R\'enyi entropies of $H_{\alpha}(\vec{p}(\mE,\mT_1))$ and $H_{\beta}(\vec{q}(\mE,\mT_2))$ under the H\"older conjugacy constraint (see Eqs.~\eqref{eq:alpha_beta}).


\subsubsection{Uncertainty of Non-Markovian Quantum Dynamics}
\label{subsec:Uncertainty_QCF}

\begin{figure*}[t]
    \centering   
    \includegraphics[width=1\textwidth]{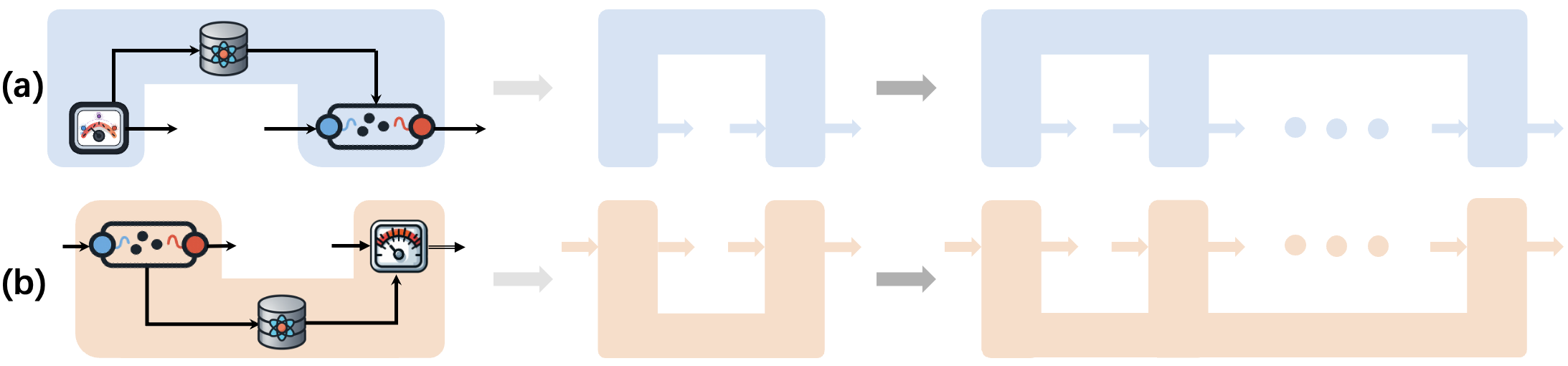}
    \caption{(Color online) \textbf{Non-Markovian Quantum Dynamics and Interactive Measurement}. In Figure (a), we illustrate the emergence of non-Markovian quantum dynamics by beginning with a simple two-process example (left): the system's history is encoded in a quantum state that feeds into a subsequent evolution described by a quantum channel. Connecting these in series yields a superchannel representation (center), which then generalizes to the fully non-Markovian case (right). In Figure (b), we show how interactive measurements probe dynamics: inserting one intervention -- modeled as a quantum channel -- followed by a final measurement defines the minimal scheme (left), which is abstracted as a one-intervention measurement (center), while permitting multiple interventions yields the most general interactive measurement (right).
   }
    \label{fig:non_Markovian}
\end{figure*}

To capture the full informational and causal structure~\cite{Rubin1974,pearl2014probabilistic,Pearl_2009} of an arbitrary quantum process, one must employ interactive measurements~\cite{PhysRevLett.130.240201} -- this involves tailored inputs and controls, actively updated based on prior measurement outcomes (see Fig.~\ref{fig:non_Markovian}(b)). Such interventions are indispensable for disentangling direct-cause and common-effect relationships from temporal correlations~\cite{Ried2015,Fitzsimons2015,MacLean2017,PhysRevX.7.031021}, and for mapping the flow of information and memory effects in genuinely non-Markovian evolutions~\cite{xiao2025superchanneltearsgeneralizedoccams}. By iteratively refining the control and measurement strategy in response to observed dynamics, interactive measurements furnish a complete reconstruction of both the mechanisms and causality underlying complex quantum processes.

The evolution of such complex quantum processes is governed by non-Markovian quantum dynamics (see Fig.~\ref{fig:non_Markovian}(a)) -- even though it goes by different names in different contexts: quantum strategies in game theory~\cite{10.1145/1250790.1250873}, quantum combs for general circuit transformations~\cite{PhysRevLett.101.060401}, process tensors in open-system modeling~\cite{PhysRevA.97.012127,PRXQuantum.2.030201,taranto2025higherorderquantumoperations}, and quantum circuit fragments in interactive-measurement and uncertainty analyses~\cite{PhysRevLett.130.240201,xing2023fundamentallimitationscommunicationquantum}. All of these formalisms employ the same operator framework to encode temporal correlations and memory effects. Consequently, any two interactive measurements probing such a process incur a fundamental trade-off: the more information one extracts, the less remains accessible to the other, giving rise to uncertainty relations for non-Markovian quantum dynamics.

Such a fundamental trade-off admits a precise expression as uncertainty relations. Let $\Theta$ denote a non-Markovian quantum process (see Fig.~\ref{fig:non_Markovian}(a)) intricately linked with two interactive measurements (see Fig.~\ref{fig:non_Markovian}(b)), $\mT_1$ and $\mT_2$, such that the output of $\Theta$ serves as the input to each interactive measurement and, conversely, the outcomes of $\mT_1$ and $\mT_2$ are fed back as inputs to $\Theta$. Writing $\vec{p}(\Theta,\mT_1)$ and $\vec{q}(\Theta,\mT_2)$ for the probability distributions generated by probing $\Theta$ with $\mT_1$ and $\mT_2$, respectively, one obtains the following majorization inequality
\begin{align}\label{eq:DUR_Maj}
    \vec{p}(\Theta,\mT_1)\oplus\vec{q}(\Theta,\mT_2)
    \prec
    \omega(\mT_1,\mT_2).
\end{align}
The left-hand side of Eq.~\eqref{eq:DUR_Maj} depends on both the dynamics and the interactive measurements, whereas the right-hand side is determined solely by the interactive measurements, independent of $\Theta$. Consequently, the bound $\omega(\mT_1,\mT_2)$ captures the intrinsic incompatibility between $\mT_1$ and $\mT_2$, and its optimal value in the direct‐sum majorization framework is established in~\cite{PhysRevLett.130.240201}.

Uncertainty relations of static quantum states quantify the fundamental trade-off between observables by characterizing the incompatibility of their associated measurements. 
In the case of interactive measurements, however, we are no longer probing static states but rather different properties of a complex quantum process, so their ``incompatibility'' reflects a trade-off between distinct dynamical features. 
For instance, in quantum causal inference~\cite{Ried2015}, one designs an interactive measurement $\mT_{\text{DC}}$ to detect a direct-cause structure and another $\mT_{\text{CC}}$ to detect a common-cause structure; these two procedures cannot both be implemented with perfect precision. 
Applying Eq.~\eqref{eq:DUR_Maj} to this pair then yields the majorization uncertainty relation that quantitatively bounds how well one can infer direct-cause versus common-cause behavior in a single experiment. 
By subsequently invoking Shannon entropy on this bound, we arrive at a causal uncertainty relation~\cite{PhysRevLett.130.240201}
\begin{align}\label{eq:CUR}
    H(\mT_{\text{DC}})+H(\mT_{\text{CC}})\geqslant H(\omega(\mT_{\text{DC}},\mT_{\text{CC}})).
\end{align}
For pure causal maps~\cite{Ried2015}, those with rank-1 Choi operators~\cite{CHOI1975285,JAMIOLKOWSKI1972275}, $H(\mT_{\text{CC}})=0$ if and only if the dynamics is purely common-cause, and likewise $H(\mT_{\text{DC}})=0$ precisely characterizes a purely direct-cause dynamics. Under these conditions, Eq.~\eqref{eq:CUR} gives $H(\mT_{\text{DC}})\geqslant H(\omega(\mT_{\text{DC}},\mT_{\text{CC}}))-H(\mT_{\text{CC}})$. Hence, for any value of $H(\mT_{\rm CC})$ with $0<H(\mT_{\rm CC})<H(\omega(\mT_{\text{DC}},\mT_{\text{CC}}))$, both pure common-cause and pure direct-cause are excluded simultaneously, implying that the underlying dynamics must embody a genuine mixture of direct-cause and common-cause. 
This result not only sharpens our operational understanding of causal structure in quantum processes but also highlights the broader potential of dynamical uncertainty relations. 
Moving beyond static scenarios, such relations offer powerful tools for certifying hybrid causal behavior~\cite{MacLean2017} and may find applications across diverse domains.


\subsection{Error Disturbance Uncertainty Relations}
\label{sec:EDUR}

Preparing i.i.d. copies of a quantum state or process, measuring them separately to collect statistical data, and analyzing this data to characterize trade-offs between different observables or properties is the essence of what is known as the preparation uncertainty relation. 
All the uncertainty relations discussed thus far fall into this category. 
While such relations are foundational to quantum theory and underpin many practical applications, they don't fully capture the spirit of Heisenberg's original idea. 
In his famous $\gamma$-ray microscope thought experiment~\cite{heisenberg1927illustrativeTranslation}, Heisenberg showed that a single measurement of position necessarily disturbs the particle's momentum. Such disturbance is not addressed by simply analyzing statistics over many independent trials. 
To capture this operational essence, two distinct frameworks have emerged. 
The noise-operator approach~\cite{PhysRevA.67.042105} defines error and disturbance using operator-based metrics, leading to state-dependent bounds that reflect how a given measurement affects a specific quantum state. 
On the other hand, the figures-of-merit approach~\cite{PhysRevLett.111.160405} compares probability distributions from ideal and actual measurements, resulting in state-independent limits that hold across all input states.
In what follows, we revisit Heisenberg's original argument in Sec.~\ref{sec:NIWD}, review the noise-operator framework in Sec.~\ref{sec:NOF}, and present the figures-of-merit approach in Sec.~\ref{sec:CEF}.


\subsubsection{No Information Without Disturbance}
\label{sec:NIWD}

To set the stage, we trace the historical roots of quantum uncertainty. As mentioned in the introduction, in 1927, while extending the kinematical and mechanical framework of classical physics into the quantum domain, Heisenberg introduced what is now called the error–disturbance relation~\cite{heisenberg1927illustrativeTranslation} -- often referred to as the measurement uncertainty relation -- to capture how the act of measuring one observable (e.g., position) inevitably disturbs its canonically conjugate counterpart (e.g., momentum). This concept stands in contrast to the preparation uncertainty relation, which bounds the intrinsic spread in outcomes when distinct ensembles of identically prepared states are measured. 

In Heisenberg's $\gamma$-ray microscope thought experiment, to locate an electron, one has to illuminate it with light; but at the very instant a photon scatters off the electron -- thereby resolving its position -- the electron receives an uncontrollable ``kick'' in momentum due to Compton effect. The mean error in position $x$ and the induced discontinuous change in momentum $p$ give rise to
\begin{align}\label{eq:Heisenberg_xp}
    (\text{Mean Error of $x$})\cdot
    (\text{Discontinuous Change in $p$})\sim h,
\end{align}
where the mean error in position is determined by the wavelength of the light. In the same paper, Heisenberg further noted that, in the Stern-Gerlach experiment, the achievable precision in energy and the duration of the atom's interaction with the magnetic field satisfy
\begin{align}\label{eq:Heisenberg_Et}
    (\text{Indeterminacy of $E$})\cdot
    (\text{Interaction Duration $t$})\sim h.
\end{align}
Together, these expressions capture the inescapable trade-off between the precisions with which observables can be determined. 
The reciprocal limitations between measurement accuracy and disturbance lies at the very heart of the uncertainty principle.

It is worth emphasizing that Heisenberg's original formulation did not involve any quantitative inequality, but rather a heuristic insight: measuring one observable inevitably disturbs its conjugate. 
Contemporary claims of ``violations'' of this principle typically arise from inconsistent or overly narrow definitions of error and disturbance, not from any actual failure of the principle itself. 
On the contrary, both advanced theoretical analyses and high-precision experiments consistently reaffirm that quantum mechanics imposes fundamental limits -- no measurement can simultaneously minimize imprecision and disturbance beyond these bounds.
For discussions of the error-disturbance trade-off considered by Heisenberg in certain special cases, where $\epsilon(x)\eta(p)=\Delta x\Delta p\geqslant\hbar/2$, see Ref.~\cite{Heisenberg1930-HEITPP}.


\subsubsection{Noise-Operator Formalism}
\label{sec:NOF}
Independent analyses by Yuen and Caves challenged the universality of the conventional Heisenberg error-disturbance relation in the context of gravitational-wave interferometry, demonstrating that it can, under certain conditions, appear to be ``violated''~\cite{PhysRevLett.51.719,PhysRevLett.54.2465,maddox1988beating}. 
In response, Ozawa developed explicit measurement models based on interaction Hamiltonians that could surpass the standard quantum limit (SQL) for free-mass position measurements~\cite{PhysRevLett.60.385}, ultimately leading to a universally valid noise-operator formalism for quantifying error and disturbance~\cite{PhysRevA.67.042105}. 
Readers interested in the conceptual foundations and practical implications of the SQL and its role in gravitational-wave detection are referred to Sec.~\ref{sec:SQL_GWD} for a more detailed discussion.

\begin{figure}[t]
    \centering   
    \includegraphics[width=0.48\textwidth]{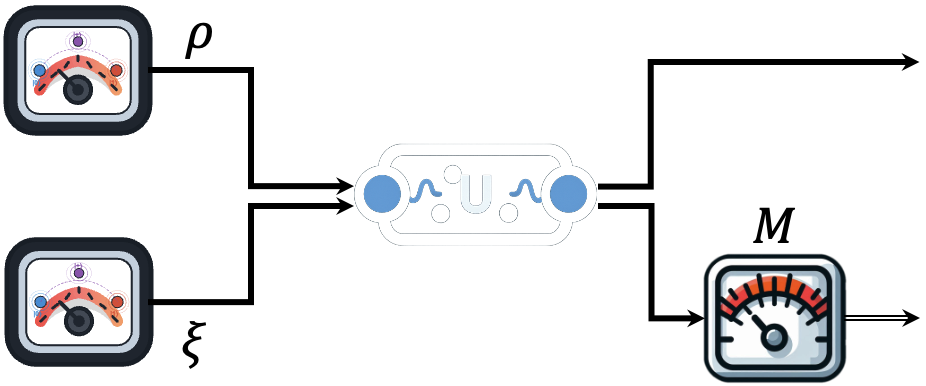}
    \caption{(Color online) \textbf{Measuring Process}. A probe, prepared in the pure state $\xi$, is coupled to the state $\rho$ by the unitary $\mU(\cdot):=U(\cdot)U^{\dagger}$; a subsequent measurement $M$ on the probe then serves to approximate the ideal observable.
   }
    \label{fig:Measuring_Process}
\end{figure}

Prior to introducing the error-disturbance uncertainty relation in terms of noise operators, we first examine the measurement process to clarify the origin of measurement-induced noise. Consider a system coupled to an auxiliary probe, which we assume, without loss of generality, to be prepared in a pure state $\xi$.  As illustrated in Fig.~\ref{fig:Measuring_Process}, the joint system undergoes a unitary interaction $\mU$. After this interaction, we perform a measurement $M$ on the probe system. In the Heisenberg picture, the noise operator $N(A)$ associated with measuring the system observable $A$ is then defined as the difference between the ideal measurement operator $A\otimes\1$ and the actually implemented operator $U^\dagger(\1\otimes M)U$~\cite{arthurs1965simultaneous,arthurs1988quantum,Ozawa2023}
\begin{align}\label{eq:EDUR_error}
    N(A):=U^\dagger(\1\otimes M)U-A\otimes\1.
\end{align}
This noisy operator originates in quantum optics; for further details, see Refs.~\cite{PhysRev.128.2407,RevModPhys.58.1001,Haus_2004,RevModPhys.82.1155}.
Meanwhile, the disturbance for observable $B$ induced by the system-probe interaction is quantified by the difference
\begin{align}\label{eq:EDUR_disturbance}
    D(B):=U^\dagger(B\otimes \1)U-B\otimes\1.
\end{align}

The measurement noise of observable $A$ on input state $\rho$, implemented by the apparatus of Fig.~\ref{fig:Measuring_Process} and denoted $\epsilon(A, M, U, \rho, \xi)$, or simply $\epsilon(A)$, is defined as the root-mean-square deviation (RMSD) of the noise operator $N(A)$~\cite{ishikawa1991uncertainty,Ozawa1991,Braginsky1992}
\begin{align}\label{eq:RMSD_error}
    \epsilon(A):=\sqrt{\langle N(A)^2 \rangle}.
\end{align}
which can be regarded as the quantum analogue of Gauss' RMSD. Similarly, the disturbance of observable $B$ induced by the same interaction is denoted $\eta(B, M, U, \rho, \xi)$, or simply $\eta(B)$, and defined as the RMSD of the disturbance operator $D(B)$
\begin{align}\label{eq:RMSD_disturbance}
    \eta(B):=\sqrt{\langle D(B)^2 \rangle}.
\end{align}
From the fact that $U^\dagger(\1\otimes M)U$ and $U^\dagger(B\otimes \1)U$ commute in Eqs.~\eqref{eq:EDUR_error} and~\eqref{eq:EDUR_disturbance}, respectively, it follows that
\begin{align}
    [N(A)+A\otimes\1,D(B)+B\otimes\1]=0,
\end{align}
which, when expanded, gives
\begin{align}
    &|\langle[N(A),D(B)]\rangle|+|\langle[N(A),B\otimes\1]\rangle|+|\langle[A\otimes\1,D(B)]\rangle|\notag\\
    &\geqslant
    |\langle[A,B]\rangle|.
\end{align}
Here, we restrict attention to the pure-state case, taking $\rho=\psi$. 
Noting that each variance is bounded above by its RMSD and invoking Robertson's relation in Eq.~\eqref{eq:VUR_Robertson} then yields Ozawa's universally valid noise-disturbance uncertainty relation~\cite{PhysRevA.67.042105}.
\begin{align}\label{eq:EDUR_Ozawa}
    \epsilon(A)\eta(B)+\epsilon(A)\Delta B+\Delta A \,\eta(B)
    \geqslant
    \frac{1}{2}|\langle[A,B]\rangle|.
\end{align}

In Ozawa's formulation, universal validity refers to the scope of the generalized noise-disturbance relation in Eq.~\eqref{eq:EDUR_Ozawa}: 
it holds for arbitrary measurement processes, arbitrary observables, and arbitrary input states, without requiring additional assumptions on the measurement interaction $\mU$ (see Eq.~\eqref{eq:EDUR_error}). 
This distinguishes it from the traditional product form (see Eq.~\eqref{eq:VUR_Robertson}), 
\begin{align}\label{eq:EDUR_Ozawa_Robertson}
    \epsilon(A)\eta(B)
    \geqslant
    \frac{1}{2}|\langle[A,B]\rangle|,
\end{align}
which is recovered only in restricted settings. 
In particular, if the measuring interaction acts as an independent intervention for the observables $A$ and $B$, so that the noise and disturbance are independent of the input states, Eq.~\eqref{eq:EDUR_Ozawa_Robertson} follows. 
For dependent interventions, however, this reduction is no longer justified, and the extra terms in Eq.~\eqref{eq:EDUR_Ozawa} are necessary to obtain a valid trade-off between noise and disturbance.

Although the bound in Eq.~\eqref{eq:EDUR_Ozawa} is state-dependent, it is not subject to the limitation discussed in Sec.~\ref{sec:VUR_PBF} and cannot be trivially tightened by the method of Eq.~\eqref{eq:VUR_trivial}. 
Indeed, the left-hand side of Eq.~\eqref{eq:EDUR_Ozawa} requires full knowledge of the coupling process $U$, the measurement $M$, the observables $A$ and $B$, the system state $\rho$, and the probe state $\xi$, whereas the right-hand side depends only on $A$, $B$, and $\rho$ -- significantly less information. 
The mismatch in informational content underscores the inherent limitations of state-dependent bounds in the context of preparation uncertainty relations, as noted in Sec.~\ref{sec:VUR_PBF}. 
For a detailed exposition of Ozawa's error-disturbance relations, including the canonical position-momentum scenario and its reformulation on a single system, see Refs.~\cite{OZAWA200321,doi:10.1142/S0219749903000437,OZAWA2004350,OZAWA2004367,PhysRevA.69.052113,Ozawa_2005,doi:10.1073/pnas.1219331110,PhysRevA.89.022124,ozawa2014errordisturbancerelationsmixedstates,Ozawa_2015,PhysRevA.102.042226,Inoue_2021}, and for their experimental verification, see Ref.~\cite{Erhart2012,baek2013experimental,PhysRevA.88.022110,PhysRevLett.112.020401,kaneda2014experimental,PhysRevLett.117.140402,Liu2019,Liu19Experimental}. 
An alternative validation via weak measurements is given by Lund and Wiseman in Ref.~\cite{Lund_2010}, with an experimental implementation reported in Ref.~\cite{PhysRevLett.109.100404}.

Subsequently, physicists recognized fundamental issues with quantifying noise and disturbance using the operator approach and RMSD~\cite{PhysRevA.89.022106}. Busch, Lahti, and Werner provided a comprehensive treatment of these problems in their review~\cite{RevModPhys.86.1261}, which we summarize in the next section. More recently, Korzekwa, Jennings, and Rudolph posed a deeper question~\cite{PhysRevA.89.052108}: if one formulates the error-disturbance relation in its product form, does the commutator still furnish a valid lower bound? They demonstrated that it does not -- that is, no universally valid relation of the form
\begin{align}\label{eq:EDUR_KJR}
    \sum_{jk}f_{jk}(A,B)\epsilon^j(A)\eta^k(B)\geqslant|\Tr[\rho[A,B]]|,
\end{align}
with $f_{00}=0$. Here, the state under consideration may be mixed, as indicated on the right-hand side of Eq.~\eqref{eq:EDUR_KJR}, and the quantities $\epsilon(A)$ and $\eta(B)$ need not be those defined in Eqs.~\eqref{eq:RMSD_error} and~\eqref{eq:RMSD_disturbance}; we continue to denote them by $\epsilon$ and $\eta$ purely for notational simplicity, with their precise definitions clear from context.

Rather than prescribing explicit formulas for $\epsilon(A)$ and $\eta(B)$, Korzekwa, Jennings, and Rudolph identify the following minimal operational requirement: if an ideal, error-free measurement $M=\{\ket{u_j}\}_j$ of $A:=\sum_j\lambda_j \ketbra{u_j}{u_j}$ leaves the statistics of any subsequent measurement $N=\{\ket{v_k}\}_k$ of $B:=\sum_k\lambda_k \ketbra{v_k}{v_k}$ completely unchanged (see Fig.~\ref{fig:Successive_Measurements}), then the disturbance to $B$ must vanish, namely $\eta(B)=0$. Mathematically, this requirement reads
\begin{align}
    \bra{v_k}
    \sum_j\ketbra{u_j}{u_j}\rho\ketbra{u_j}{u_j}
    \ket{v_k}
    =
    \bra{v_k}
    \rho
    \ket{v_k},
    \quad\forall\,k.
\end{align}
Here the left‐hand side is the probability of obtaining outcome $k$ after first measuring observable $A$, while the right‐hand side is the probability without that measurement. They further show that there exist state $\psi$ which are unbiased with respect to both observables
\begin{align}
    |\braket{\psi}{u_j}|^2=|\braket{\psi}{v_k}|^2=\frac{1}{d},
\end{align}
so that
\begin{align}
    \epsilon(A)=\eta(B)=0,
\end{align}
yet still
\begin{align}\label{eq:non_zero_comm}
    |\bra{\psi}[A,B]\ket{\psi}|\neq0.
\end{align}
In other words, although measuring observable $A$ incurs no error and produces no disturbance to $B$, one can still have Eq.~\eqref{eq:non_zero_comm}, demonstrating that no universally valid error-disturbance relation can rest solely on the expectation value of the commutator as in Eq.~\eqref{eq:EDUR_KJR}. Importantly, this impossibility hinges on the bound's state dependence, whereas Heisenberg's original insight -- though heuristic rather than formulaic -- appealed to the state‐independent constant. Ozawa subsequently redefined the RMSD of error and disturbance and incorporated them into the uncertainty relation to yield a universally valid error-disturbance bound~\cite{Ozawa2019}, which has since been experimentally validated~\cite{Sponar2021}. Nevertheless, a truly state-independent bound remains elusive within this framework.


\subsubsection{Figure-of-Merit Formalism}
\label{sec:CEF}

To quantify a measurement device's overall performance and characterize its quality with figures of merit, Busch, Lahti, and Werner introduced a state-independent framework. Their approach reformulates the measurement uncertainty relation and provides an alternative perspective on the error-disturbance trade-off in quantum mechanics~\cite{RevModPhys.86.1261}. 

\begin{figure}[t]
    \centering   
    \includegraphics[width=0.48\textwidth]{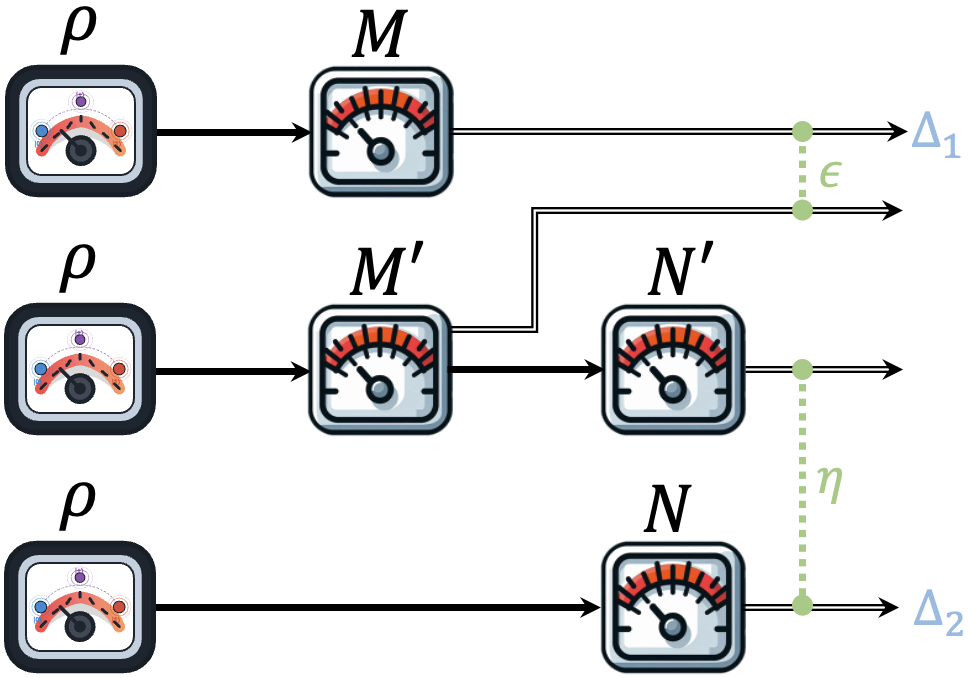}
    \caption{(Color online) \textbf{Preparation and Measurement Uncertainty}. Consider observables $A=\sum_j\lambda_j \ketbra{u_j}{u_j}$ and $B=\sum_k\lambda_k \ketbra{v_k}{v_k}$ with ideal projective measurements $M=\{\ket{u_j}\}_j$ and $N=\{\ket{v_k}\}_k$, performed in the first and third rows, respectively. Denote by $\Delta_1$ and $\Delta_2$ (not necessarily variances) the uncertainties of their outcomes; the trade-off between them yields the preparation uncertainty relation in variance-based (Sec.~\ref{sec:VUR}), entropic (Sec.~\ref{sec:EUR}), or majorization (Sec.~\ref{sec:MUR}) forms. In a real experiment, one implements an approximate measurement $M'$, whose deviation from $M$ is quantified by an error parameter $\epsilon$.  Applying an approximate measurement $N'$ to the post-$M'$ state produces statistics that differ from those of $N$ on the original state $\rho$; this disturbance is measured by $\eta$. The trade-off between $\epsilon$ and $\eta$ therefore encapsulates the measurement uncertainty relation, capturing the fundamental error–disturbance in the measuring process.
   }
    \label{fig:Measurement_Uncertainty}
\end{figure}

The framework begins with apparatus $M'$, which approximates the ideal measurement $M$ of observable $A$, followed by apparatus $N'$, which approximates the ideal measurement $N$ of observable $B$ (see Fig.~\ref{fig:Measurement_Uncertainty}). By comparing the outcome statistics of $M'$ and $N'$ with those of $M$ and $N$, it establishes state-independent bounds on both the measurement error $\epsilon(M,M')$ for $A$ and the disturbance $\eta(N,N')$ imparted to $B$.

For position and momentum, Busch, Lahti, and Werner~\cite{PhysRevLett.111.160405} establish a rigorous product-type error-disturbance relation, formally identical to Eq.~\eqref{eq:VUR_Kennard}, 
\begin{align}\label{eq:EDU_BLW_xp}
    \epsilon(x,x')\eta(p,p')\geqslant\frac{\hbar}{2},
\end{align}
where both the state-independent calibration error $\epsilon(x,x')$ of the position measurement and the disturbance $\eta(p,p')$ of the momentum distribution are quantified by their maximal Wasserstein-$2$ deviations from the ideal probability distributions; notably, in general neither metric coincides with the standard deviation of the corresponding observable, the complete proof is given in Ref.~\cite{busch2014measurement} and the conceptual origin traces back to Werner's earlier work in Ref.~\cite{10.5555/2011593.2011606}. In the qubit setting, let $A=\vec{a}\cdot\vec{\sigma}$ and $B=\vec{b}\cdot\vec{\sigma}$ be dichotomic observables with eigenvalues $\pm1$, associated ideal measurements $M$ and $N$, and denote their approximate joint implementations by $M'$ and $N'$, where they can measure jointly. We define the state-independent Wasserstein deviations as $\epsilon(M,M')$ and $\eta(N,N')$, the resulting measurement–uncertainty relation can be written as~\cite{PhysRevA.89.012129}
\begin{align}\label{eq:EDU_BLW_qubit}
    \epsilon(M,M')+\eta(N,N')\geqslant
    \sqrt{2}\left(\|\vec{a}+\vec{b}\|+\|\vec{a}-\vec{b}\|-2\right).
\end{align}
Remark that in~\cite{PhysRevA.89.012129}, the authors also introduce the preparation uncertainty bounds of Eqs.~\eqref{eq:VUR_Sum_qubit_SD} and~\eqref{eq:VUR_Sum_qubit_two_V}.
Here, measurements are jointly measurable if and only if $\|\vec{a}+\vec{b}\|+\|\vec{a}-\vec{b}\|\leqslant2$, a condition that characterizes the degree of measurement incompatibility and thus quantifies the associated measurement uncertainty.
For further background on joint measurability, or measurement incompatibility, we refer to Refs.~\cite{Heinosaari_2016,guhne2023colloquium}. 
A more comprehensive discussion of their connection to measurement uncertainty can be found in the dedicated review in Ref.~\cite{Bullock_2018}.

In Ref.~\cite{PhysRevA.96.022137}, the experimental implementation and direct comparison of Ozawa's error-disturbance relations with the Busch-Lahti-Werner framework for qubits are presented. Two key motivations underlie the latter approach. 
First, numerous subsequent studies and experiments have mischaracterized Ozawa's findings as violations of ``Heisenberg's uncertainty relation''~\cite{Inoue_2021,baek2013experimental,PhysRevA.88.022110,PhysRevLett.109.100404}, even though Heisenberg himself never formulated a precise error-disturbance bound;
accordingly, these so-called violations do not contradict any relation that Heisenberg established. Second, Ozawa's definition of measurement error involves the operator difference $U^\dagger(\1\otimes M)U-A\otimes\1$, where the observable $A\otimes\1$ and the measurement operator 
$U^\dagger(\1\otimes M)U$ need not commute; this non-commutativity renders both the physical meaning and the statistical interpretation of $U^\dagger(\1\otimes M)U$ ambiguous~\cite{RevModPhys.86.1261}. For further details and related developments, see~\cite{BUSCH2007155,busch2016quantum,PhysRevLett.116.160405,math4020038,Werner2019,PhysRevA.108.042208,PhysRevLett.131.150203} and references therein.
Beyond the formalism discussed here, the error-disturbance trade-off can also be expressed in entropies through an information-theoretic framework; see~\cite{PhysRevLett.112.050401,COLES2015105} for related developments.


\subsection{Time-Energy Uncertainty Relations}
\label{sec:TEUR}

Returning to Heisenberg's original work, he introduced both the position-momentum and the time-energy uncertainty relations~\cite{heisenberg1927illustrativeTranslation}. 
The former expresses the fundamental incompatibility of simultaneously measuring position and momentum -- observables that do not commute -- while the latter is less straightforward, since time in quantum mechanics enters as a parameter rather than an operator. 
This naturally raises the question of how one might operationally define and measure time and energy in an experiment. 
Einstein famously challenged the energy-time uncertainty relation with a thought experiment involving a weighing box in a gravitational field, and Bohr replied with a counterargument~\cite{hilgevoord1996uncertainty,hilgevoord1998uncertainty,hilgevoord2002time}. 
A decisive advance came from Mandelstam and Tamm~\cite{mandelstam1945uncertainty,mandelstam1991uncertainty}, who clarified that the product $\Delta t\Delta E$ does not pertain to simultaneous measurements but rather to the intrinsic timescale of unitary quantum evolution. 
Their insight underpins the modern formulation of the time-energy uncertainty relation and gives rise to the quantum speed limit, a fundamental bound with wide-ranging implications in quantum control~\cite{PhysRevLett.103.240501,Deffner_2014,PhysRevLett.118.100601,PhysRevLett.118.100602}, quantum metrology~\cite{doi:10.1126/science.1104149,giovannetti2006metrology,giovannetti2011advances,PhysRevA.85.052127,PhysRevLett.119.010403}, and the thermodynamics of computation~\cite{Lloyd2000,PhysRevLett.125.100602,PhysRevLett.127.190602,PhysRevLett.129.120603}. 
We will review these developments briefly below and direct the interested reader to Ref.~\cite{Deffner_2017} for a more comprehensive treatment. Specifically, we discuss quantum speed limits in Sec.~\ref{sec:QSL}, and explore time-energy uncertainty relations from the perspective of quantum clocks in Sec.~\ref{sec:QCTO}.


\subsubsection{Quantum Speed Limits}
\label{sec:QSL}

By substituting the system Hamiltonian $H$ for the observable $B$ in Robertson's uncertainty relation (see Eq.~\eqref{eq:VUR_Robertson}), one obtains an energy-observable trade-off, which can be written as
\begin{align}\label{eq:ETUR_MT_1}
    \Delta A\Delta H\geqslant
    \frac{1}{2}|\langle[A,H]\rangle|
    =
    \frac{\hbar}{2}\left|\langle\frac{\partial A}{\partial t}\rangle\right|,
\end{align}
where the equation in Eq.~\eqref{eq:ETUR_MT_1} follows directly from the Liouville-von-Neumann equation. 
Integrating this inequality under the requirement that the final state be orthogonal to the initial state leads to the Mandelstam-Tamm time-energy uncertainty relation, 
\begin{align}\label{eq:ETUR_MT}
    \tau\geqslant\frac{\pi\hbar}{2\Delta H},
\end{align}
whose physical interpretation in terms of minimal evolution time was later formalized by Aharonov and Bohm~\cite{PhysRev.122.1649}.
Uffink has further noted that employing the standard deviation as a measure of uncertainty in Eq.~\eqref{eq:ETUR_MT} can be problematic, since the bound may be driven arbitrarily close to zero even for states of finite mean energy~\cite{uffink1993rate}. To address this issue, Margolus and Levitin derived an alternative quantum speed limit~\cite{MARGOLUS1998188}
\begin{align}\label{eq:ETUR_ML}
    \tau\geqslant\frac{\pi\hbar}{2\langle H\rangle},
\end{align}
in which the lower bound is set by the system's mean energy rather than by its energy variance. An elementary proof of Eq.~\eqref{eq:ETUR_ML} appears in Ref.~\cite{PhysRevA.73.024303}, and Refs.~\cite{Brody_2003,Giovannetti_2003,Zander_2007} extend these quantum speed limits to mixed and entangled states. By taking the larger of Eq.~\eqref{eq:ETUR_MT} and~\eqref{eq:ETUR_ML}, 
\begin{align}\label{Eq:QSL_pure}
    \tau_{\text{QSL}}=\max\left\{\frac{\pi\hbar}{2\Delta H},\frac{\pi\hbar}{2\langle H\rangle}\right\}.
\end{align}
one obtains a tight bound on the evolution time required for a pure initial state to reach a target state orthogonal to it under a time-independent Hamiltonian~\cite{PhysRevLett.103.160502}. Alternative results for systems with a bounded energy spectrum are discussed in Ref.~\cite{PhysRevLett.129.140403}.

Any comprehensive extension of Eq.~\eqref{Eq:QSL_pure} must address quantum speed limits for arbitrary initial and final mixed states evolving under general, including time-dependent, Hamiltonians~\cite{PhysRevLett.110.050402,PhysRevX.6.021031}. To this end, one requires a canonical metric on states. For two pure states $\psi_1$ and $\psi_2$, one adopts the Fubini–Study angle $\theta(\psi_1,\psi_2):=\arccos{(|\braket{\psi_1}{\psi_2}|)}$. For mixed states $\rho_1$ and $\rho_2$, this notion generalizes via the Uhlmann (or Uhlmann-Jozsa) fidelity $F(\rho_1,\rho_2):=(\Tr[\sqrt{\sqrt{\rho_1}\rho_2\sqrt{\rho_1}}])^2$~\cite{UHLMANN1976273,jozsa1994fidelity,Liang_2019}, from which one defines the Bures angle $\theta(\rho_1,\rho_2):=\arccos{(\sqrt{F(\rho_1,\rho_2)})}$. With these ingredients in place, the quantum speed limit for mixed-state evolution can be expressed as~\cite{braunstein1994statistical,PhysRevA.51.1820,braunstein1996generalized}
\begin{align}\label{eq:ETUR_MT_mix}
    \tau\geqslant\frac{\hbar\,\theta(\rho_1,\rho_2)}{\Delta E_{\tau}},
\end{align}
where $\Delta E_{\tau}$ denotes the Hamiltonian's time-averaged energy variance. Alternative formulations and derivations appear in Refs.~\cite{Bhattacharyya_1983,PhysRevLett.70.3365,PhysRevA.67.052109,10.1117/12.507486,Giovannetti_2004,PhysRevA.82.022107,PhysRevA.86.016101,PhysRevLett.111.010402,Deffner_2013}, including mixed-state extensions of Eq.~\eqref{Eq:QSL_pure} in Refs.~\cite{PhysRevA.67.052109,10.1117/12.507486,Giovannetti_2004} and treatments of non-Markovian dynamics in Ref.~\cite{PhysRevLett.111.010402}.

Quantum speed limits have far-reaching implications in both physics, information theory, and computation. Bremermann fused Shannon's foundational information-theoretic framework~\cite{Shannon1948} with quantum speed limits~\cite{bremermann1967quantum}. Subsequently, Bekenstein demonstrated that the average energy cost required to transmit $k$ bit in a time $\tau_{\text{QSL}}$ is constrained by~\cite{PhysRevLett.46.623}
\begin{align}\label{eq:ETUR_Bekenstein}
    \frac{\langle H\rangle}{k}\geqslant\frac{\hbar}{\pi\tau_{\text{QSL}}}.
\end{align}
This inequality not only establishes a direct relationship between energy expenditure and the ultimate speed of information processing but also reveals a deep connection to the second law of thermodynamics and Bekenstein-Hawking entropy~\cite{PhysRevD.7.2333,Hawking1975}, particularly within the context of black hole thermodynamics. These intricate connections are detailed in Refs.~\cite{PhysRevD.9.3292,doi:10.1142/S0129183190000207}. On the other hand, it is unsurprising that the quantum speed limit relates directly to computation: Margolus and Levitin originally interpreted their bound in terms of the maximum number of gate operations a machine can execute per second~\cite{MARGOLUS1998188}. Lloyd later demonstrated that, if $\Delta t_i$ represents the time per logic gate $i$, the overall operations per second are then bounded by~\cite{Lloyd2000}
\begin{align}\label{eq:ETUR_Lloyd}
    \frac{2\langle H\rangle}{\pi\hbar}\geqslant\sum_i\frac{1}{\Delta t_i}.
\end{align}

More recently, quantum speed limits have driven progress in classical Hamiltonian dynamics~\cite{PhysRevLett.120.070401,PhysRevLett.120.070402}, stochastic processes~\cite{PhysRevLett.121.070601}, weak values~\cite{PhysRevA.99.012108}, quantum sensing~\cite{PhysRevLett.133.210802}, open quantum system~\cite{PhysRevLett.110.050403,PhysRevLett.115.210402}, and many‐body system -- including the advent of topological speed limits~\cite{PhysRevLett.130.010402}. Although space constraints preclude a comprehensive review, we pause to ask: do these advances challenge state‐independent uncertainty bounds (see Sec.~\ref{sec:VUR_PBF})? Recall that Eq.~\eqref{eq:ETUR_MT_1} follows directly from Robertson's uncertainty relation, underscoring its foundational importance. At its core, the quantum speed limit emerges from the Cauchy-Schwarz inequality, quantifying the minimum evolution time under energy constraints rather than probing measurement incompatibility. Thus, while rooted in the time-energy uncertainty relations, quantum speed limits serve a purpose fundamentally distinct from that of uncertainty relations.


\subsubsection{Quantum Clock and Time Operator}
\label{sec:QCTO}

The fundamental distinction between quantum speed limits and time-energy uncertainty relations lies in the role of time: in quantum speed limits, time enters the formalism as an parameter~\cite{PhysRevA.31.2078} rather than as an operator~\cite{PhysRev.122.1649}, whereas uncertainty relations concern the intrinsic indeterminacy in simultaneously measuring observables, which are represented by operators. Consequently, any rigorous time-energy uncertainty relation must first confront the absence of a self‐adjoint time operator in standard quantum theory~\cite{PhysRevD.7.359,PhysRevA.50.933,PhysRevLett.124.110402}. Specifically, Pauli demonstrated that the existence of such an operator would force the Hamiltonian to be unbounded from below, and hence unphysical~\cite{Pauli1933}; see also Ref.~\cite{muga2007time,butterfield2013time,Dodonov_2015} for a comprehensive overview of time in quantum theory and its implications for uncertainty. This absence sets modern operator-based uncertainty frameworks apart from both semiclassical treatments~\cite{RevModPhys.67.759} and the parameter-centric perspective of quantum speed limits (see Sec.~\ref{sec:QSL}).

Nevertheless, a viable path forward emerges by invoking quantum clocks~\cite{PhysRevD.27.2885,PhysRevD.30.368,PhysRevD.92.045033}. Just as preparation- and measurement-uncertainty relations admit both standard-deviation and entropy-based formulations, time-energy uncertainty relations can also be cast in multiple forms. In a quantum clock conditioned framework, Maccone and Sacha define a time-of-arrival operator~\cite{PhysRevLett.124.110402}, and within this setting Fadel and Maccone establish the trade-off for quantum event~\cite{PhysRevA.104.L050204}
\begin{align}\label{eq:ETUR_FM}
    \Delta T_{\text{event}}\Delta E_{\text{event}}\geqslant\frac{\hbar}{2},
\end{align}
where $T_{\text{event}}$ denotes the instant at which a specified event occurs and $E_{\text{event}}$ is the system's energy conditioned on that occurrence.

Building on the Feynman-Kitaev history‐state formalism~\cite{Feynman:85,kitaev2002classical} and the quantum time framework of~\cite{PhysRevD.92.045033}, Coles {\it et al.} formulate an entropic uncertainty relation in the presence of quantum memory~\cite{PhysRevLett.122.100401}. They obtain a Maassen-Uffink type uncertainty relations (see Eq.~\eqref{eq:EUR_Maassen_Uffink_Renyi}) in terms of R\'enyi conditional entropies $H_{\alpha}(\cdot|\cdot)$ by introducing a register $T$ to record temporal information. For any quantum state $\rho_A$, the associated temporal uncertainty is quantified by the following classical-quantum state
\begin{align}
    \sigma_{TA}:=\frac{1}{d}\sum_{j=1}^{d}\ketbra{t_j}{t_j}_{T}\otimes e^{-iHt_j}\rho_A e^{iHt_j},
\end{align}
where $\{\ket{t_j}\}_j$ forms an orthonormal basis of the reference clock -- thereby fixing the time resolution -- and 
$H$ is a finite‐dimensional Hamiltonian acting on system $A$. On the other hand, if system $A$ is entangled with a memory $B$, denoted by the joint state $\rho_{AB}$, then an energy measurement $H=\sum_k e_k E_k$ on $A$ projects the joint state into a corresponding post‐measurement state
\begin{align}
    \tau_{EB}:=\sum_k\ketbra{e_k}{e_k}\otimes\Tr_A[E_k\rho_{AB}],
\end{align}
where the operators $\{E_k\}_k$ are the orthogonal projectors onto energy eigenstates $e_k$. The pair $\sigma_{TA}$ and $\tau_{EB}$ then satisfy the following entropic time-energy uncertainty relation
\begin{align}\label{eq:ETUR_Coles}
    \widetilde{H}_{\alpha}(T|A)_{\sigma}
    +
    \widetilde{H}_{\beta}(E|B)_{\tau}\geqslant\log d.
\end{align}
As in Eq.~\eqref{eq:EUR_Maassen_Uffink_Renyi}, the orders here satisfy the H\"older conjugacy constraint given in Eq.~\eqref{eq:alpha_beta}. 
The sandwiched R\'enyi relative entropy $\widetilde{H}_{\alpha}$ appearing in Eq.~\eqref{eq:ETUR_Coles} is defined in Appendix~\ref{appendix:Entropies}. 
Noted that Gao, Junge, and LaRacuente's algebraic framework for unifying entanglement and uncertainty~\cite{gao2019unifyingentanglementuncertaintysymmetries} was later refined by Bertoni, Yang, and Renes~\cite{Bertoni_2020}, yielding an improved result of Eq.~\eqref{eq:ETUR_Coles}. 
Readers interested in related topics may refer to Hall's pioneering work~\cite{Hall_2008,Hall_2018}, examine applications of time-energy uncertainty relations, particularly their relevance to quantum key distribution~\cite{PhysRevLett.112.120506,PhysRevA.94.052323,Qi:06}, and survey recent developments in quantum metrology~\cite{PRXQuantum.4.040336}.
For readers interested in this broader perspective, we also point to the close connection between quantum chaos and quantum uncertainty~\cite{chirikov1993uncertainty}.


\subsection{Extended Themes}
\label{sec:ET}

Over the past century, Heisenberg's uncertainty principle has inspired a rich and continually evolving research landscape, encompassing a wide array of formulations, conceptual frameworks, and practical applications. 
While a comprehensive survey of all developments is beyond the scope of this review, we'll highlight several recent advancements that have substantially improved our understanding of quantum uncertainty and are not covered in previous sections. 
These include the complementary information principle, which offers an operationally meaningful and unifying framework for characterizing uncertainty region (see Sec.~\ref{sec:CIP}), and recent advances in stochastic thermodynamics, which have uncovered fundamental trade-offs between fluctuations and dissipation, leading to the formulation of thermodynamic uncertainty relations (see Sec.~\ref{sec:TUR}). 
Additional notable developments -- though not detailed here -- include the use of uncertainty relations in foundational studies, such as explaining Bell nonlocality via fine-grained uncertainty relations, see Ref.~\cite{doi:10.1126/science.1192065} and subsequent works~\cite{PhysRevA.85.024103,PhysRevLett.110.020402,PhysRevA.87.012120,PhysRevA.90.052110,Xiao_2020,PhysRevA.104.032424,PhysRevA.109.022408}, the decomposition of total uncertainty into classical and quantum components~\cite{Luo2005,PhysRevA.72.042110,PhysRevA.89.042122}, tolerant testing of stabilizer states~\cite{10.1145/3717823.3718201}, kinetic uncertainty relation~\cite{PhysRevResearch.5.023155,PhysRevLett.134.020401,rpls-mp8z,kvdn-skn1}, as well as generalizations of uncertainty relations to non-inertial reference frames~\cite{Huang2018,https://doi.org/10.1002/andp.201900014,https://doi.org/10.1002/andp.201900386,Wang2020,Shahbazi_2020,PhysRevD.102.096009,Li2021,Li2022,Dolatkhah2024,zhang2025quantumnessentropicuncertaintypair}, which further extend their conceptual reach; see the references therein for more information.


\subsubsection{Complementary Information Principle}
\label{sec:CIP}

Uncertainty relations offer only a coarse characterization of the fundamental trade-offs imposed by measurement incompatibility. Take entropic uncertainty relations as an example: they typically provide a lower bound on the sum $H(M)+H(N)$, aiming to constrain the joint uncertainty of two measurements. Geometrically, the optimal bound corresponds to a tangent line touching the lower-left corner of the uncertainty region in the $(H(M), H(N))$ plane. However, such a single-point characterization fails to fully capture the structure of complementarity between $M$ and $N$. A more complete understanding requires analyzing the entire uncertainty region -- a task that remains challenging. Even for standard entropic uncertainty relations, optimal bounds are often unknown, and determining the full boundary of the region remains an open problem. More broadly, if distinct uncertainty measures are used for different observables -- represented by non-negative Schur-concave functions $f$ and $g$ -- the relevant uncertainty region becomes $\mR(f,g):=\cup_{\rho}(f(\vec{p}),g(\vec{q}))$, where $\vec{p}$ and $\vec{q}$ are the probability distributions resulting from measuring a quantum state $\rho$ with respect to $M$ and $N$, respectively.

The first systematic construction of an outer approximation -- that is, a region enclosing the true uncertainty region $\mR(f,g)$ -- was proposed by Xiao, Fang, and Gour in Ref.~\cite{xiao2019complementaryinformationprinciplequantum}, based on the complementary information principle (CIP). This principle exploits temporal ordering: if measurement $M$ has already been performed, then even without complete knowledge of the measured state, one can derive majorization-based upper and lower bounds on the probability distribution $\vec{q}$ associated with a subsequent measurement $N$. This temporal correlation enables the construction of an outer approximation to the uncertainty region, which is provably optimal in the qubit case. Within this framework, numerous strengthened entropic uncertainty relations have been derived~\cite{xiao2019complementaryinformationprinciplequantum}; see also Ref.~\cite{doi:10.1142/S0219749915500458} for complementary results. By linking past and future measurements, CIP reveals a form of temporal correlation that may have broader implications for quantum information theory and related fields.

Alongside formulations based on entropies and other non-negative Schur-concave functions, uncertainty regions may also be formulated in terms of variances~\cite{PhysRevLett.119.170404,Szymanski_2020,Sehrawat_2020}, giving rise to sum-of-variances uncertainty relation (see Sec.~\ref{sec:VUR_SBF}), including those characterizing the joint uncertainty of angular momentum.
For example, tight state-independent bounds for variance-based uncertainty relations can be derived through convex optimization and geometric outer-approximation~\cite{PhysRevLett.119.170404}, with corresponding experimental demonstrations reported in~\cite{PhysRevLett.122.220401}.


\subsubsection{Thermodynamic Uncertainty Relations}\label{sec:TUR}
The notion of an uncertainty relation between energy and temperature dates back to Bohr and Heisenberg~\cite{bohr1985foundations}, who suggested an analogy with the position-momentum uncertainty principle. The central idea is the complementarity between energy and temperature~\cite{BOHR1928}: a system can possess a well-defined temperature only when it is coupled to a heat bath, which inherently induces fluctuations in its energy. Conversely, a sharply defined energy requires thermal isolation, rendering the concept of temperature inapplicable. This mutual exclusivity suggests a fundamental trade-off between the precisions with which energy and temperature can be simultaneously defined. For early developments, see Refs.~\cite{DELAPENAAUERBACH197265,golin1985uncertainty,de1986lesson,SCHLOGL1988679,Uffink1999,nelson2020dynamical}. A particularly well-known semiclassical formulation was provided by Rosenfeld, who quantified this trade-off in terms of the standard deviations of energy and temperature~\cite{rosenfeld1961ergodic}
\begin{align}\label{eq:TUR_Rosenfeld}
    \Delta E\Delta T\geqslant k_B T_{0}^2,
\end{align}
where $k_{B}$ is Boltzmann's constant, and $T_0$ denotes the temperature of the heat bath. Remark that here $T$ represents temperature, in contrast to its usage in Sec.~\ref{sec:TEUR}, where it denotes time.

Over the past decade, a new class of preparation uncertainty relations -- termed thermodynamic uncertainty relations (TURs) -- has emerged, characterizing fundamental trade-offs between current fluctuations and entropy production in nonequilibrium systems~\cite{horowitz2020thermodynamic}. These relations trace their origin to the study of biochemical signaling networks~\cite{PhysRevLett.113.148103}, where the accuracy of estimating the concentration of an external signal is intrinsically limited by energy consumption. This connection unveils fundamental thermodynamic constraints on statistical inference or learning, with applicability extending beyond quantum systems. In particular, Lang {\it et al.} have demonstrated that enhancing estimation accuracy inherently demands greater energy expenditure. Related insights were also developed in the pioneering work of Mehta and Schwab on the thermodynamics of cellular computation~\cite{doi:10.1073/pnas.1207814109}.

The first thermodynamic uncertainty relation was formulated in the context of a simple nonequilibrium system~\cite{PhysRevLett.114.158101}: an enzyme-catalyzed chemical reaction, modeled as a biased random walk. In this setting, the fluctuations of currents are quantified by their relative uncertainty $\mF$, given by
\begin{align}
    \mF:=\frac{\Delta X^2}{\langle X\rangle^2}.
\end{align}
Meanwhile, the thermodynamic dissipation is captured by the total entropy production $\mD$,
\begin{align}
    \mD:=T_0\sigma_{\text{epr}} t,
\end{align}
where $T$ denotes the temperature of the external environment, $t$ is the time duration, and $\sigma_{\text{epr}}$ represents the entropy production rate. The trade-off between fluctuations and dissipation is then governed by the following inequality
\begin{align}
    \mF\mD\geqslant2k_{B}T_0.
\end{align}
This foundational result was subsequently generalized by Gingrich {\it et al.} to encompass arbitrary currents in Markov jump processes~\cite{PhysRevLett.116.120601}.

In contrast to the conventional uncertainty relations focused on measurement incompatibility across different observables -- TURs reveal a fundamental trade-off between fluctuations and dissipation in nonequilibrium systems. At their core, TURs assert that entropy production imposes constraints on the precision of thermodynamic currents. However, unlike their quantum counterparts, TURs are not encapsulated by a single universal expression; instead, they emerge in dynamics-specific forms, resulting in a growing collection of related inequalities~\cite{PhysRevE.93.052145,PhysRevE.96.012101,PhysRevLett.119.160601,PhysRevLett.119.170601,PhysRevLett.120.190602,Seifert2019,PhysRevE.99.062126,PhysRevLett.125.050601,Nicholson2020,PhysRevLett.125.140602,PhysRevLett.125.260604,PhysRevLett.126.010602,PhysRevE.108.054126}. A typical formulation takes the form 
\begin{align}
    \mF\geqslant f(\mD),
\end{align}
where $\mF$ denotes the relative uncertainty, or fluctuations, and functional $f(\mD)$ quantifies the corresponding dissipation $\mD$. For a recent overview, see~\cite{horowitz2020thermodynamic} and references therein. Parallel to these developments, an alternative research direction seeks to derive TURs from foundational principles in stochastic thermodynamics, particularly through fluctuation theorem~\cite{PhysRevLett.123.090604,PhysRevLett.123.110602}.

As the study of TURs remains in its early stages, discussions and investigations in this area have largely proceeded independently from the conventional theory of quantum uncertainty relations. This raises a fundamental question: can the TURs framework -- centered on fluctuations and dissipation~\cite{6nww-8wcp} -- offer new insights into the traditional uncertainty principle? Conversely, can established concepts from conventional uncertainty relations help advance the understanding of TURs? Encouraging progress has already been made in this direction; for instance, entropic measures -- central to information processing -- have recently been employed to formulate TURs~\cite{hasegawa2025thermodynamicentropicuncertaintyrelation}, potentially enriching them with operational interpretations rooted in quantum information theory. Beyond these initial steps, it remains an open and intriguing challenge whether measurement uncertainty relations can be systematically embedded within the framework of stochastic thermodynamics, or whether multi-property trade-offs can be formalized. We anticipate that further exploration at the intersection of these two paradigms will yield fruitful developments in both fields.


\subsection{Experimental Demonstrations}
\label{sec:experimentaltestsUC}

As Peres noted~\cite{peres2002quantum}, ``Quantum phenomena do not occur in a Hilbert space. They occur in a laboratory.'' 
Uncertainty relations are not just mathematical expressions -- they reflect fundamental constraints that shape what we can observe and know about physical systems. 
Having surveyed the many theoretical formulations of quantum uncertainty, we now close Part~\ref{sec:Uncertainty_Relations} by shifting our focus to the experiments that bring these ideas to life.
Experimental tests of uncertainty relations have been essential for grounding quantum theory in reality. 
From early, indirect demonstrations using matter waves to today's finely controlled experiments with photons, atoms, and solid-state qubits, the field has steadily advanced toward more direct and precise probes of uncertainty.
In the sections that follow, Sec.~\ref{sec:Exp_PUR} reviews key experiments testing preparation uncertainty relations, while Sec.~\ref{sec:Exp_MUR} focuses on measurement uncertainty relations, highlighting experimental setups that push the boundaries of what quantum systems allow us to observe.


\subsubsection{Testing Preparation Uncertainty}
\label{sec:Exp_PUR}

Early tests of the uncertainty principle focused on the experimental verification of the position-momentum uncertainty relation (see Eq.~\eqref{eq:VUR_Kennard}). However, due to experimental limitations, initial tests only probed this relationship indirectly. One of the early experimental investigations into the wave nature of matter and its implications for the uncertainty principle involved the diffraction of neutrons. In a seminal study conducted by Shull in the 1960's, the single-slit diffraction of monochromatic slow neutrons was meticulously observed~\cite{shull1969single}. By associating the slit width with $\Delta x$ and the corresponding angular broadening effect of the neutron beam with $\Delta p$, it is possible to test Eq.~\eqref{eq:VUR_Kennard}. The experiment utilized a high-resolution double-crystal spectrometer, to analyze the passage of neutrons with a wavelength of \mbox{4.43 \r{A}} through fine slits with widths ranging from 21 to \mbox{4.1 $\mu$m}. Two gadolinium-edged slits were employed due to the high neutron absorption cross-section of gadolinium, ensuring a well-defined aperture. The angular broadening effect of the neutron beam was measured examining angles up to 20 seconds of arc, and the observed broadening agreed with the predictions derived from wave diffraction theory. This can be understood through the lens of uncertainty principle -- the narrower the slit, the greater the spatial confinement, and consequently, the larger the spread in the transverse momentum, leading to a wider diffraction pattern. Shull's experiment provided compelling evidence for the wave-like behavior of neutrons and offered a quantitative confirmation of the relationship between spatial confinement and momentum spread as predicted by the position-momentum uncertainty relation. This work has been recognized as a crucial step in understanding the quantum nature of matter and the uncertainty principle~\cite{kaiser1983direct,klein1983longitudinal,nairz2002experimental}. 

Further investigations into the quantum properties of neutrons were carried out through neutron interferometry. Two notable studies in 1980's by Kaiser et al. and Klein et al. focused on the longitudinal coherence length of neutron beams~\cite{kaiser1983direct,klein1983longitudinal}. Kaiser and colleagues performed an experiment using a perfect-silicon-crystal neutron interferometer where a series of aluminum slabs were placed in one arm of the interferometer~\cite{kaiser1983direct}. The interaction of the neutron wave packet with the aluminum slabs effectively increased the path length difference between the two arms of the interferometer beyond the coherence length of the neutrons, leading to a diminished interference effect. This experiment provided a means to directly measure the longitudinal coherence length of the neutron beam. Complementary work by Klein et al. explored the proposition that the coherence length of de Broglie wave packets remains unchanged even as the physical length of the packets increases during propagation~\cite{klein1983longitudinal}. Their findings, also based on neutron interferometry, supported this idea, contributing to a deeper understanding of the nature of neutron wave packets and their coherence properties. These experiments, while not directly measuring position and momentum uncertainty simultaneously, offered valuable insight into the wave packet characteristics of neutrons, particularly the concept of coherence, which is intrinsically linked to the uncertainties in momentum and energy through the de Broglie relations~\cite{klein1983longitudinal}. The ability of a wave to interfere with itself is fundamentally limited by its coherence length, which reflects the range of momenta present in the wave packet~\cite{clark2015controlling,cappelletti2018intrinsic}.

The verification of the position-momentum uncertainty relation was subsequently extended to the realm of more massive and complex objects such as fullerene molecules. In the early 2000's, Nairz, Arndt, and Zeilinger reported the diffraction of C70 fullerene molecules through a single slit~\cite{nairz2002experimental}. These molecules, with a mass of 840 amu and emerging from an oven at a temperature of 900 K, represent a significant step towards macroscopic quantum phenomena. The experiment involved passing a beam of these molecules through a narrow slit with a variable width, down to 70 nm, and observing the resulting diffraction pattern~\cite{nairz2002experimental}. The researchers found that the momentum spread of the molecules increased after passing through the slit, and this increase was inversely proportional to the width of the slit. Importantly, the experimental results showed good quantitative agreement with the theoretical predictions, specifically Eq.~\eqref{eq:VUR_Kennard}, where $\Delta x $ is the slit width and $\Delta p $ is the momentum transfer required to deflect the fullerene to the first interference minimum~\cite{nairz2002experimental}. This experiment provided compelling evidence that the uncertainty principle is not limited to microscopic particles but extends to objects of considerably larger mass, further solidifying its universality as a fundamental principle of physical sciences. 

While the experiments involving neutron and Fullerene diffraction, and neutron interferometry provided strong support for the position-momentum uncertainty relation, a closer examination reveals that these tests were not strictly direct demonstrations of Eq.~\eqref{eq:VUR_Kennard} at a single instant~\cite{kaiser1983direct,klein1983longitudinal,nairz2002experimental}. These experiments typically involved inferring the position uncertainty from the physical dimensions of the apparatus, such as the width of a slit or the separation between slits. Similarly, the uncertainty in momentum was not directly measured at the moment of interaction with the slit. Instead, it was often inferred from the spatial distribution of the particles at the detector, which was located at a considerable distance from the slit and thus at a later time. This inference relies on the understanding of how the wavefunction of the particle evolves in time according to the Schr\"{o}dinger equation and the connection between momentum and spatial distribution provided by the Plancherel's theorem. Thus, these tests cannot be thought of as a direct test of the uncertainty principle.

Modern experiments are able to perform more direct tests of the uncertainty principle. The field of quantum optics offers direct tests of the preparation uncertainty principle through the measurement of amplitude ($x$) and phase ($p$) quadratures that are analogous to position and momentum variables~\cite{shapiro1984phase}. It should be noted that we use the same symbols for the amplitude and phase quadratures as we do for position and momentum, as the meaning should be clear from context. These quadratures are mathematically defined as $x=a+a^\dagger$ or $p=-\mathrm{i}(a-a^\dagger)$, where $a^\dagger$ and $a$ represent the creation and annihilation operators respectively. These operators obey 
\begin{align}
    \Delta x^2 \Delta p^2
    \geqslant1,
\end{align}
which has been directly confirmed in numerous quantum optics experiments. In particular, recent experiments with squeezed light demonstrate this fundamental trade-off~\cite{steinlechner2013strong}.


\subsubsection{Testing Measurement Uncertainty}
\label{sec:Exp_MUR}

As outlined in Sec.~\ref{sec:EDUR}, measurement uncertainty relations can be broadly classified into two formulations. 
The first, proposed by Ozawa, is based on the noise-operator formalism (see Sec.~\ref{sec:NOF}). 
The second, developed by Busch, Lahti, and Werner, adopts a figure-of-merit approach rooted in operational criteria (see Sec.~\ref{sec:CEF}). 
In this section, we explore the experimental efforts that have tested these formulations, and begin with the verification of Eq.~\eqref{eq:EDUR_Ozawa} using neutron spin measurements by Erhart \textit{et al.}~\cite{Erhart2012}. 
The experimental setup consisted of a beam of monochromatic neutrons, polarized along a defined direction, passing sequentially through two successive measurements. 
The observables $A$ and $B$ were chosen as the $\sigma_x$ and $\sigma_y$, which are non-commuting. 
The first measuring apparatus was configured to perform a projective measurement of an observable $\sigma_\phi = \sigma_x \cos \phi + \sigma_y \sin \phi$, rather than directly measuring $A=\sigma_x$. 
This introduced a controllable error $\epsilon(A)=\|(\sigma_\phi-\sigma_x)\ket{\psi}\|=2\sin(\phi/2)$ (see Eq.~\eqref{eq:RMSD_error}), and in the process, disturbed observable $B=\sigma_y$ by an amount $\eta(B)=\sqrt{2}\|[\sigma_\phi,\sigma_y]\ket{\psi}\|=\sqrt{2}\cos\phi$ (see Eq.~\eqref{eq:RMSD_disturbance}).
By controlling $\phi$, Erhart \textit{et al.} systematically varied the error $\epsilon(A)$ in the measurement of $\sigma_x$ and demonstrated the resulting disturbance $\eta(B)$ on $\sigma_y$ using the second measuring apparatus. 
Successive projective measurements were performed, and the intensities of the different measurement outcomes were recorded to determine the error and disturbance. 

Following suit, Rozema \textit{et al.}~\cite{PhysRevLett.109.100404} used the three-qubit protocol proposed by Lund and Wiseman~\cite{Lund_2010} to measure the precision and disturbance associated with $\sigma_x$ and $\sigma_z$ for a single qubit, thereby verifying Eq.~\eqref{eq:EDUR_Ozawa}. 
A key advantage of this method is that it operates without requiring any prior knowledge of the system's initial state or the internal details of the measurement apparatus.
The experiment targeted the measurement of one photon polarization component, $\sigma_z$, and quantified the induced disturbance on the conjugate component, $\sigma_x$, using weak measurements~\footnote{Weak measurements are designed to extract a minimal amount of information about a quantum system without causing significant disturbance to its state, a characteristic that contrasts sharply with traditional strong measurements that lead to the collapse of the wave function.}. 
To evaluate the disturbance, $\sigma_x$ was weakly measured before the system interacted with the measurement apparatus, followed by a strong measurement of $\sigma_x$ afterward. The change in outcomes reflected the measurement-induced disturbance. 
Likewise, the precision of the $\sigma_z$ measurement was determined by weakly probing $\sigma_z$ before the apparatus and post-selecting on the probe's recorded outcome, enabling a direct estimate of the measurement error. 
To realize a measurement of variable strength, the researchers entangled the photon with an ancillary probe and performed a strong measurement on the probe. 
Entangled two-photon states were generated, with the path degrees of freedom serving to implement both the weak pre-measurement and the subsequent measurement apparatus.

While these experiments affirm the mathematical validity of Eq.~\eqref{eq:EDUR_Ozawa}, they do not fully capture the physical essence of the error-disturbance trade-off for incompatible observables (see Sec.~\ref{sec:CEF}). 
For example, the observables $\sigma_\phi$ and $\sigma_x$ employed in~\cite{Erhart2012} are not jointly measurable from a physical perspective.
Establishing a general theoretical framework that more accurately encapsulates the physical content of error-disturbance relations remains an open and pressing challenge.

Beyond noise-operator (see Sec.~\ref{sec:NOF}) and figure-of-merit (see Sec.~\ref{sec:CEF}) formulations, experimental tests of the measurement uncertainty have also been expressed in terms of fidelity-based metrics that quantify how closely a measurement reproduces the ideal outcomes. Two demonstrations published side-by-side used this approach for different physical systems~\cite{PhysRevLett.96.020408,PhysRevLett.96.020409}.

In one experiment, Sciarrino \textit{et al.} implemented a minimal-disturbance measurement on polarization-encoded qubits using linear optics, postselection, and classical feed-forward~\cite{PhysRevLett.96.020408}. They quantified the measurement process in terms of two state-dependent figures of merit: the classical estimation fidelity $G$, describing how well the unknown input state can be inferred from the measurement outcome, and the quantum fidelity $F$, expressing how closely the post-measurement state resembles the original state. By continuously tuning the measurement strength, the experiment demonstrated that the achievable pairs $(F,G)$ saturate the theoretical quantum boundary derived by Banaszek~\cite{PhysRevLett.86.1366}, thereby realizing the optimal balance between information gain and disturbance for both universal and phase-covariant qubit ensembles. A generalization of this trade-off was later derived in~\cite{PhysRevLett.96.220502}.

In a parallel experiment, Andersen \textit{et al.} extended this concept to continuous-variable systems, performing a minimal-disturbance measurement on optical coherent states using linear optics and homodyne detection~\cite{PhysRevLett.96.020409}. Here, the same fidelity-based quantities were employed to characterize the performance: the estimation fidelity quantified the accuracy with which the coherent amplitude could be inferred, while the output fidelity quantified how little the optical field was disturbed by the measurement. The results verified the theoretically predicted optimal trade-off between these fidelities under Gaussian operations (see~\cite{PhysRevA.73.032335} for non-Gaussian operations), showing that the limit is fundamentally state-dependent, set by the input coherent-state distribution and optical loss. Note that the theoretical analysis has been extended in~\cite{PhysRevA.74.012301}.

Taken together, they enrich the experimental landscape of measurement-uncertainty testing discussed in Sec.~\ref{sec:EDUR}, complementing the experimental demonstrations based on noise-operator and figure-of-merit approaches.


\section{Part 2 - The Uncertainty Principle in Practice: Implications and Applications}
\label{sec:Metrology}

As evidenced throughout history, advances in the physical sciences have often been driven by the increasing precision with which physical quantities can be measured~\cite{cavendish1798xxi,michelson1887relative,aad2012observation,abbott2016observation}. At the heart of this quest for precision lies the uncertainty principle. Often thought of as merely a constraint, the uncertainty principle establishes the fundamental structure of quantum metrology, the study of how precisely physical quantities can be estimated using quantum systems. In this sense, quantum metrology formalizes one common interpretation of the uncertainty principle -- a joint measurement uncertainty principle, as discussed in Sec.~\ref{sec:EDUR}. Quantum metrology reveals that although the uncertainty principle limits measurement precision, it also enables scientists to surpass classical bounds on precision (Sec.\ref{sec:SQLHS}). Conversely, quantum metrology also captures the trade-offs and ultimate limits involved in estimating non-commuting observables (Sec.\ref{subsec:multiparameter}), elegantly illustrated by squeezed states (Sec.\ref{section:squeezedlight}). The uncertainty principle thereby defines both the obstacles and the opportunities in quantum metrology, and can be viewed as a deep and natural manifestation of the uncertainty principle in an operational setting.


\subsection{Heisenberg Scaling}
\label{sec:SQLHS}
It is of great importance in quantum sensing to achieve the maximum precision possible given finite resources. In this context, the scaling of the measurement error with the number of quantum probe states used in the sensing protocol is a fundamental quantity. From the central limit theorem and classical statistical inference, it is known that repeating an experiment $M$ times gives rise to a $\sqrt{M}$-fold improvement in the statistical measurement error (standard deviation). We call this scaling the \textit{standard quantum limit} (SQL). Thus, there was great excitement when it was observed that by using quantum resources the error can scale as $1/M$, a $\sqrt{M}$ improvement. This scaling is known as the \textit{Heisenberg scaling} or the \textit{Heisenberg limit}, for reasons that shall become evident. In this section we shall discuss the origins and implications of this quantum improvement. We begin by introducing some preliminary material in Sec.~\ref{sec:metPrelims}. In Sec.~\ref{sec:QFIuncpr}, we show how quantum metrology reveals a fundamental uncertainty relation, which in turn defines the SQL (Sec.~\ref{subsubsecSQL}) and Heisenberg scaling (Sec.~\ref{subsubsecHS}). We then examine how Heisenberg scaling is affected by loss (Sec.~\ref{subsubsecHSloss}), evaluate claims of surpassing this limit (Sec.~\ref{subsubsecSurpass}), and finish by recalling a famous debate about the SQL in the context of gravitational wave detection (Sec.~\ref{sec:SQL_GWD}).


\subsubsection{Quantum Metrology}
\label{sec:metPrelims}
Before proceeding we shall present several definitions which will be useful going forward. In classical estimation theory the aim is to learn a parameter $\theta$, or a vector of parameters $\theta=(\theta_1,\theta_2,\hdots \theta_n)$, that are encoded into some probability distribution $p$. Unless specified otherwise, below we consider the local estimation setting, where the parameters $\theta$ are known to be close to $\theta_0$, and we are interested in sensing changes in $\theta$. Information can be obtained about $\theta$ by sampling random variables $X$ from the distribution with probability $p(X|\theta)$. This allows an estimated value of $\theta$, $\tilde{\theta}$, to be obtained.


It is of great importance to determine the most efficient estimator $\tilde{\theta}$ possible. However, in order to do so, we must first define our figure of merit. We shall use the mean squared error (MSE) as our measure of error, and define the MSE matrix as
\begin{equation}
\label{eq:MSEdefinition}
V(\theta)_{i,j}=E\left[(\tilde{\theta}_i-\theta_i)(\tilde{\theta}_j-\theta_j)\right]\;.
\end{equation}
 Note that, as is conventional, $V$ denotes the MSE matrix, since for unbiased estimators the MSE is equal to the variance. In order to obtain a single number from the MSE matrix, we define the quantity $v=\text{Tr}[WV(\theta)]$, where $W$ is some positive definite weight matrix, that assigns differing degrees of importance to different parameters. For example, in two parameter estimation, choosing $W=\1$, the figure of merit becomes $v=V_{11}+V_{22}$, assigning equal importance to both parameters. In contrast, choosing $V=\text{diag}(1,\epsilon)$ ($\epsilon\ll1$), the figure of merit becomes $v=V_{11}+\epsilon V_{22}$, assigning greater importance to the first parameter.
 
 The most efficient estimator is the one that minimises $v$ for a given $W$. For mathematical simplicity, we restrict to locally unbiased estimators, that satisfy
\begin{equation}
    E_{\theta_0}[\tilde{\theta}]=\theta_0,
\qquad
\left.\frac{\partial}{\partial \theta_j}E_{\theta}[\tilde{\theta}_i]\right|_{\theta=\theta_0}
=\delta_{ij}\;.
\end{equation}
This local-unbiasedness assumption is standard in local quantum estimation theory and underlies the Cramér--Rao bounds discussed below.

To determine a fundamental limit on how small $v$ can be, we turn to the classical Fisher information (CFI) matrix, with elements defined by
\begin{align}
\label{eq:CFImat}
J(\theta)_{i,j}&=E\bigg[\left(\frac{\partial}{\partial\theta_i}\log p(X;\theta)\frac{\partial}{\partial\theta_j}\log p(X;\theta)\right)\bigg\rvert\theta\bigg]
.
\end{align}
Note that here the expectation value is taken over all possible values $X$ conditioned on the true value $\theta$. In order to connect this to the number of samples needed to achieve some error, we turn to the Cram{\'{e}}r-Rao inequality~\cite{rao1992information,rao1947minimum}, which states
\begin{equation}
\label{eq:classicalCR}
V(\theta)\geqslant \frac{J^{-1}}{M},
\end{equation}
where $M$ is the number of samples, or equivalently repetitions of the experiment. This factor is a simple consequence of the central limit theorem. Multiplying by $W$ and taking the trace on both sides gives a bound on $v$ in terms of the CFI and number of samples. This classical Cram\'er-Rao inequality is attainable (i.e. the inequality in Eq.~\eqref{eq:classicalCR} becomes equality) in the limit where $M\to\infty$~\cite{rao1992information,rao1947minimum}.

In quantum estimation theory, the parameters $\theta$ are encoded in a quantum state rather than a probability distribution. By performing a specific measurement, a probability distribution is obtained, from which the CFI can be calculated. This raises a new question: which measurement should be implemented to maximize the corresponding CFI? In general, this is a very complex, non-convex optimization problem. However, for estimating a single parameter, the solution is known. In 1967 Helstrom introduced the quantum Fisher information (QFI)~\cite{helstrom1967minimum} as a quantum generalization of the CFI that places a lower bound on the MSE attainable.
Helstrom achieved this by introducing the symmetric logarithmic derivative (SLD) operator, defined by the following equation
\begin{equation}
\label{eq:SLDopBG}
\frac{\partial\rho}{\partial\theta_k}=\frac{1}{2}[\rho L_k+L_k\rho].
\end{equation}
An explicit form for $L$ is presented in appendix~\ref{QFIpropapen}, Eq.~\eqref{eq:apen:LDef}. $L$ is the quantum generalisation of the log derivative term required to compute the CFI in Eq.~\eqref{eq:CFImat}. For estimating a single parameter, the QFI can be calculated as 
\begin{equation}
\label{eq:SLDQFIBG}
J_{\text{S}}=\text{Tr}[\rho L^2].
\end{equation}
The QFI satisfies a similar inequality as Eq.~\eqref{eq:classicalCR}, that can be used to compute a bound on the MSE for estimating a single parameter
\begin{equation}
\label{eq:sldbound}
V(\theta)\geqslant\mathcal{C}_{\text{S}}\coloneqq\frac{1}{MJ_{\text{S}}}.
\end{equation}
This bound is sometimes called the Helstrom bound, the SLD Cram\'er-Rao bound (SLDCRB), or quantum Cram\'er-Rao bound~\cite{helstrom1967minimum}. As we present several other quantum Cram\'er-Rao bounds in Sec.~\ref{subsec:multiparameter}, we will use the term SLDCRB for Eq.~\eqref{eq:sldbound}. 

It is worth noting that while the QFI is most commonly introduced for 
parameters encoded in quantum states, analogous 
bounds can be formulated when the unknown parameters are encoded in quantum channels~\cite{fujiwara2008fibre}, or in quantum measurements themselves~\cite{das2026precision}.

\begin{figure*}
    \centering
 \includegraphics[width=0.95\textwidth]{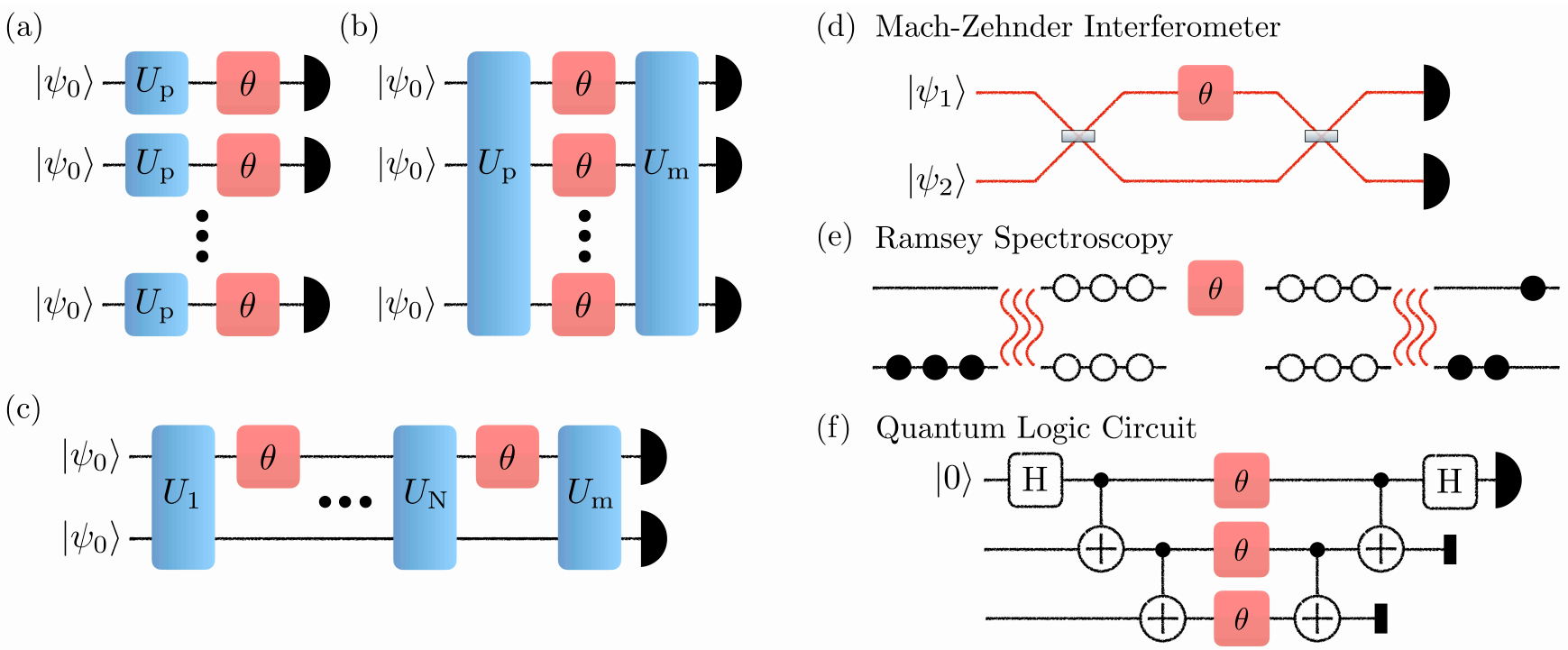}
    \caption{\textbf{Quantum Metrology with $N$ Resources.} (a)Parallel, unentangled sensing. The $N$ quantum states interact with the parameter to be measured independent of one another. (b) Parallel, entangled sensing. The $N$ quantum states may be entangled before interacting with the parameter to be measured. (c) Sequential sensing. The probe and ancilla may be prepared in any quantum state and operations can be performed on the joint state in between interacting with the parameter to be measured $N$ times. $U_\text{p}$ and $U_\text{m}$ denote preparation and measurement unitary matrices respectively. (d), (e), (f) Following Ref.~\cite{lee2002quantum}, we present the equivalence between three different physical systems (optical interferometry, Ramsey interferometry, and quantum circuits) for quantum enhanced interferometry. }
    \label{fig:RosettaStone}
\end{figure*}


\subsubsection{The Quantum Fisher Information Uncertainty Relation}
\label{sec:QFIuncpr}
The QFI can be used to derive a particular form of the uncertainty principle, which defines the quantum and classical limits of how precisely an unknown parameter can be measured. Consider estimating a single unknown parameter $\theta$ using a pure quantum state $\ket{\psi_0}$, where the parameter is encoded through some Hamiltonian $H$, as $\ket{\psi_\theta}=\mathrm{e}^{-\mathrm{i}H\theta}\ket{\psi_0}$. For this problem, the SLD operator is given by $L_\theta=2(\ket{\psi_\theta}\bra{\psi'_\theta}+\ket{\psi'_\theta}\bra{\psi_\theta})$, and the SLDCRB (Eq.~\eqref{eq:sldbound}) becomes (see appendix~\ref{apenQFIuncerP} for derivation)
\begin{equation}
\label{eq:SLDCRB:UP}
    \delta\theta^2\geqslant\frac{1}{4M(\Delta H)^2},
\end{equation}
where $\delta\theta^2$ denotes the MSE in estimating $\theta$. This relation resembles an uncertainty principle, positing a fundamental trade-off between how well a parameter can be estimated and the variance of the encoding Hamiltonian with respect to the input state. We shall refer to this as the QFI uncertainty principle. 

In order to use this uncertainty principle to determine the quantum and classical limits of estimation, it is necessary to first distinguish precisely quantum and classical protocols. Fig.~\ref{fig:RosettaStone} (a) shows a classical sensing protocol, where the $N$ sensing resources are unentangled when they interact with the parameter of interest. Fig.~\ref{fig:RosettaStone} (b) and (c) show quantum enhanced sensing protocols, whereby either the $N$ sensing resources are entangled before the interaction, or a single sensing resource interacts with the parameter of interest $N$ times sequentially. It is worth noting that protocol (c) contains protocol (b) which contains protocol (a). However, protocol (c) relies on being able to interact with the parameter of interest at $N$ different times, which may not be possible in all scenarios.

We have deliberately avoided specifying the exact nature of our quantum sensing resource above. The reason for this is that there is a formal equivalence between
the Mach-Zehnder interferometer, the Ramsey spectroscope, and a generic quantum logic circuit for sensing~\cite{lee2002quantum}. Thus, the schemes shown in Fig.~\ref{fig:RosettaStone} (d) - (f) are equivalent, but their measure of resource is different. In the optical interferometer the resource is the number of photons used in the two modes. In Ramsey spectroscopy the resource is the number of two level systems, and also the total time taken by the measurement. In quantum logic circuit sensing the resource is the number of qubits used. Additionally, in the setting of Fig.~\ref{fig:RosettaStone} (c) the number of channel uses is the resource. Counting resources in a unified manner is necessary to make a fair comparison between protocols. The different methods of counting resources depends on whether mode or particle entanglement is utilized in the underlying physical system. This in turn depends on whether the particles are distinguishable (e.g. qubits) or indistinguishable (e.g. photons).

It is now possible to see the origin of quantum advantage in sensing. For unentangled (classical) protocols, of the form in Fig.~\ref{fig:RosettaStone} (a), $\Delta H^2$ for a $N$-mode system is simply $N$ times that of a single mode system (see appendix~\ref{apenQFISQl}). The MSE in estimating $\theta$ scales inversely with $N$ and $M$ -- precisely the central limit theorem. On the contrary, when using entangled states $\Delta H^2$ can scale as $N^2$ (see appendix~\ref{apenHscale2}), giving rise to improved scaling in the estimation of $\theta$.

While the above is a very general argument as to the origin of quantum advantage, for clarity, in what follows we shall focus on optical interferometry, with the reader asked to bear in mind that the SQL and Heisenberg limit apply to other systems. For optical interferometry, the phase shift operator is written as
\begin{equation}
    \text{R}(\theta)=\text{e}^{\mathrm{i}n\theta},
\end{equation}
where $n$ is the number operator. Thus, the QFI uncertainty principle becomes
\begin{equation}
\label{eq:QFIphaseuncertainty}
    \delta\theta^2\geqslant\frac{1}{4M(\Delta n)^2}.
\end{equation}
This relation is closely related to that of Dirac's seminal paper where he wrote that $[\theta,E]=\mathrm{i}\hbar$ where $E$ is the energy of the system~\cite{dirac1927quantum}. (For more general uncertainty relations for any parameter see Refs.~\cite{braunstein1994statistical,braunstein1996generalized}.)

To understand the implications of this equation for quantum enhanced sensing, it is useful to consider some instructive quantum states. Fock states $\ket{N}$ have definite photon number, and so accordingly have completely undefined phase references, see Fig.~\ref{fig:interferometry} (d). Conversely an infinitely squeezed state has a perfectly defined phase reference, but requires a superposition of infinitely many different photon number states, see Fig.~\ref{fig:interferometry} (e). Finally, coherent states lie between these two extremes, having neither definite photon number, nor a definite phase reference, see Fig.~\ref{fig:interferometry} (f). It is important to note that for states with indefinite photon number, such as squeezed and coherent states, the measure of resources is the mean photon number, $\langle n\rangle=N$.


\subsubsection{Standard Quantum Limit}
\label{subsubsecSQL}

\begin{figure*}
    \centering
\includegraphics[width=1.8\columnwidth]{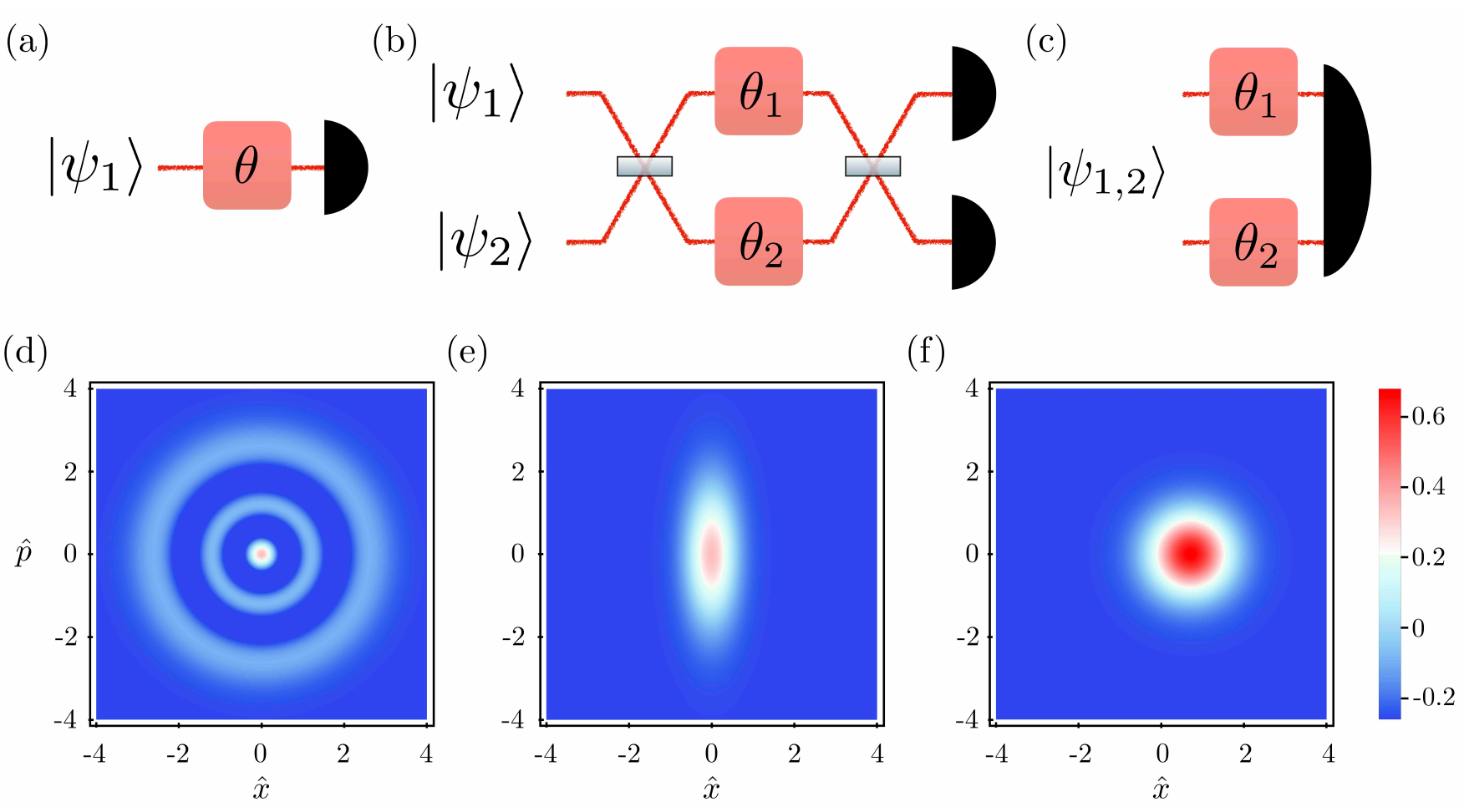}
    \caption{\textbf{Quantum Enhanced Optical Interferometry.} (a) Optical phase estimation with an implicitly assumed phase reference. (b), (c) Optical interferometry without an assumed phase reference for separable and not necessarily separable input states respectively. (d), (e) and (f) show heat maps for the Wigner function of a 4-photon Fock state, a squeezed state ($r=0.5$), and a coherent state ($\alpha=0.5$) respectively. }
    \label{fig:interferometry}
\end{figure*}

The limit on the attainable precision using classical resources only is often called the  (SQL).
This term takes a slightly different meaning depending on the context. For optical interferometry, it can be derived by considering using coherent states or single photon states in the setting shown in Fig.~\ref{fig:RosettaStone} (a). Note that this setting can also be thought of as $N$ repetitions of the same experiment. Before proceeding, it is important to point out a subtlety arising in optical interferometry. There are several different ways to carry out this task, shown in Fig.~\ref{fig:interferometry}. For example, in Fig.~\ref{fig:interferometry} (a) $\theta$ is estimated under the assumption that a fixed phase reference is incorporated into the measurement. In contrast in Fig.~\ref{fig:interferometry} (b) and (c) this assumption is dropped and the reference phase is generated via an interferometer. In this setting, the parameter of interest is the relative phase difference between the two arms. These different approaches give rise to subtle differences in the attainable precision, reflecting the different underlying assumptions.

One way to understand the SQL is to consider using $N$ uncorrelated single photon states in the configuration shown in Fig.~\ref{fig:RosettaStone} (a). Allowing for $M$ repetitions of this experiment, the total number of photons can be denoted as $N_\text{T}=NM$. Note that because the QFI is additive for tensor product states, $J_\text{S}(\rho^{\otimes N})=NJ(\rho)$ it is immediately clear that for this configuration the error scaling will be $\delta\theta\sim1/\sqrt{NM}$. To derive an explicit measurement strategy achieving this scaling, consider the Mach-Zehnder interferometer shown in Fig.~\ref{fig:interferometry} (b) with a single photon state as input ($\ket{\psi_1}\ket{\psi_2}=\ket{1}\ket{0}$) to measure the phase shift. The state immediately inside the interferometer is $(\ket{01}+\ket{10})/\sqrt{2}$, after the phase shift this becomes $(\text{e}^{\mathrm{i}\theta}\ket{01}+\ket{10})/\sqrt{2}$, and after the second beam splitter the state is $((1+\text{e}^{\mathrm{i}\theta})\ket{01}+(1-\text{e}^{\mathrm{i}\theta})\ket{10})/2$. Therefore, the probability of detecting a photon on either arm is given by $p_\pm=(1\pm\text{cos}(\theta))/2$. Assume that $\theta$ is small, so that $p_+\approx1-\theta^2/ 4$ and $p_-\approx\theta^2/4$. Therefore, we can estimate $\theta$ using the following estimator
\begin{equation}
    \tilde{\theta}=\sqrt{\frac{4N_-}{NM}},
\end{equation}
where we use the tilde symbol to denote an estimator, and $N_-$ is the total number of times the photon corresponding to $p_-$ is measured in the total $N_\text{T}$ experiments. The variance in $N_-$ is $N_\text{T}p_-p_+$. Using the standard error propagation formula we can calculate the standard deviation in our estimate of $\theta$ from this method as (appendix~\ref{apensubsubsinglephoton})
\begin{equation}
\label{Eq:SQLsinglePhotons}
    \delta\theta=\frac{\sqrt{1-p_-}}{\sqrt{NM}}=\frac{\sqrt{1-p_-}}{\sqrt{N_\text{T}}},
\end{equation}
which shows the expected SQL scaling.

To emphasize the subtle difference between the different optical interferometry configurations shown in Fig.~\ref{fig:interferometry} (a) - (c), we now consider the same task using the state $\ket{+}=(\ket{0}+\ket{1})/\sqrt{2}$ for Fig.~\ref{fig:interferometry} (a). It must be stressed that, while assuming a phase reference is reasonable in some settings (e.g. for atomic or superconducting systems), for measuring the relative phase between two photon number states it has less physical relevance. Nevertheless, we consider it here as it shows the important role an assumed phase reference plays in resource characterization. $\Delta n^2=1/4$ for the state $\ket{+}$, therefore Eq.~\eqref{eq:QFIphaseuncertainty} becomes $\delta_\theta\geqslant1/\sqrt{NM}$. Observe, however, that because $\langle n\rangle=1/2$ for this example, the average number of photons used is $N_\text{T}/2$, meaning the same precision as Eq.~\eqref{Eq:SQLsinglePhotons} has been achieved using half as many photons. This demonstrates an important difference between using channel uses (Fig.~\ref{fig:RosettaStone} (c)) and photons as the measure of resources used.

The SQL can also be derived using classical light (a coherent state $\ket{\alpha}$) to estimate $\theta$. We consider $M$ repetitions of the experiment and using a coherent state where the average photon number is $N$, $\bra{\alpha}  n\ket{\alpha}=\alpha^2=N$. After experiencing the optical phase shift, the coherent state evolves as $\ket{\alpha}\rightarrow\ket{\text{e}^{\mathrm{i}\theta}\alpha}$. By interfering this state with a bright local oscillator $\ket{\beta}$ homodyne detection can be carried out. (Note that this explicitly assumes that we have a well defined phase reference, i.e. the setting of Fig.~\ref{fig:interferometry} (a).) Homodyne detection measures either the $x$ or $p$ quadratures of an optical field. For a coherent state
\begin{equation}
\begin{split}
     \langle x\rangle&=2\text{Re}(\text{e}^{\mathrm{i}\theta}\alpha) \\
     \langle p\rangle&=2\text{Im}(\text{e}^{\mathrm{i}\theta}\alpha).
\end{split}
\end{equation}
Similarly $\Delta x^2=\Delta p^2=1$ (note that here we take the convention $\hbar=2$). By actively controlling the phase of the local oscillator $\ket{\beta}$, $\theta$ can always be  assumed to be close to 0. Measuring the $p$ quadrature gives $\langle p\rangle\approx2\theta\alpha$ with a variance of 1. Therefore, an unbiased estimator of $\theta$ is $\tilde{\theta}=\langle p\rangle/2\alpha=\langle p\rangle/2\sqrt{N}$. The standard deviation of this estimate is
\begin{equation}
\label{eq:SQLcoherent1}
    \delta\theta =\frac{1}{2\sqrt{NM}}=\frac{1}{2\sqrt{N_\text{T}}}.
\end{equation}
In the specific context of optical interferometry, this limit is often referred to as the shot noise limit. Thus far, we have only considered a single measurement strategy, it could be argued that there may exist alternative measurement strategies which may perform better and surpass the SQL. However, observe that $\Delta  n=\sqrt{\langle n^2\rangle-\langle n\rangle^2}=\alpha=\sqrt{N}$, so that for $\ket{\alpha}$ Eq.~\eqref{eq:QFIphaseuncertainty} becomes $\delta\theta\geqslant1/2\sqrt{N_\text{T}}$, which coincides with the limit derived above. In appendix~\ref{apen:cohQFI}, we calculate the QFI for this example using techniques specific to Gaussian states.

As with single photons, removing the implicitly assumed phase reference for coherent state interferometry results in a small change. Consider the configuration in Fig.~\ref{fig:interferometry} (b) driven by the quantum state $\ket{\psi_1}\ket{\psi_2}=\ket{\alpha}\ket{0}$.
One can then explicitly calculate the photocurrent and work out the variance directly, see e.g. chapter 2 in Ref.~\cite{conlon2023quantum}. This results in an estimation error of
\begin{equation}
\label{eq:SQLcoherent2}
    \delta\theta =\frac{1}{\sqrt{NM}}=\frac{1}{\sqrt{N_\text{T}}}.
\end{equation}

The importance of the presence or absence of a fixed external phase reference has been discussed in detail in Ref.~\cite{jarzyna2012quantum}. While an external phase reference results in small constant factor changes, the scaling of the SQL is of far greater importance. Indeed the $\delta\theta\sim1/\sqrt{N_\text{T}}$ scaling is fundamental, arising from the additive property of the QFI.


\subsubsection{Heisenberg Scaling}
\label{subsubsecHS}
Eq.~\eqref{eq:QFIphaseuncertainty} suggests that the SQL can be surpassed by choosing a state with large photon number variance~\footnote{Conversely states with small photon number variance are not useful for phase estimation. For example, consider a Fock state $\ket{n}$, so that $\Delta  n=0$ -- such a state is completely useless for phase estimation}. For the interferometry configuration shown in \mbox{Fig.~\ref{fig:interferometry} (b),} entangled states have a large photon number variance in one arm, suggesting they may be useful for this task. The first proposals for using entangled definite photon number states to achieve the Heisenberg limit were made by Yurke and collaborators~\cite{yurke19862, dowling1998correlated} (see also Ref.~\cite{yurke1986input} for a proposal based on fermions). The use of a maximally entangled state for sensing was first proposed in 1996~\cite{bollinger1996optimal}, and this has since become synonymous with Heisenberg scaling. Given $N$ photons to be used in the interferometry configuration shown in Fig.~\ref{fig:interferometry} (b), the state which maximizes photon number variance is the celebrated $N00N$ state, \mbox{$(\ket{N0}+\ket{0N})/\sqrt{2}$}, which has $\Delta n^2=N^2/4$. Therefore, the corresponding error scales as
\begin{equation}
\label{eq:heisenbergN00N}
    \delta\theta =\frac{1}{\sqrt{M}N}.
\end{equation}
At first glance, it appears that if all $N_\text{T}$ photons can be made into a single $N00N$ state, i.e. \mbox{$(\ket{N_\text{T}0}+\ket{0N_\text{T}})/\sqrt{2}$}, then the Heisenberg scaling is achieved exactly:
\begin{equation}
\label{eq:heisenberg2}
    \delta\theta =\frac{1}{N_\text{T}}.
\end{equation}
However, this implies that the experiment is performed exactly once. The bound in Eq.~\eqref{eq:sldbound} is attainable when asymptotically many repetitions of the experiment are considered, and deviations from this bound in the finite sample regime are known~\cite{genoni2012optical,tsang2012ziv}. Thus, the advantage of using entanglement is not quadratic with respect to all resources, only with respect to photon number.
 
The reason for this enhanced scaling is that a phase shift acting on one arm of the $N00N$ state transforms this state to
\begin{equation}
\label{eq:phaseshiftedN00N}
    \text{R}(\theta)(\ket{N0}+\ket{0N})/\sqrt{2}=(\text{e}^{N\mathrm{i}\theta}\ket{N0}+\ket{0N})/\sqrt{2}.
\end{equation}
The phase shift is encoded onto the state $N$ times faster than with single photon state. By measuring the observable $A_N=\ket{0,N}\bra{N,0}+\ket{N,0}\bra{0,N}$, it can be verified that the appropriate scaling is achieved (see appendix~\ref{apenHscale1}). Note that several other papers around the time proposed the use of other photonic states~\cite{holland1993interferometric,boto2000quantum} or measurement schemes~\cite{sanders1995optimal,ou1996complementarity} saturating the Heisenberg limit. One drawback of using $N00N$ states for phase estimation is that they are only sensitive to phases smaller than $2\pi/N$. However, this can be avoided by the scheme presented in Ref.~\cite{berry2009perform}.

One practical challenge for sensing with $N00N$ states is their experimental generation. 
As $N$ increases, state preparation and detection typically become increasingly demanding. 
Nevertheless, there have been many important experimental demonstrations of sensitivities beyond 
the standard quantum limit using $N00N$ states~\cite{mitchell2004super,walther2004broglie,
nagata2007beating,daryanoosh2018experimental,slussarenko2017unconditional,
gao2010experimental,wang2016experimental,you2021scalable}. 
When interpreting these demonstrations, it is important to distinguish conditional sensitivities, 
obtained after successful state preparation or detection, from unconditional scalings in which all 
trials and resource costs are included. In post-selected implementations, the success probability 
can reduce the overall scaling once these resources are accounted for. Notably, 
Refs.~\cite{slussarenko2017unconditional,you2021scalable} avoid this qualification by demonstrating 
unconditional quantum-enhanced sensing without post-selection.

A more experimentally feasible approach to achieving Heisenberg scaling is to use Gaussian quantum states. As these states do not have definite photon number, to compare with previous protocols, we fix the mean photon number, $\langle  n\rangle=N$. The first proposal for surpassing the SQL involved Gaussian quantum states. In the early 1980s Caves proved that using squeezed light enables one to estimate the position of mirrors in an interferometer with better precision than the SQL (see Sec.~\ref{section:squeezedlight} for a detailed introduction to squeezed light)~\cite{caves1981quantum}. This first proposal used squeezed light in conjunction with a bright coherent state, i.e. the set-up shown in Fig.~\ref{fig:interferometry}(b) with $\ket{\psi_1}\ket{\psi_2}=\ket{\alpha}\ket{r}$, where $\ket{r}$ is a squeezed state. Squeezed light has altered quadrature variances compared to coherent states of light, with $\langle x^2\rangle-\langle x\rangle^2=\text{e}^{2r}$ and $\langle p^2\rangle-\langle p\rangle^2=\text{e}^{-2r}$.
Using a displaced squeezed state we can calculate the QFI and the corresponding uncertainty in $\theta$ is given by~\cite{pezze2008mach}
\begin{equation}
    \delta\theta\geqslant\frac{1}{\sqrt{\alpha^2\text{e}^{2r}+\text{sinh}(r)^2}}
\end{equation}

For practical applications such as gravitational wave detection~\cite{tse2019quantum}, it is necessary to have $\alpha\gg \text{r}$, as large $r$ values are difficult to generate experimentally. Therefore, with this approach we only get a constant improvement by a factor $\text{e}^{r}$ over the SQL. In this scenario, where we have a bright coherent state injected into one arm of an interferometer, it is known that a squeezed vacuum state is the optimal state to inject into the other arm, subject to a constraint on the number of photons~\cite{lang2013optimal}. Nevertheless, allowing for infinite squeezing $r\to\infty$, then the mean photon number of a squeezed state is $\langle  n\rangle\to\text{e}^{2r}/4$. Assuming our photons are equally split between the coherent and squeezed state we have that $\alpha^2=\text{e}^{2r}/4=N/2$, so that finally we arrive at $\delta\theta\sim1/N$~\cite{pezze2008mach}.


Clearly, the above proposal does not represent a feasible approach to saturate the Heisenberg limit, requiring infinite squeezing. However, there are realistic scenarios where it is known that squeezed light can saturate the Heisenberg scaling with finite levels of squeezing~\cite{monras2006optimal,anisimov2010quantum}. 
This is particularly important for applications with a natural restriction on the number of photons used such as biomedical imaging, and for fundamental research. 
Using Gaussian states, it has also been demonstrated that the Heisenberg limit can be saturated using a displaced squeezed state as the input to both arms of the interferometer in Fig.~\ref{fig:interferometry}(b)~\cite{olivares2007optimized}.

In the context of Fig.~\ref{fig:interferometry}(a), where an external phase reference is assumed, using a squeezed vacuum state, Heisenberg limited sensitivity has been experimentally demonstrated recently~\cite{nielsen2023deterministic}. With an ideal squeezed state, it is possible to achieve a phase sensitivity of~\cite{monras2006optimal}
\begin{equation}
\label{eq:heisenbergSQ}
    \delta\theta =\frac{1}{\sqrt{8(\langle n\rangle^2+\langle n\rangle)}}=\frac{1}{\sqrt{8(N^2+N)}}.
\end{equation}
This phase sensitivity appears to violate the uncertainty principle and surpass the Heisenberg limit. Indeed, this is correct, however, the scaling is the same, and so this only surpasses the Heisenberg limit by a constant factor. A detailed discussion of exactly how the Heisenberg limit is surpassed will be presented later (see Sec.~\ref{sec:surpassHSindefinite} for the impatient reader).

At this point, the astute reader will have  observed that the use of squeezed light in optical interferometry appears to fall into the framework of Fig.~\ref{fig:RosettaStone}(a), in that it requires unentangled single-mode states, suggesting it should only achieve the SQL. Indeed, in both Fig.~\ref{fig:RosettaStone}(a) and Fig.~\ref{fig:interferometry}(a), there is no Heisenberg scaling with respect to the number of modes. But with squeezed light it is possible to have a Heisenberg scaling with respect to the number of photons in a single mode. This is because squeezed light collectively uses the $N$ available photons in a manner that suppresses the fluctuations associated with a coherent state, showing a fundamental distinction between Fig.~\ref{fig:RosettaStone}(a) for distinguishable and indistinguishable particles. 
A second way of understanding the advantage of squeezed light in this context is by considering Fig.~\ref{fig:interferometry}(b). Observe that if squeezed light and coherent states are the inputs to the interferometer shown in Fig.~\ref{fig:interferometry}(b), then mode entanglement is generated inside the interferometer~\cite{pezze2008mach}, reflecting the subtle differences between the three equivalent approaches shown in Fig.~\ref{fig:RosettaStone}(d) - (f)~\cite{li2023equivalence,huntington2005demonstration}. See Ref.~\cite{sahota2015quantum} for another viewpoint.

Thus far in our discussion on the Heisenberg scaling we have considered only photonic quantum states and optical interferometry. However, the advantage demonstrated in this setting immediately applies to other settings also, as shown in Fig.~\ref{fig:RosettaStone}. For example, the Heisenberg scaling can be achieved in the spectroscopy of two level systems such as atoms or ions~\cite{bollinger1996optimal}. In this setting, we have $N$ particles that can be in the ground $\ket{\text{g}}$ or excited $\ket{\text{e}}$ state, and it has been shown that the state $(\ket{\text{g}}^{\otimes N}+\ket{\text{e}}^{\otimes N})/\sqrt{2}$ can achieve the Heisenberg scaling. The GHZ state~\cite{greenberger1989going} which can be generated on quantum computers using only Hadamard and CNOT gates is also able to achieve the Heisenberg scaling. Also note that while entanglement is useful for achieving the Heisenberg scaling~\cite{huang2024entanglement}, it is not necessary~\cite{braun2018quantum,fiderer2018quantum}. For example, a single photon state acted on by $N$ sequential rotations is able to achieve Heisenberg scaling~\cite{higgins2007entanglement} (Fig.~\ref{fig:RosettaStone}(c)).  

There have been many applications of beyond SQL performance, including gravitational wave detection~\cite{tse2019quantum},  microscopy~\cite{casacio2021quantum}, and atomic clocks~\cite{cao2024multi,finkelstein2024universal}. Additionally, scaling at the Heisenberg limit has been demonstrated on multiple platforms including photonic~\cite{kacprowicz2010experimental,slussarenko2017unconditional,guo2020distributed}, trapped ion~\cite{mccormick2019quantum,leibfried2004toward}, superconducting~\cite{wang2019heisenberg} and atomic~\cite{muessel2014scalable,gross2010nonlinear} systems. 

For reviews of optical interferometry, $N00N$ states in quantum metrology, and a tutorial on optical interferometry we recommend Refs.~\cite{demkowicz2015quantum,datta2025sensing}, Refs.~\cite{kapale2007quantum,dowling2008quantum} and Ref.~\cite{barbieri2022optical} respectively. 


\subsubsection{Heisenberg Scaling with Loss}
\label{subsubsecHSloss}

While the above sounds very promising, in reality there are many obstacles in the way of achieving the Heisenberg limit in practical applications. For instance, the techniques mentioned above, i.e. using a squeezed or $N00N$ state as opposed to a coherent state only reduces noise associated with photon number fluctuations. This is important in the presence of other noise sources.For example, consider Fig.~\ref{fig:interferometry}(b) with a significant temperature difference between the upper and lower arms. This could cause the mirrors directing the light to expand or contract adding an artificial additional phase shift - different from the true phase shift we are trying to measure. Indeed, when Carlton Caves first proposed to use squeezed light in gravitational wave detection, shot noise was far from the limiting factor~\cite{caves1981quantum}, as is true in many other settings~\cite{conlon2022enhancing}. When quantum noise is not dominant, the above techniques will not be beneficial.

The second and perhaps most significant reason that the Heisenberg limit may not be achieved in practice is that the advantage provided by using \textquote{quantum} states, such as squeezed or $N00N$ states, is very sensitive to loss. To see this, let us consider a very simple example comparing how a coherent and squeezed state evolve under photon loss. We model photon loss as interaction with a beamsplitter of transmissivity $\eta$ with a vacuum state entering the other beamsplitter port. In this model, a coherent state will evolve from $\ket{\alpha}$ to $\ket{\sqrt{\eta}\alpha}$. Thus, the effective mean of the coherent state is reduced by a factor $\sqrt{\eta}$, and Eq.~\eqref{eq:SQLcoherent1} increases by a factor $1/\sqrt{\eta}$ accordingly. In contrast, a squeezed state has a variance $\text{e}^{-2r}$ in one quadrature, which after the loss transforms to $\eta\text{e}^{-2r}+1-\eta$. Consider now the scenario where $\text{e}^{-2r}\ll1$ -- small amounts of loss result in huge increases in the variance and accordingly a significant reduction in precision. For example, if $\text{e}^{-2r}=0.01$ and $\eta=0.5$, then this loss results in an increase in variance by approximately a factor of 50. This simple example serves to demonstrate that quantum states are more susceptible to loss.

To mitigate this effect, one can carefully engineer the quantum state used for estimating $\theta$ to minimize the MSE for a given level of loss. For example, Ref.~\cite{ono2010effects} analyzed the setting shown in Fig.~\ref{fig:interferometry}(b) with $\ket{\psi_1}\ket{\psi_2}=\ket{\alpha}\ket{r}$ in the presence of noise. As a greater fraction of the total photons are lost it becomes optimal to allocate more and more photons to the coherent state $\ket{\alpha}$. This is in stark contrast to the lossless case where the photons should be equally split between $\ket{\alpha}$ and $\ket{r}$. When twin Fock states are used as the input to the interferometer~\cite{holland1993interferometric}, the loss of Heisenberg scaling has been shown in the presence of various imperfections~\cite{kim1998influence,dunningham2002interferometry,meiser2009robustness,datta2011quantum}. Similarly, refs.~\cite{dorner2009optimal,demkowicz2009quantum} calculate the optimal definite photon number state in the presence of loss. With no loss, this reduces to the $N00N$ state. In the presence of decoherence, Ref.~\cite{shaji2007qubit} investigated the attainable precision for qubit states. Refs.~\cite{huelga1997improvement,andre2004stability} investigate the improvement in Ramsey spectroscopy over the SQL in the presence of different types of noise. The difficulty of achieving the Heisenberg scaling in the presence of noise has been experimentally verified~\cite{kacprowicz2010experimental}. 

The above results raise the question as to whether or not it is possible to attain the Heisenberg scaling in the presence of any loss. For lossy optical interferometry, the scaling of the measurement error with $N$ was investigated in Refs.~\cite{knysh2011scaling,kolodynski2010phase}. Here it was found that for sufficiently large $N$, the error scales proportional to $1/\sqrt{N}$, i.e. the improvement over the SQL is at most a constant factor. These results were later extended to a more general setting, where it was demonstrated that for any estimation problem where decoherence is present, in the large $N$ limit, the error scales proportional to $1/\sqrt{N}$~\cite{escher2011general,demkowicz2012elusive}. Although these results appear very negative and pessimistic at first glance, they can be flipped on their head to place limits on how much decoherence can be tolerated while achieving beyond-SQL scaling for a given value of $N$.

As such, it appeared that the presence of any noise would place fundamental limitations on sensitivity. However, shortly after the above results appeared, it emerged that quantum error correction, a tool from quantum computing, could be used to restore Heisenberg scaling. Using this idea, it was shown that for certain specific noise and signal models, Heisenberg scaling can still be achieved in the presence of noise~\cite{ozeri2013heisenberg,kessler2014quantum,arrad2014increasing,dur2014improved}. These results were later extended, providing general conditions for when error correction allows Heisenberg scaling to be achieved~\cite{lu2015robust,sekatski2017quantum,zhou2018achieving}. Subsequently there has been much theoretical work on quantum error correction for restoring the Heisenberg scaling, including ancilla free codes~\cite{layden2019ancilla}, Markovian noise~\cite{demkowicz2017adaptive},  practical limitations~\cite{shettell2021practical} and many other results~\cite{reiter2017dissipative,layden2018spatial,zhuang2020distributed,gorecki2020optimal,omanakuttan2024quantum}. There have even been some initial experimental results along this line~\cite{unden2016quantum,niroula2024quantum}.


Before finishing this section, we would like to point out that techniques other than error correction can be used to overcome noise in metrology. For example, error mitigation can be used to remove systematic noise~\cite{yamamoto2022error,conlon2023approaching} and certain noise types can be learned~\cite{ijaz2025more}.


\subsubsection{Surpassing the Heisenberg Limit}
\label{subsubsecSurpass}

Owing to the fundamental nature of the QFI uncertainty principle, Eq.~\eqref{eq:SLDCRB:UP}, one might imagine that Heisenberg scaling cannot be surpassed. Indeed, for many years the Heisenberg limit was thought to be the fundamental limit on interferometric precision~\cite{bollinger1996optimal,ou1997fundamental}. However, in recent years, there have been many proposals which surpass even this Heisenberg scaling. The reason for this is a series of (very reasonable and natural) assumptions which go into the Heisenberg limit. These assumptions are~\cite{garbe2022critical}
\begin{enumerate}
    \item  The unitary matrix encoding the dynamics, $U=\text{e}^{-\mathrm{i}H\theta}$, is constructed from a Hamiltonian $H$, that acts independently on each quantum state or mode, i.e. $H=\sum_{j=1}^{N}h^{(j)}$ where $h^{(j)}$ acts on a single mode. We also assume the Hamiltonian is bounded.
    \item $H$ is time-independent.
\item $H$ is independent of $\theta$.
\item The number of quantum systems used, $N$, for example photons or qubits, is fixed.
\end{enumerate}
With these assumptions, one may observe that the Heisenberg limit may be more aptly called the linear quantum limit. In what follows we shall describe several proposals in brief which typically break one of the assumptions above and hence can be thought of as surpassing the Heisenberg limit. 


Before beginning, there are two important points to note. The first is that, while above, we have considered primarily the Heisenberg limit for optical quantum states, this limit can take a slightly different form in other settings. For atomic sensing protocols, the Hamiltonian will act for some total time $T$ using $N$ probe states, giving the Heisenberg limit in this setting as
\begin{equation}
    \delta\theta\geqslant\frac{1}{NT}.
\end{equation}
The corresponding SQL is $\delta\theta\geqslant1/\sqrt{NT}$. Secondly, it should be mentioned that there were some early proposals that appeared to surpass the Heisenberg scaling~\cite{shapiro1989ultimate,shapiro1991quantum,dowling1991quantum}. However, upon more careful analysis, this was shown to be false, and an artifact of the figure of merit used~\cite{braunstein1992maximum,braunstein1992quantum,lane1993maximum,braunstein1994some,hradil1992quantum,hradil1992performance}.

\paragraph{Many-body Hamiltonians}
\label{section:manybody}
By relaxing the assumption that the unitary matrix encoding the dynamics we wish to sense acts separably on the $N$ different modes, i.e. assumption 1 above, we can surpass the Heisenberg limit. For instance, in Ramsey spectroscopy given $N$ modes, we typically implement the operator $R_z(\theta)$ on each mode to derive the Heisenberg limit whereby $h^{(j)}$ corresponds to the $\sigma_z$ operator acting on the $j$th qubit (see appendix~\ref{apenHscale2} for a detailed derivation). Early proposals showed that nonlinear or entangling dynamics could break this scaling~\cite{luis2004nonlinear,beltran2005breaking}; this was later formalized by considering entangling unitary operations, e.g., $H=\sigma_z\otimes\sigma_z$, which can surpass the usual Heisenberg limit~\cite{roy2008exponentially}.

Shortly after Ref.~\cite{roy2008exponentially}, Boixo \textit{et al} formalized this, showing that $k$-body Hamiltonians can achieve a scaling of~\cite{boixo2007generalized}
\begin{equation}
    \delta\theta\geqslant\frac{1}{N^{k}}.
\end{equation}
Beyond Heisenberg scaling was subsequently demonstrated experimentally~\cite{napolitano2011interaction,sewell2014ultrasensitive,hou2021super}. As a result of this progress, Ref.~\cite{zwierz2010general} carefully analyzed all of the resources used in such experiments leading to an alternative formulation of the Heisenberg limit, valid for all Hamiltonians. 

\paragraph{Time Dependent Hamiltonians}
\label{section:Timedependent}
For general time dependent protocols, the unitary matrix that encodes the dynamics of interest is given by $U=\text{e}^{-\mathrm{i}H\theta T}$, where $T$ is the total time the Hamiltonian acts for. By relaxing the second assumption above, and allowing $H$ to be time dependent, it is possible to surpass the Heisenberg limit in $T$. At first glance this appears trivial. From Eq.~\eqref{equation:HamiltonianEigs}, we know that if $H$ depends exponentially on $t$, then we can get an exponential growth in the QFI with $t$. This fact is noted in Ref.~\cite{pang2017optimal}. However, crucially, it was noted in Ref.~\cite{pang2017optimal} that by using optimal quantum controls a scaling of 
 \begin{equation}
     \delta\theta\sim\frac{1}{T^2},
 \end{equation}
can be achieved even for Hamiltonians where the difference in the maximum and minimum eigenvalue does not grow in time. This idea has been extended to distinguishing different frequency signals~\cite{gefen2019overcoming}, estimating frequencies~\cite{yang2017quantum,gefen2017control}, and has been experimentally demonstrated~\cite{naghiloo2017achieving}. Experiments with a scaling of $\delta\theta\sim1/T^{3/2}$ have also been demonstrated ~\cite{jordan2017classical,boss2017quantum,schmitt2017submillihertz}.

\paragraph{Indefinite Causal Order}
In a standard quantum experiment things happen in a definite temporal order. For example, in an optical experiment a single photon may first interact with a beam-splitter and then a phase shifter before interacting with another beam splitter. In a quantum circuit gates are implemented on a qubit one by one. Indefinite causal order throws away this idea. By using a control qubit, it is possible to implement gates in an indefinite causal order, implementing operations such as $\ket{0}\bra{0}\otimes U_1U_2+\ket{1}\bra{1}\otimes U_2U_1$. For this operation, if the control qubit is in the state $\ket{0}$, $U_2$ acts first followed by $U_1$, and the reverse is true if the control is in the state $\ket{1}$. However, if the control qubit is in the state $\ket{+}$, then the target mode experiences a superposition of gate orderings. This idea has found application in several areas of quantum information, see Ref.~\cite{rozema2024experimental} for a review.

Indefinite causal order was first applied to quantum metrology in Ref.~\cite{zhao2020quantum}. The task considered was to estimate the product of $N$ unknown displacements in both the $x$ and $p$ quadratures. Depending on the control qubit, the target qubit either experienced the $x$ quadrature displacements followed by the $p$ quadrature displacements or vice versa. Using this scheme, the product of the average displacements, $A=\bar{x}\bar{p}=(\sum_{j=1}^{N}x_j)(\sum_{j=1}^{N}p_j)/N^2$, can be estimated with a MSE that scales as $\delta A=1/N^2$. In contrast, any scheme that does not use indefinite causal order is limited to $\delta A=1/N$~\footnote{Here we use the number of operations as our measure of resources used, drawing on the equivalence in Fig.~\ref{fig:RosettaStone}.}. 
This protocol has recently been experimentally demonstrated, verifying the advantage of indefinite causal order~\cite{yin2023experimental}. At first glance, this protocol appears to violate the Heisenberg limit. To further investigate, let us explicitly construct the unitary operator implemented for this task:
\begin{equation}
    U=\Pi_{j=1}^ND_{x}(x_j)\Pi_{j=1}^ND_{p}(p_j)=\mathrm{e}^{\mathrm{i}(\sum_jp_jx-\sum_jx_jp)}.
\end{equation}
Observe that this cannot be written in the form \mbox{$U=\mathrm{e}^{\mathrm{i}HA}$}, where $H$ is some linear Hamiltonian for our parameter of interest $A$. Thus, although schemes that do not use indefinite causal order are limited to $\delta A=1/N$, this cannot be thought of as the Heisenberg limit in the usual sense, and therefore it is inaccurate to state that such schemes surpass the Heisenberg limit. 

This original indefinite causal order proposal and experiment relied on estimating displacements in an infinite dimensional Hilbert space. Around the same time, the idea of using indefinite causal order was been extended to other aspects of quantum sensing~\cite{frey2019indefinite,mukhopadhyay2018superposition,chapeau2021noisy,ban2023quantum,an2024noisy,goldberg2023evading}. However, it is worth emphasizing that Ref.~\cite{zhao2020quantum} is the only paper to prove a scaling advantage arising from indefinite causal order~\cite{mothe2024reassessing}. In fact, it has been proven that asymptotically (i.e. sufficiently large $N$, where $N$ is the number of channel uses) there is no advantage of using indefinite causal order in quantum metrology~\cite{kurdzialek2023using}. This appears to directly contradict the results of Ref.~\cite{zhao2020quantum}. However, recall that Ref.~\cite{zhao2020quantum} involves the estimation of displacements in an infinite dimensional Hilbert space. This difference between finite and infinite dimensional systems is striking. For results on the theory of indefinite causal order with a finite number of channel uses, see Ref.~\cite{liu2023optimal}.

\paragraph{Post-selection Based Metrology}
Post-selection of measurement results in quantum mechanics can lead to many surprising phenomena, including improved teleportation~\cite{zhao2023enhancing,shajilal2024improving}, generation of non-Gaussian states~\cite{jeng2024strong,larsen2025integrated}, and values of observables that lie outside the corresponding eigenspectrum~\cite{aharonov1988result,duck1989sense,ferrie2014result}. From a fundamental quantum sensing viewpoint, one might expect that post-selecting certain measurement results (and discarding the rest) can only decrease the precision in a given experiment. This idea follows directly from the data processing inequality~\cite{beaudry2011intuitive}. Early discussions on whether or not post-selection could enhance the accuracy with which we can estimate some unknown parameter focused mainly on weak values~\cite{aharonov1988result,aharonov1990properties}, defined by postselecting measurement results on specific initial and final states. This technique has been used to measure the spin Hall effect of light~\cite{hosten2008observation}, reconstruct Bohmian trajectories~\cite{kocsis2011observing}, determine beam deflections~\cite{dixon2009ultrasensitive,starling2009optimizing} among many other applications~\cite{brunner2010measuring,xu2013phase,gorodetski2012weak,feizpour2011amplifying,xu2020approaching}. Intriguingly, by appropriately choosing the initial and final states, the weak value can be significantly amplified. Weak value amplification has been proven to have an advantage over conventional measurements in some noise models~\cite{jordan2014technical,harris2017weak,xu2020approaching}, however, there is no advantage under other noise models~\cite{knee2013quantum,knee2014amplification}. In the case of ideal detectors, it is known that weak value amplification and more generally postselection offers no advantage over conventional measurements when all measurement results (not just the postselected ones) are taken into account~\cite{knee2014amplification,zhang2015precision,tanaka2013information,alves2015weak,ferrie2014weak,combes2014quantum}. More recently it has been demonstrated that when a cost is assigned to the final measurement, postselection can improve the Fisher information per trial~\cite{assad2017phase,arvidsson2020quantum}. This advantage was proven to arise from negative quasi-probability distributions. This has been demonstrated experimentally, in a set-up offering an advantage per detected photon~\cite{lupu2022negative}. However, to conclude, we must remember that in the scenario where detection is ideal, and no cost is assigned to performing the measurement, then postselection does not improve performance. This does raise the interesting question as to whether postselection can enable measurement of minuscule physical effects in scenarios where quantum noise is not dominant~\cite{mohageg2022deep,zych2012general,schaferfeasibility}.

\paragraph{Critical Point Scaling}
A quantum critical point is the point at which a phase transition occurs in some quantum system as a model parameter $\lambda$ is varied. At such a point, one would expect a correspondingly large change in the fidelity. For a Hamiltonian $H=H_0+\lambda H_1$, we know from Sec.~\ref{section:manybody}, that when $H_1$ consists of only single body terms the best estimate for $\lambda$ will have a precision bounded by the Heisenberg limit. Defining the fidelity between two states acted upon by $H$ as 
\begin{equation}
    F(\lambda,\delta\lambda)=|\braket{\psi(\lambda)}{\psi(\lambda+\delta\lambda)}|,
\end{equation}
allows us to define the fidelity  susceptibility as~\cite{you2007fidelity}
\begin{equation}
    F(\lambda,\delta\lambda)=1-\frac{1}{2}\chi_F\delta\lambda^2+O(\delta\lambda^3).
\end{equation}
Given that we expect the estimation of $\lambda$ to be at the Heisenberg limit, the connection between the QFI and the fidelity of a quantum state (see appendix~\ref{QFIpropapen}), dictates that we should expect a fidelity susceptibility scaling of $\chi_F\sim N^{-2}$. However, in Refs.~\cite{gong2008fidelity,gu2008fidelity,greschner2013fidelity} it was observed that the fidelity susceptibility between two subsystems can scale as $N^{-l}$, where $2\leqslant l\leqslant3$, leading to apparent super-Heisenberg scaling. However, this was resolved in Ref.~\cite{rams2018limits}, by properly accounting for the time needed for the entire sensing protocol (see also Ref.~\cite{garbe2020critical}). In Refs.~\cite{garbe2022critical,garbe2022exponential} it was demonstrated that a precision scaling in time significantly beyond the Heisenberg limit can be achieved. However, these protocols relied upon both time dependent Hamiltonians (Sec.~\ref{section:Timedependent}) and critical points to achieve this effect. Additionally, these Hamiltonians actually changed $N$ throughout the protocol, thus breaking assumption 4 that goes into the Heisenberg scaling. Finally, we note that experimental critical point metrology at the Heisenberg limit has been demonstrated~\cite{liu2021experimental}.

\paragraph{Indefinite Photon Number States}
\label{sec:surpassHSindefinite}
In Sec.~\ref{subsubsecSQL}, we derived the SQL by considering the mean photon number, $\langle n\rangle$ in a coherent state $\ket{\alpha}$. This is fine for classical states, however, can lead to apparent violations of the Heisenberg scaling when we allow \textquote{more quantum} states. For example, squeezed states of light and entangled coherent states are able to surpass the Heisenberg limit by a constant factor~\cite{anisimov2010quantum,nielsen2023deterministic,joo2011quantum}. The reason for this, as we shall see, is that $\langle n\rangle$ is not an accurate quantifier of the resources used in metrology for states with an indefinite photon number. For squeezed states of light and entangled coherent states, this is not very important as they do not achieve any improvement in scaling over the Heisenberg limit. However, for more exotic states with fixed $\langle n\rangle$ this is not the case.

In Ref.~\cite{rivas2012sub}, the following state was proposed $\ket{\psi}=\mu\ket{0}+\nu\ket{\epsilon}$, where $\ket{\epsilon}$ represents a displaced squeezed state. By allowing $\nu\to0$ it was shown that an arbitrarily large QFI can be obtained for finite mean photon number. The same result was shown for a superposition of all possible (i.e. infinitely many) $N00N$ states with an appropriate coefficient~\cite{zhang2012unbounded}. In Ref.~\cite{zhang2012unbounded} it was noted that although the QFI can go to infinity for finite $\langle n\rangle$, for their proposed state $\langle n^2\rangle\to\infty$. This suggests a new limit, proposed by Hoffman, may represent the true limit of interferometry~\cite{hofmann2009all}
\begin{equation}
    \delta\theta\geqslant\frac{1}{\sqrt{\langle n^2\rangle}}.
\end{equation}


However, before completely abandoning the Heisenberg limit, let us first return to the reason for considering $\langle n\rangle$ to be a good estimate of the resources needed for quantum sensing. For most quantum states, $\langle n\rangle$ is a reasonable measure of how difficult the state is to experimentally generate: for coherent states increasing $\langle n\rangle$ requires us to build more powerful lasers, for squeezed states of light increasing $\langle n\rangle$ requires more optical non-linearity or input power and for $N00N$ states increasing $\langle n\rangle$ requires far more complex interferometric sequences or lower probability of successfully producing the state. Thus, it appears reasonable at first glance to assume that increasing QFI for a fixed $\langle n\rangle$ as in the proposals of Refs.~\cite{rivas2012sub,zhang2012unbounded} is a good thing. However, in both proposals increasing QFI is only obtained by states that are increasingly complex to generate experimentally. 

The second argument typically given for desiring increased QFI at a given value of $\langle n\rangle$, is that quantum states may be used to sense fragile samples, that would be damaged if the photon number is too large. However, as is highlighted in Ref.~\cite{branford2021average}, for the states described above~\cite{rivas2012sub,zhang2012unbounded}, a small $\langle n\rangle$ does not imply that a fragile sample would not be damaged. Indeed, it is demonstrated in Ref.~\cite{branford2021average} that the results of Refs.~\cite{rivas2012sub,zhang2012unbounded} arise from the probabilistic implementation of states that do achieve Heisenberg scaling, i.e. the probabilistic implementation of high photon number states which would damage a fragile sample.

Even more crucially, there is a fundamental flaw in the proposals of Refs.~\cite{rivas2012sub,zhang2012unbounded}. In a number of works it was demonstrated that these proposals only work if one already knows the correct value of $\theta$ to within a certain accuracy. Thus, these proposals can be thought of as taking advantage of prior information about $\theta$~\cite{hall2012universality,berry2012optimal,tsang2012ziv,hall2012heisenberg,giovannetti2012sub,jarzyna2015true}. Indeed, it has even been shown that with the amount of prior information required to achieve beyond-Heisenberg scaling, one can simply randomly guess $\theta$ within this prior range and achieve a similar precision~\cite{giovannetti2012sub}. These works collectively restore the validity of the Heisenberg limit in the face of challenges from Refs.~\cite{rivas2012sub,zhang2012unbounded}, and highlight the importance of properly accounting for all resources to recover the Heisenberg scaling. 


\subsubsection{Gravitational Wave Detection}
\label{sec:SQL_GWD}

Finally, we note that surpassing the standard quantum limit in gravitational wave detection (equivalently measuring the position of a test mass) was the cause of much debate in the past century~\cite{maddox1988beating,doi:10.1126/science.1104149}. The original SQL in gravitational wave detection, can be understood as follows~\cite{braginskiui1975quantum}. Suppose a particle of mass $m$ is prepared at $t=0$ with position and momentum spreads of $\delta x_0$ and $\delta p_0$ respectively. Measuring the position of the particle at a time $t=\tau$, the total uncertainty in its position is the sum of the original position uncertainty and the propagated uncertainty due to the momentum spread. Optimizing $\delta x_0$ and $\delta p_0$ subject to the constraints of the uncertainty principle, one arrives at $(\delta x)^2\geqslant \hbar\tau/2m$.

In the early 1980's several papers demonstrated that with a more appropriate experimental configuration, this limit can be surpassed. This involved a seminal proposal by Caves to use squeezed light~\cite{caves1981quantum}, which decades later was implemented to great effect~\cite{aasi2013enhanced}. Other proposals included quantum non-demolition measurements for resonant bar gravitational wave detectors~\cite{braginsky1980quantum}, and for determining the speed of a test mass~\cite{khalili1996speed}. Importantly, these papers did not claim there was any fundamental flaw in the SQL argument above.

However, by considering states with correlated position and momentum, Yuen demonstrated that the conventional SQL ($(\delta x)^2\geqslant \hbar\tau/2m$) is not universally valid~\cite{PhysRevLett.51.719}. Yuen's work proved controversial, with several authors responding~\cite{wodkiewicz1984comment,yuen1984yuen,lynch1984comment,yuen1984yuen2,lynch1985repeated}. Caves attempted to put the discussion on a precise mathematical footing by
reformulating the SQL to consider two measurements made with identical apparatus a time $\tau$ apart~\cite{PhysRevLett.54.2465}. The debate was finally put to bed when Ozawa constructed explicit models that surpass the standard quantum limit~\cite{PhysRevLett.60.385}, that ultimately led to his universally valid noise-operator formalism for the error-disturbance uncertainty relation~\cite{PhysRevA.67.042105} discussed in detail in Sec.~\ref{sec:NOF}. This debate was instrumental in improving our understanding of the ultimate attainable limits in quantum metrology, and in the development of many of the quantum-enhanced sensing platforms available today. 


\subsection{Quantum Multiparameter Estimation}
\label{subsec:multiparameter}
In his original work, Heisenberg himself was concerned with
``the simultaneous determination of two canonically conjugate quantities''~\cite{heisenberg1927illustrativeTranslation}. Estimating multiple unknown physical parameters is thus intimately connected with the uncertainty principle, and can be thought of as a special case of the relations discussed in Sec.~\ref{sec:EDUR}. To see this let us briefly return to Heisenberg's microscope thought experiment. Assume that we wish to measure both the position and momentum of an electron. Measuring the position first will introduce an additional error into any subsequent measurement of momentum and vice versa. This introduces a non-trivial trade-off when one tries to minimize the (possibly weighted) total MSE. This minimization becomes even trickier when techniques from quantum information are utilized as, for example, if we had two identical copies of the electron we wish to measure we can further reduce the MSE by entangling the electrons before measuring them~\cite{conlon2023approaching}. 
Owing to these inherently quantum mechanical features which emerge when simultaneously estimating multiple parameters, quantum multiparameter estimation is a topic that has generated much interest in recent years~\cite{demkowicz2020multi,liu2020quantum,albarelli2020perspective,pezze2025advances}. Finding the minimum possible MSE in this setting is thus a task of both fundamental and practical importance. In this section we describe the consequences of the uncertainty relation for multiparameter estimation, describing theoretical tools used to evaluate the MSE in Sec.~\ref{subsubsecMultiParCR}, and the trade-offs in MSE presented in Sec.~\ref{subsubStradeoff}.


\subsubsection{Cram\'er-Rao Bounds}
\label{subsubsecMultiParCR}
We shall begin by describing various theoretical tools for determining limits on how small the total MSE can be when simultaneously measuring multiple physical quantities. As in the single parameter case, these bounds are commonly called Cram\'er-Rao (CR) bounds after the author who discovered the classical equivalent~\cite{rao1947minimum,rao1992information}. For multiparameter estimation, there are several different CR bounds which apply in different settings and make different assumptions. As such, great care must be taken with regards what can be deduced about the uncertainty principle from these bounds. After introducing each CR bound, we shall discuss their implications. Throughout this section, we shall drop any explicit dependence on the number of experimental repetitions $M$, as CR bounds are defined to hold in the regime of asymptotically large $M$. As such, all of the bounds can be understood as being normalized by the number of repetitions. This does not affect the relative ratios of the different CR bounds with each other.

Formally, in quantum multiparameter estimation we consider a density matrix that contains information about $n$ parameters $\theta=(\theta_1,\hdots,\theta_n)$, $\rho=\rho(\theta)$. By acting on $\rho$ with some measurement we aim to construct unbiased estimators of the different $\theta_i$ with the minimum possible variance. A measurement in quantum mechanics is represented by a positive operator valued measure (POVM), $\{\Pi_{k}\}$. 
The $k$-th measurement outcome occurs with probability
$p_k=\text{Tr}[\rho(\theta)\Pi_{k}]$, allowing an unbiased estimator for the parameters $\theta$, $\tilde{\theta}$, to be constructed from a set of measurement outcomes.

In what follows we shall assume, without loss of generality, that $\theta_0=0$, i.e. the true values are close to 0. Given the restriction that the estimator is unbiased, the goal in quantum multiparameter estimation is then to minimize the weighted trace of the MSE matrix,
$V(\theta)$, with elements given by Eq.~\eqref{eq:MSEdefinition}.

\paragraph{Classical Fisher Information}
The CFI was described in Sec.~\ref{sec:metPrelims}, as a mathematical function calculated based on a probability distribution. Essentially, it quantifies the amount of information that a random variable contains about some unknown parameter of the underlying probability distribution. However, in quantum mechanics we only obtain a probability distribution once a measurement has been specified. Hence, we can only compute the CFI for a specific measurement. 

Assume we specify a measurement $\{\Pi_k\}$ which given the quantum state $\rho(\theta)$, gives rise to a probability distribution $p(\theta)$. The CFI matrix can then be calculated based on Eq.~\eqref{eq:CFImat}. For unbiased estimators, the classical Cram\'er-Rao inequality is then given by Eq.~\eqref{eq:classicalCR}. Denoting by $\delta\theta_i^2$ the MSE with which we can estimate parameter $\theta_i$, we arrive at the following bound
\begin{equation}
\sum_i\delta\theta_i^2=\text{Tr}[V(\theta)]\geqslant \text{Tr}[J^{-1}],
\label{eq:cfivarianceunweighted}
\end{equation}
This equation is now starting to resemble an uncertainty principle of sorts. The resemblance becomes more striking when we introduce a positive-definite weight matrix $W$, such that
\begin{equation}
\text{Tr}[WV(\theta)]\geqslant \text{Tr}[WJ^{-1}].
\end{equation}
By changing $W$, we obtain a family of equations similar in form to Eq.~\eqref{eq:cfivarianceunweighted}. Each equation restricts the space of allowed simultaneous values of $\delta\theta_i$. The overlap of all these spaces forms an uncertainty relation.

We now consider a simple example to demonstrate the CFI and how it connects to the uncertainty principle. We consider trying to estimate two qubit parameters $c_x$ and $c_z$ in the quantum state
 \begin{equation}
 \label{simpleCRexample}
     \rho(c_x,c_z)=\frac{\1}{2}+\frac{c_x}{2}\sigma_x+\frac{c_z}{2}\sigma_z.
 \end{equation}
 This is a variant of the well studied quantum state tomography problem~\cite{wootters1989optimal,vrehavcek2004minimal,li2023optimal}. Intuitively for measuring $c_x$, one might think that the optimal measurement is to project along the $x$-axis of the Bloch sphere (and indeed this is optimal). Such a measurement corresponds to the POVM
\begin{equation}
    \Pi_{x\pm}=\frac{1}{2}\begin{pmatrix}
1&\pm1\\
\pm1&1
    \end{pmatrix},
\end{equation}
that, from Born's rule, gives $p(X;\theta)$ as
\begin{equation}
    p(\pm;c_x,c_z)=\frac{1\pm c_x}{2}.
\end{equation}
Therefore, we can see that $\partial_{c_z} p(\pm;c_x,c_z)=0$, meaning that the CFI matrix is singular for this example, i.e. the optimal measurement for estimating $c_x$ reveals no information about $c_z$. Similarly one could also construct an optimal measurement for estimating $c_z$ corresponding to measuring along the $z$-axis of the Bloch sphere $\Pi_{z\pm}$. Analogously, this measurement reveals no information about $c_x$.
 
At this stage, if for a classical object like a coin, we could perform one optimal measurement (e.g. measure its diameter), and then later implement the second optimal measurement (e.g. measure its weight). However, for quantum objects, this is not possible, rigorously captured by the fact that $\Pi_{x\pm}$ and $\Pi_{z\pm}$ do not commute. In this case, we can consider instead implementing a probabilistic mixture of the two optimal POVMs $\{\Pi_{x\pm}/2,\Pi_{z\pm}/2\}$. One can calculate the CFI for this measurement to find that, as a result of this incompatibility, there is a factor of 2 increase in the MSE for measuring $c_x$ and $c_z$ together compared to measuring either quantity alone. This factor of 2 increase in MSE is a direct manifestation of the uncertainty principle in multiparameter estimation.

\paragraph{Most Informative Cram\'er-Rao Bound}
We begin by introducing the most informative CR bound, defined as the precision attained by the optimal separable measurement, 
\begin{equation}
    \mathcal{C}_\text{MI}\coloneq\min\text{Tr}[WV],
\end{equation}
where the minimisation is over all unbiased measurement and estimator combinations. As mentioned above, this is an extremely complex and non-convex optimisation problem. Indeed, the difficulty in calculating the most informative bound is the motivation for introducing the following CR bounds. In the special case of two parameter estimation with pure states a simple expression for $\mathcal{C}_\text{MI}$ is known~\cite{yung2025most}.

\paragraph{The quantum Fisher Information}
\label{paraQFI}
When attempting to define a quantum mechanical equivalent of the CFI, we are faced with an immediate problem - quantum systems are described by density matrices not probability distributions. One could specify a measurement to implement which would give rise to a probability distribution, however, there is no guarantee this measurement is optimal. In Sec.~\ref{sec:metPrelims}, we presented the QFI as a quantum analogue of the CFI for estimating a single parameter. For estimating multiple parameters, the QFI matrix is defined as
\begin{equation}
J_{\text{S},i,j}=\text{Tr}[\rho \frac{L_iL_j+L_jL_i}{2}].
\end{equation}
As the QFI matrix satisfies a similar inequality as Eq.~\eqref{eq:classicalCR}, this can be used to compute a bound on the total MSE
\begin{equation}
\text{Tr}[WV(\theta)]\geqslant\mathcal{C}_{\text{S}}\coloneqq\frac{\text{Tr}[W(J_{\text{S}})^{-1}]}{M}.
\end{equation}

The QFI and SLD operators actually specify the theoretically optimal measurement to implement when estimating a single parameter. Additionally, if the optimal measurements for each individual parameter commute with one another then it is known that the SLDCRB is attainable. This is known as the locally quasi-classical model and illustrates the fundamental importance of the QFI~\cite{suzuki2019information}. For an explicit measurement saturating the SLDCRB see Ref.~\cite{paris2009quantum}. However, for estimating multiple parameters, if the optimal measurements do not commute with one another then they cannot be simultaneously implemented and the MSE given by the QFI cannot be reached by any physical measurement. In this manner, the uncertainty principle manifests itself in quantum multiparameter estimation. In this setting other CR bounds become important. The condition for saturating the SLDCRB is known to be
\begin{equation}
\label{eq:compcond}
\text{Tr}[\rho[L_i,L_j]]=0,\;\forall i,j,
\end{equation}
when we allow for entangling measurements acting on infinitely many copies of our quantum state~\cite{ragy2016compatibility}. (More detail on entangling measurements is provided in Sec.~\ref{subsubStradeoff}.) However, in the more relevant setting of separable measurements general conditions for saturating the multiparameter SLDCRB are unknown. Indeed, this was recently listed as one of five open problems in quantum information~\cite{horodecki2022five}. The solution to this problem is known in some special cases, such as when the density matrix is a direct sum of pure states~\cite{matsumoto2002new}, when the density matrix is full rank~\cite{yang2019optimal}, and when the density matrix has a 1-dimensional kernel space~\cite{conlon2025role}. Note that Refs.\cite{matsumoto2002new,yang2019optimal,conlon2025role} together therefore completely solves the question of attainability for qutrits. However, a general solution remains elusive~\cite{imai2026hierarchy}.

For the simple problem specified above of estimating $c_x$ and $c_z$ in Eq.~\eqref{simpleCRexample}, we can use the QFI to reveal a number of interesting properties. When $c_x, c_z\approx0$, we can easily calculate the SLD operators as $L_x=\sigma_x$ and $L_z=\sigma_z$. This verifies that, for this pair of $c_x, c_z$ values, the measurements $\Pi_{x,\pm}$ and $\Pi_{z,\pm}$ are optimal. Also note that, although the SLD operators do not commute, the weak commutativity condition, stated in Eq.~\eqref{eq:compcond} is satisfied. Finally, for the curious reader, we note that the QFI matrix is easily calculated with known expressions, see Ref.~\cite{vsafranek2018simple} and appendix~\ref{QFIpropapen}.

Before concluding the discussion on the QFI and SLDCRB we highlight a few final points of interest. First, to further emphasize the fundamental importance of the QFI, we note that the QFI can be used to derive the standard quantum limit, Eq.~\eqref{eq:SQLcoherent1}, and Heisenberg scaling, Eq.~\eqref{eq:heisenbergN00N}. This is shown in appendices~\ref{apenQFISQl} and \ref{apenHscale2} respectively. Secondly, we point out that the QFI has a number of simple, but useful mathematical properties listed in appendix~\ref{QFIpropapen}. For two parameter estimation, the conditions for exactly saturating the QFI for one parameter are known~\cite{yung2025saturating}. Penultimately, we note that for Gaussian quantum states and single qubit states, the QFI has a particularly simple form, which can be found in Refs.~\cite{pinel2013quantum} and \cite{liu2020quantum} respectively. Finally, it is important to observe that the QFI is intimately connected to the distinguishability of neighbouring quantum states~\cite{braunstein1994statistical}.

\paragraph{Right Logarithmic Derivative Cram\'er-Rao Bound}
In order to derive the QFI and corresponding SLDCRB Helstrom introduced the SLD operators, Eq.~\eqref{eq:SLDopBG}, as a quantum analogue of the log derivative. However, there is no unique way to define this quantity quantum mechanically. In fact, different ways of defining the log derivative in quantum metrology correspond to different ways of defining the product of two operators~\cite{hayashi2017quantum}. As such, Yeun and Lax were able to define an alternative quantum logarithmic derivative as the solution to the following equation~\cite{yuen1973multiple}
\begin{equation}
\label{RLDsolve}
\frac{\partial\rho}{\partial\theta_k}=\rho \tilde{L}_k.
\end{equation}
Unsurprisingly, this operator is called the right logarithmic derivative (RLD) operator. The RLD operators give rise to the RLD QFI matrix with the following elements
\begin{equation}
J_{\text{R},i,j}=\text{Tr}[\rho \tilde{L}_i\tilde{L}_j^\dagger].
\end{equation}
As above, this Fisher information matrix can be used to define a CR bound, in this case the RLD CR bound (RLDCRB)
\begin{align}
&\text{Tr}[WV(\theta)]\notag\\
\geqslant&\mathcal{C}_\text{RLD}\coloneqq\text{Tr}[W\text{Re}(J_\text{R}^{-1})]+\text{TrAbs}[W\text{Im}(J_\text{R}^{-1})].\label{eqRLDlim}
\end{align}
Like the SLDCRB, the RLDCRB is in general not a tight bound, meaning there does not always exist a measurement which saturates the RLDCRB~\footnote{By a tight bound we mean one that is attainable by some measurement and estimator.}. However, the RLDCRB is asymptotically attainable for a particular class of quantum statistical models, namely D-invariant models~\cite{suzuki2016explicit}. Recently, it has been demonstrated that the RLDCRB is connected to the purification of Gaussian states~\cite{yadavalli2024optimal}.

One might also define a left logarithmic derivative (LLD) operator by placing the logarithmic derivative on the other side of the density operator. However, the LLD is the adjoint analogue of the RLD and leads to the same metric, so it does not provide an additional CR bound distinct from the RLDCRB~\cite{hayashi2017quantum}. Related to this, Yamagata introduced a logarithmic derivative operator that lies between the SLD and RLD operators~\cite{yamagata2021maximum}, characterized by a parameter $\beta$. Remarkably, for qubit systems by optimizing over $\beta$, one arrives at the next bound we describe.


\paragraph{Holevo Cram\'er-Rao Bound}
In spite of their historical significance, both of the bounds presented so far suffer from a major flaw -- namely that there exists many physical settings where these bounds cannot be saturated, i.e. they are not tight bounds. It was thus a major advancement when Holevo introduced a new CR bound which unified the SLDCRB and RLDCRB and which was always attainable. This bound has subsequently become known as the Holevo Cram\'er-Rao bound (HCRB)~\cite{holevo1976noncommutative, holevo2011probabilistic}. Loosely speaking, as the HCRB is always attainable, whereas the SLDCRB is not, we can think of the HCRB as incorporating the uncertainty principle whereas the SLDCRB does not. 

Although the HCRB is a tighter bound than the SLDCRB or RLDCRB, a drawback is that it is not as simple to calculate as it involves a non-trivial minimisation. The HCRB can be written as a minimisation over Hermitian matrices $X$, subject to certain constraints
\begin{align}
&\text{Tr}[WV(\theta)]\notag\\
\geqslant&\mathcal{C}_\text{H} \coloneqq  \min_{X} \text{Tr}[W Z_\theta[X]] +\text{TrAbs} [W\Im Z_\theta[X]],\label{eq_hol2}
\end{align}
where
\begin{align}
\label{eq_zmat}
Z_\theta[X]_{jk} \coloneqq  \text{Tr}[ \rho X_j X_k],
\end{align}
and  $\text{TrAbs}[\text{Im}Z_{\theta}[X]]$ is the sum of the absolute values of the eigenvalues of the matrix $\text{Im}Z_{\theta}[X]$. The matrices $X$ are estimator matrices, and for an unbiased estimator we require the following conditions to be satisfied
\begin{equation}
\text{Tr}[\rho X_k]=\theta_k,
\label{eq:unb1}
\end{equation}
and
\begin{equation}
\text{Tr}[\frac{\partial\rho}{\partial\theta_j} X_k]=\delta_{j,k}.
\label{eq:unb2}
\end{equation}
Note that the difficulty in computing the HCRB has partly been alleviated by new computational techniques~\cite{albarelli2019evaluating,sidhu2021tight}.

The HCRB in this form was introduced by Nagaoka~\cite{nagaoka2005new}. Recently, it has been proven that the ratio of the HCRB to the SLDCRB is at most a factor of two~\cite{carollo2019quantumness,tsang2020quantum}. This result can be used to place upper limits on the HCRB in an efficient manner. 

The HCRB has become the quantum Cram\'er-Rao bound which is in most widespread use. This is primarily because it is the bound which gives the best possible measurement precision. By this statement, we mean that there does exist, in theory, a measurement saturating this bound. However, the measurement required to saturate the HCRB may be an entangling measurement on infinitely many copies of the probe state. Indeed the claim that the HCRB is the tightest attainable bound, is based on the fact that the HCRB is asymptotically attainable~\cite{kahn2009local, yamagata2013quantum, yang2019attaining}. However, it has recently been proven that this asymptotic attainability is of no practical relevance~\cite{conlon2022gap}. Thus, although the HCRB improves upon the SLDCRB by incorporating the uncertainty principle, for an experimentalist who wishes to know about how the uncertainty principle impacts quantum multiparameter estimation in a practical setting the HCRB is of no use.

\paragraph{Nagaoka Cram\'er-Rao Bound}
\label{secNCRB}
If we were to describe an ideal CR bound, we would probably describe a bound that is always attainable (unlike the SLDCRB) with practical measurements that requires entanglement on only a finite number of copies of the quantum state (unlike the HCRB)~\footnote{It would be remiss of us not to reemphasize that in some scenarios the SLDCRB and HCRB are attainable with separable measurements}. This is where the Nagaoka Cram\'er-Rao bound (NCRB) comes in.

In the same paper where he wrote the HCRB in the form presented in Eq.~\eqref{eq_hol2}, Nagaoka introduced a new bound~\cite{nagaoka2005new} that applies when restricted to separable measurements. A major limitation of the NCRB is that it is only applicable for estimating two parameters. The NCRB can be written as
\begin{align}
&\text{Tr}[WV(\theta)]\notag\\
\geqslant 
&\mathcal{C}_\text{N} \coloneqq  \min_{X} \text{Tr}[ W Z_\theta[X]]+\sqrt{\text{det}(W)}\text{TrAbs} [ \rho [X_1,X_2]],\label{eq_hol2NBG}
\end{align}
where the matrices $X_1$ and $X_2$ are subject to the same unbiased constraints as for the HCRB, Eqs.~\eqref{eq:unb1} and \eqref{eq:unb2}, and det$(W)$ denotes the determinant of $W$. The NCRB is a tighter bound than the HCRB. For qubit
systems Nagaoka proved that the NCRB is attainable~\cite{nagaoka2005new}, i.e. there exists a separable measurement which can reach the NCRB. However, for higher dimensional systems, explicit counter-examples to this claim have been presented~\cite{hayashi2023tight,conlon2025role}. Nevertheless, the fact that the NCRB applies to separable measurements implies that an experimentalist can use this bound to investigate the uncertainty principle in multiparameter estimation in a practical setting. If a separable measurement saturating the NCRB is found, then this measurement is optimal.

The fact that the NCRB applies to separable measurements gives rise to an intriguing property that the NCRB is subadditive. Simply put, when calculating the NCRB for two copies of a quantum state, $\mathcal{C}_\text{N}(\rho\otimes\rho)$, one can find a MSE which is less than half that of the single-copy bound, i.e. $\mathcal{C}_\text{N}(\rho\otimes\rho)\leqslant\mathcal{C}_\text{N}(\rho)/2$. This is simply a reflection of the fact that when performing a measurement on $\rho\otimes\rho$, i.e. on the joint Hilbert space, an entangling measurement can be used which improves the measurement sensitivity. This type of entangling measurement has an important role to play in saturating the ultimate limits allowed by quantum mechanics. 

For estimating more than 2 parameters, the NCRB has been extended to the Nagaoka-Hayashi Cram{\'{e}}r-Rao bound (NHCRB) by Conlon \textit{et al}~\cite{conlon2021efficient}, which we shall denote as $\mathcal{C}_\text{NH}$. As it is not essential for understanding the connection with the uncertainty principle, we do not present the explicit form of the NHCRB here, and instead refer the reader to Ref.~\cite{conlon2021efficient} for further details. The NHCRB also applies when restricted to separable measurements and has also been shown not to always be a tight bound~\cite{hayashi2023tight,conlon2025role}. However, it is worth emphasizing that the NHCRB is a tight bound in many physically motivated settings~\cite{conlon2021efficient,conlon2023multiparameter,yung2026beating}. As we shall discuss in detail below, CR bounds for estimating 2 or more parameters can be used to construct trade-off surfaces which represent the uncertainty principle in quantum metrology.

\paragraph{Hayashi-Ouyang Cram\'{e}r-Rao Bound}
Recently, Hayashi and Ouyang have demonstrated that the aforementioned bounds can be formulated as
conic programs with the same objective and constraint functions~\cite{hayashi2023tight}. By appropriately choosing the cone to optimize over, the most informative CR bound, the SLDCRB, the HCRB and the NCRB can all be recovered. Crucially, optimizing over the cone of all separable matrices, gives the most informative bound
\begin{equation}
\mathcal{C}_\text{MI} \coloneqq \min_{\mathcal{X}\in \mathcal{S}_\text{SEP}}\text{Tr}[(W\otimes \rho)\mathcal{X}],
\end{equation}
where $\mathcal{S}_\text{SEP}$ is the cone of separable matrices on a joint quantum-classical Hilbert space. The optimization is subject to $\mathcal{X}$ being unbiased which takes a slightly different form compared to what was used above in the HCRB and NCRB (Eqs.~\eqref{eq:unb1} and \eqref{eq:unb2}). Unfortunately, optimization over the separable cone is not easy in practice (see theorem 16 of Ref.~\cite{hayashi2023tight}), hence the authors introduced a semi-definite programming relaxation of the above conic program, which we shall refer to as the Hayashi-Ouyang Cram\'{e}r-Rao Bound (HOCRB). As the exact formulation of the bound is quite complex, we shall not detail it here, and refer the interested reader to Ref.~\cite{hayashi2023tight}. We shall denote the HOCRB as $\mathcal{C}_{\text{HO}}$.

\paragraph{Gill-Massar Bound}
Another bound which applies when restricted to separable measurements is the Gill-Massar bound (GMB)~\cite{gill2005state}. The GMB is not of exactly the same form as typical CR bounds, hence we do not refer to it as a CR bound. The GMB states
\begin{equation}
    \text{Tr}[J_\text{S}^{-1} (NV(\theta)^N)^{-1}]\leqslant d-1
\end{equation}
where $V(\theta)^N$ denotes the MSE matrix when performing a separable measurement on $N$ copies of the quantum state and $d$ is the dimension of the Hilbert space. The GMB is a tight bound for estimating parameters in a qubit system, similar to the NCRB. An advantage of the GMB over the NCRB is that it can be used to obtain bounds on the MSE for estimating 3 parameters of a qubit state. Nevertheless, in this case, we can simply use the NHCRB.

Additionally, the GMB suffers from a significant drawback. For $d>2$ it is not a tight bound. Indeed, if $d>2$ and $n\leqslant d-1$ the SLDCRB provides a tighter bound. Thus, when bounding the MSE using separable measurements we shall consider the NCRB or NHCRB going forward.

\paragraph{Ordering of the Different Cram\'{e}r-Rao Bounds}
Having described the different CR bounds, we now wish to comment briefly on their ordering. For estimating two parameters
\begin{equation}
    \mathcal{C}_\text{MI}\geqslant\mathcal{C}_\text{HO}\geqslant\mathcal{C}_\text{N}\geqslant\mathcal{C}_\text{H}\geqslant\text{max}\{\mathcal{C}_\text{S},\mathcal{C}_\text{R}\},
\end{equation}
with no ordering in general between the SLDCRB and RLDCRB. For estimating multiple parameters we simply switch $\mathcal{C}_\text{N}$ for $\mathcal{C}_\text{NH}$. It is also known that~\cite{carollo2019quantumness}
\begin{equation}
   2\mathcal{C}_\text{S} \geqslant \mathcal{C}_\text{H}\geqslant\mathcal{C}_\text{S}.
\end{equation}

\begin{figure}[t]
    \centering
\includegraphics[width=0.9\columnwidth]{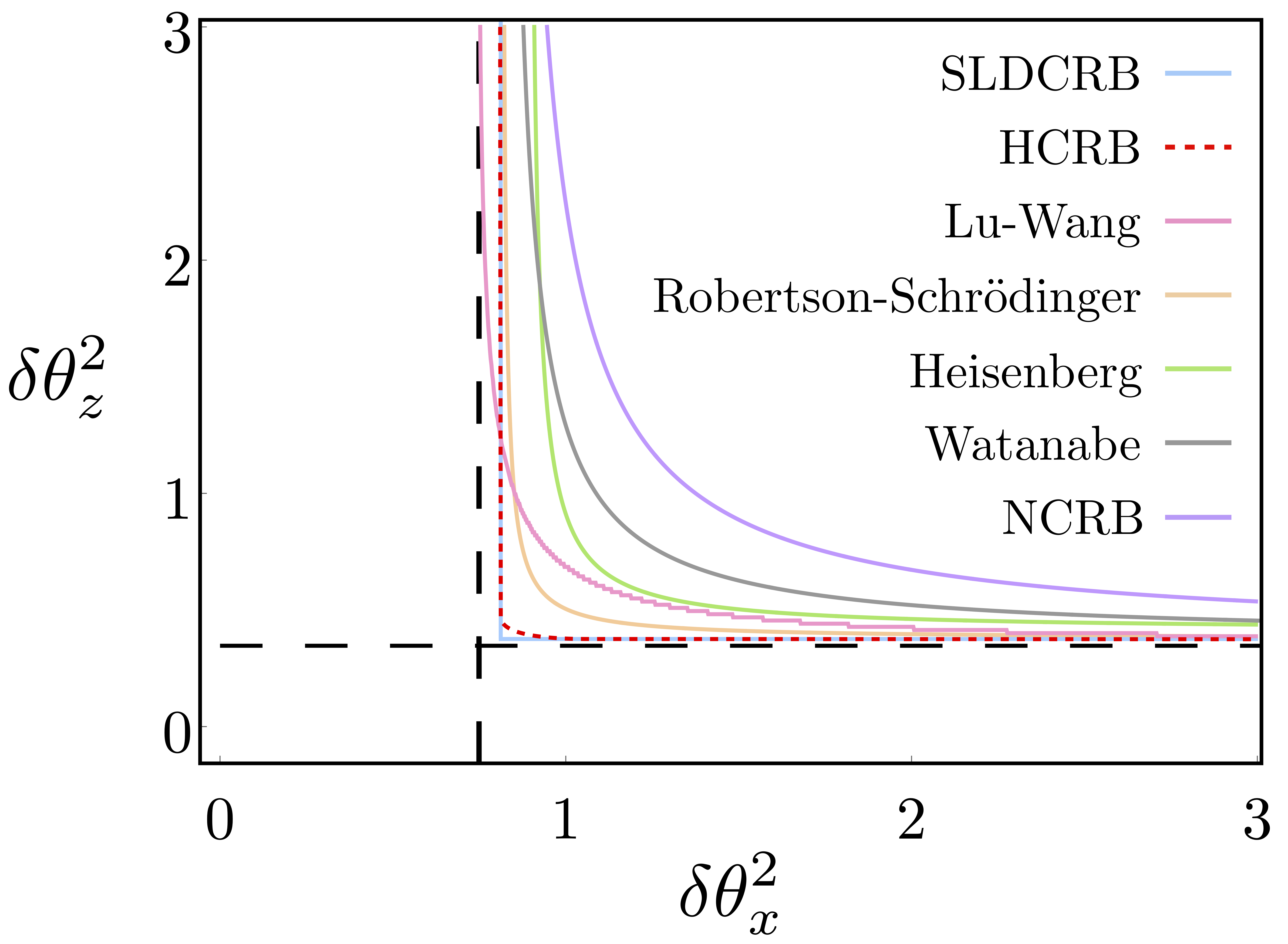}
    \caption{\textbf{Trade-off Relations for MSE when Estimating $c_x$ and $c_z$ in Eq.~\eqref{simpleCRexample2}.} $\delta\theta_x^2$ and $\delta\theta_z^2$, represent the MSE in estimating $c_x$ and $c_z$ respectively. Different ways of determining trade-off relations in quantum mechanics. SLDCRB~\cite{helstrom1967minimum}, HCRB~\cite{holevo2011probabilistic} and NCRB~\cite{nagaoka2005new} denote the symmetric logarithmic derivative, Holevo and Nagaoka Cram\'er-Rao bounds respectively. These bounds are determined by computing the corresponding Cram\'er-Rao bound for many different weight matrices. Other parameter estimation uncertainty relations include the Lu-Wang~\cite{lu2021incorporating}, Robertson-Schrödinger~\cite{schrodinger1930heisenbergschen}, Heisenberg~\cite{heisenberg1927illustrativeTranslation}, and Watanabe~\cite{watanabe2011uncertainty} relations.
    The dashed black lines correspond to the MSE attainable when estimating each parameter individually. The discontinuous nature of the Lu-Wang bound comes from the finite resolution used in its calculation.}
\label{fig:tradeoffplots}
\end{figure}

\paragraph{Example Illustrating the Different Cram\'er-Rao Bounds}
We now examine a simple example that demonstrates the connection between the different CR bounds and the uncertainty principle. For this we consider a problem similar to that in Eq.~\eqref{simpleCRexample}. We consider trying to estimate $c_x$ and $c_z$ simultaneously for the following qubit state
\begin{equation}
 \label{simpleCRexample2}
     \rho(c_x,c_z)=\frac{\1}{2}+\frac{c_x}{2}\sigma_x+\frac{c_z}{2}\sigma_z+\frac{c_z-c_y}{2}\sigma_y.
 \end{equation}
(The rather strange nature of this problem was chosen to ensure all of the bounds differ from one another.) Note that the condition to make this a valid density matrix is $c_x^2+c_z^2+(c_z-c_y)^2\leqslant1$. The derivatives of this qubit with respect to the parameters of interest $c_x$ and $c_z$ are denoted by $\dot{\rho}_x$ and $\dot{\rho}_z$ respectively and are given by
\begin{equation}
    \dot{\rho}_x=\frac{1}{2}\sigma_x\qquad\text{and}\qquad\dot{\rho}_z=\frac{1}{2}(\sigma_z+\sigma_y).
\end{equation}
From this we are able to compute the various CR bounds mentioned above for different weight matrices which gives rise to a trade-off curve shown in Fig.~\ref{fig:tradeoffplots} (computed for $c_x=0.4, c_z=0.6$ and $c_y=0.55$). This is done via the technique developed by Yung \textit{et al}~\cite{yung2024comparison}. Note that for this problem we have a  gap between the SLDCRB, HCRB and NCRB. Several other trade-off curves are also shown in Fig.~\ref{fig:tradeoffplots}, which do not come from Cramér-Rao bounds and are described below.

While the gap between the NCRB and SLDCRB does not seem overly large for this problem, this is partly because we have considered a simple two parameter estimation problem. For more general $d$-parameter estimation problems the ratio between the two can scale as $O(\sqrt{d})$, as noted by Das \textit{et al}~\cite{das2024holevo}. Finally, we note that CR bounds have been used to determine trade-offs in various multiparameter estimation problems, e.g. in quantum superresolution~\cite{hervas2024optimizing}, and in magnetic field sensing~\cite{baumgratz2016quantum} to name but a few.

\paragraph{Experiments Saturating Cram\'er-Rao Bounds} 

We conclude this subsection with a brief discussion on experiments saturating the different CR bounds. As the HCRB in general requires an entangling measurement on infinitely many copies of the quantum state to be saturated, experiments saturating the HCRB have only been demonstrated for specific scenarios. For simultaneously estimating displacements in the $x$ and $p$ quadratures, the optimal measurement is known~\cite{bradshaw2018ultimate,bradshaw2017tight}. Experiments verifying the presence of steering for two mode squeezed vacuum states, can be interpreted as saturating the HCRB for this problem~\cite{steinlechner2013strong}.

In contrast, as the NCRB is a bound on separable measurement precision, it is more experimentally accessible. There are many experiments saturating this bound, for example Ref.~\cite{li2023optimal}. More interestingly, there are several experiments which have demonstrated that entangling measurements can allow an accuracy greater than that predicted by the single-copy NCRB (i.e. the NCRB evaluated for only a single copy of the probe state), such as Refs.~\cite{roccia2017entangling,parniak2018beating,hou2018deterministic}. However, it was not until recently that the two-copy NCRB was saturated~\cite{conlon2023approaching}, marking the implementation of a theoretically optimal entangling measurement. Since this, experiments going beyond two-copy entangling measurements have been demonstrated~\cite{yi2025optimal,zhou2025experimental}.


\subsubsection{Multiparameter Trade-offs}
\label{subsubStradeoff}
It has been demonstrated that the simultaneous estimation of multiple parameters can outperform individual estimation~\cite{baumgratz2016quantum}, leading to even greater interest in quantum multiparameter estimation. However, in many other cases, the simultaneous estimation of multiple parameters offers no advantage over individual estimation~\cite{ragy2016compatibility}. In this case, as simultaneous estimation is experimentally more complex than individual estimation, in all likelihood the experimentalist will choose to measure the parameters individually. In this setting a simple-to-evaluate trade-off relationship when estimating multiple parameters simultaneously offers critical insight. In recent years these relationships have been established in several different forms. In what follows we shall describe how such trade-off relations have evolved in detail.

\paragraph{Informal Connection with Uncertainty Principle}
The original Robertson uncertainty relation, Eq.~\eqref{eq:VUR_Robertson}, technically refers to a relationship between the standard deviation (SD) of two observables. Loosely speaking, one might anticipate that if the product of the SD of two observables is above a certain threshold then the product of the SD when measuring these observables must be above the same threshold. Borrowing notation from Sec.~\ref{sec:NOF}, we shall use the notation $\epsilon_{A}$ to denote the measurement noise -- a measure of the difference between measuring the ideal observable $A$ and the measurement actually implemented $\tilde{A}$. We shall denote by $\eta_{B}$ the disturbance of $B$ introduced by the approximate measurement of $A$. With this notation, the Robertson uncertainty relation may be reinterpreted as
\begin{align}\label{eq:RobertsonReInterp}
    \epsilon_A\cdot\eta_B
    \geqslant
    \frac{1}{2}|\bra{\psi}[A,B]\ket{\psi}|.
\end{align}
This is arguably the form of the uncertainty relation that Heisenberg first intended. Similarly, for the joint measurement of two observables, we have
\begin{align}\label{eq:RobertsonReInterp2}
    \epsilon_A\cdot\epsilon_B
    \geqslant
    \frac{1}{2}|\bra{\psi}[A,B]\ket{\psi}|.
\end{align}
The stronger Sch\"odinger version of this is
~\cite{schrodinger1930heisenbergschen} 
\begin{equation}
\label{eq:RobSchro}
\epsilon_A^2\cdot\epsilon_B^2\geqslant\bigg|\frac{1}{2}\langle\{A,B\}\rangle-\langle A\rangle\langle B\rangle\bigg|^2+\frac{1}{4}|\langle[A,B]\rangle|^2.
\end{equation}

In the 1960s, Arthurs and Kelly further developed the application of uncertainty relations to practical measurements~\cite{arthurs1965simultaneous,arthurs1988quantum,ishikawa1991uncertainty}. However, broadly speaking, the Sch\"odinger uncertainty relation reflected the widely accepted limitations of joint estimation until Ozawa's seminal work described in Sec.~\ref{sec:NOF}. However, there remained no direct way of applying these uncertainty relations to multiparameter estimation until recently~\cite{lu2021incorporating}. The formalism derived in Ref.~\cite{lu2021incorporating}, allows us to use the above uncertainty relationships to determine trade-off curves for quantum multiparameter estimation, as shown in Fig.~\ref{fig:tradeoffplots}.

\paragraph{Formal Connection with Uncertainty Principle}

All of the above uncertainty relations are not directly related to estimating physical quantities, making it hard to connect multiparameter estimation to the uncertainty principle. In this section, we describe several works which make a formal connection between multiparameter estimation and uncertainty relations. This connection was made by Watanabe \textit{et al} in 2011~\cite{watanabe2011uncertainty} (see also Refs.~\cite{hofmann2003uncertainty,PhysRevA.89.022106} for related but distinct work). In this work, the quantity $\epsilon_{A}$ was related to the difference between the variance of an estimator and the minimum possible variance. 

In recent years, the connection  between multiparameter estimation and the uncertainty principle has been further developed. Kull \textit{et al} computed trade-off curves based on different CR bounds for the MSE attainable~\cite{kull2020uncertainty}. A similar approach using the NCRB was developed in Ref.~\cite{yung2024comparison}, and a bound on the simultaneous measurement error based on the NHCRB was developed in Ref.~\cite{chen2024simultaneous}. In 2021, Lu and Wang established an analytical bound on the MSE attainable, derived directly from Ozawa's uncertainty principle~\cite{lu2021incorporating}. All of these uncertainty relations are shown in Fig.~\ref{fig:tradeoffplots}. From this it is clear that the NCRB gives rise to the tightest uncertainty relation~\cite{yung2024comparison}. This is unsurprising, given that the NCRB is a tight CR bound in many scenarios. However, given that the NCRB is not always tight, it may be possible to derive a tighter uncertainty relation in certain settings~\cite{hayashi2023tight,conlon2025role}.

\paragraph{Entangling Measurements - Saturating Bounds \& Violating Uncertainty Relations}

As Fig.~\ref{fig:tradeoffplots} shows, the NCRB is the tightest known uncertainty relation~\footnote{The HOCRB is a tighter bound than the NCRB and as such would give rise to a tighter uncertainty relation. However, this bound is very close to the NCRB and considerably more computationally complex, hence we do not include it here.}. The reason for this is that the NCRB is a tight multiparameter estimation bound in most physical scenarios. However, by measuring two copies of the quantum state in an entangling basis, the error per quantum state measured can be reduced~\cite{conlon2023approaching}. In this setting, it has been observed that the Lu-Wang~\cite{lu2021incorporating}, Robertson-Schrödinger and Heisenberg uncertainty relationships can all be violated~\cite{conlon2023approaching}, i.e. they intrinsically assume that quantum states are measured individually. It is not clear whether these uncertainty relations can be tightened to apply to all measurement types. Thus, the field remains highly active and there are many open questions.

The tightness of an CR-type uncertainty relation (i.e. how close it is to the true limit) is ultimately related to whether that CR bound is attainable or not. It is known that the NCRB is not always attainable when measuring quantum states individually~\cite{hayashi2023tight,conlon2025role}, and the HCRB is attainable when measuring an asymptotically large number of quantum states simultaneously~\cite{kahn2009local, yamagata2013quantum, yang2019attaining}. Additionally, the conditions under which the SLDCRB is equal to the HCRB are known~\cite{ragy2016compatibility}. However, general conditions for when the SLDCRB can be saturated by measurements on a single copy of a quantum system remain unknown~\cite{horodecki2022five}. In spite of some recent progress~\cite{pezze2017optimal,yang2019optimal,chen2022information,chen2022incompatibility,amari2000methods,suzuki2020quantum,conlon2025role}, this question remains unanswered in all generality. Answering this question is not only of practical importance to the quantum information community~\cite{horodecki2022five}, as we have discussed here, it has fundamental implications for uncertainty relations in quantum mechanics.


\subsection{Extended Themes}
\label{sec:ET2}
The preceding sections, have elaborated on how quantum metrology and the uncertainty principle are fundamentally intertwined with one another. To do so, we focused on two classes of problems that directly illustrate this connection. In this section, we wish to highlight several other considerations that are of fundamental importance. In Sec.~\ref{subsec:EntAndMet}, the connection between entanglement and quantum metrology is described. In Sec.~\ref{subsec:FinePrint}, we detail some of the (often unmentioned) issues with using CR bounds as a fundamental limit on precision. Finally, Sec.~\ref{subsec:Relativistic} examines how the uncertainty principle changes in relativistic quantum systems.


\subsubsection{Entanglement and Metrology}
\label{subsec:EntAndMet}

The estimation of any single physical parameter is fundamentally limited by the eigenvalues of the generator encoding the parameter into the quantum state, generating an uncertainty trade-off between the two (Eq.~\eqref{eq:SLDCRB:UP}). As a direct consequence of this we obtain the standard quantum limit and the Heisenberg limit. By relaxing various assumptions, it is possible to either surpass the Heisenberg limit or achieve the Heisenberg limit without entanglement~\cite{braun2018quantum}. However, generally speaking (excluding configurations like that shown in Fig.~\ref{fig:RosettaStone} (c)), entangled quantum states are required to achieve the Heisenberg limit. When we consider estimating multiple physical parameters simultaneously, the situation becomes even more interesting. In this setting, there are fundamental trade-offs to how accurately the multiple parameters can be estimated. While entangled states are required to saturate the Heisenberg scaling, entangling measurements are needed to saturate fundamental limits in multiparameter estimation. Taken together, these statements highlight the important connection between entanglement and uncertainty in quantum mechanics. See Ref.~\cite{marciniak2022optimal}, for an experiment using entanglement in both the state preparation and measurement stages.

Conversely to above, where entanglement is a resource for quantum metrology, quantum metrology is also useful for entanglement witnessing. There have been a number of theoretical studies into how quantum metrology acts as a witness of non-classical behavior~\cite{strobel2014fisher,hauke2016measuring,pezze2017multipartite,frowis2019does,yadin2021metrological}. 


\subsubsection{Unspoken Assumptions}
\label{subsec:FinePrint}
It is important to briefly mention some key points which we have thus far neglected to discuss in detail. All of the above results assume that the parameter to be estimated is approximately known, and the task is to measure small changes in the value of $\theta$ within this neighbourhood. Although it is of great interest to estimate parameters without this prior information~\cite{rubio2019quantum,rubio2021global,suzuki2024bayesian}, the local approach used above is well justified assuming that we have a sufficiently large number of quantum probe states available. It should be noted, however, that there are some fundamental differences between global and local estimation. This can be particularly important when discussing the Heisenberg limit~\cite{berry2000optimal,gorecki2020pi}. 

Closely related, but distinct, is that all of the CR bounds presented apply when considering asymptotically many repetitions of the experiment. When only a finite number of samples are used, these bounds need not apply~\cite{genoni2012optical,tsang2012ziv}. Another restriction of CR bounds is that they only allow for unbiased estimation. While unbiased estimators are known to be optimal with asymptotically many repetitions, in the finite sample regime, biased estimators may achieve a lower MSE. This effect is also present in classical statistics~\cite{stein1956inadmissibility}. 

The discussion in Sec.~\ref{subsec:multiparameter} is restricted to trade-off's when estimating multiple parameters using the same quantum probe state. There exist additional trade-off's between choosing the optimal probe state for each parameter~\cite{albarelli2022probe}. Related to this, for some problems, classically correlated quantum states and measurements can match the performance of using entangled states~\cite{park2022optimal}.

Finally, we want to reemphasize two points from above. First, it is crucial to clearly define what constitutes a resource and derive the Heisenberg limit appropriately (Sec.~\ref{subsubsecSurpass}). Secondly, it is important to use an appropriate CR bound, as naively calculating the QFI or HCRB will not produce a useful bound in many multiparameter estimation problems (Sec.~\ref{subsubsecMultiParCR}).


\subsubsection{Relativistic Quantum Metrology}
\label{subsec:Relativistic}
Thus far, our review has been restricted to considering the uncertainty principle in non-relativistic quantum mechanics. Here we briefly describe a few examples of the uncertainty principle in relativistic quantum systems. In Sec.~\ref{sec:ET} we discussed the problem of defining a time operator in quantum mechanics. In relativistic quantum field theory (QFT), rather than promoting time to an operator, symmetry between space and time is achieved by demoting position to a parameter \cite{Birrell1984}. Hence the type of uncertainty relations derived in QFT are of the form discussed immediately above. In this section, we describe several simple examples, to illustrate the role of the uncertainty principle in QFT and beyond.

The simplest relativistic particle is the photon. The annihilation operator describing a single, plane-wave, spatio-temporal optical mode, $\hat a_g$, can be written \cite{PhysRevD.105.056023}:
\begin{equation}
\hat a_g = \int_{-\infty}^{\infty} d \omega f_g(\omega) \hat a_{\omega}
\end{equation}
where the $\hat a_{\omega}$ are single frequency continuum operators obeying the commutator $[\hat a_{\omega},\hat a^{\dagger}_{\omega'}] = \delta(\omega - \omega')$. At optical frequencies it is usually true that the pulse length is much longer than the period ($2 \pi/\omega_o$). If this is the case then we can take $f_g$ to be a Gaussian 
\begin{equation}
f_g = (2 \pi \sigma^2)^{-1/4}  e^{-i t_o(\omega-\omega_o) - \frac{(\omega-\omega_o)^2}{4 \sigma^2}},
\end{equation}
with negligible support on negative frequencies. This represents a mode with central frequency $\omega_o$ and temporal variance (or pulse length) $1/(4 \sigma^2)$ whose maximum amplitude crosses the origin of the $x-$axis at time $t_o$ (where propagation is in the positive $x$ direction). Under these conditions there is a Fourier transform relationship between time and frequency. If a single photon is prepared in this mode then its energy uncertainty is given by $\Delta \hat H^2 = \hbar^2 \Delta \omega^2 = \hbar^2 \sigma^2$ and hence we find:
\begin{equation}
\Delta \hat H^2 \Delta t^2 = \frac{\hbar^2}{4},
\label{TUC}
\end{equation}
saturating the usual time energy uncertainty relation. If, instead of a single photon pulse, we consider a coherent state, $|\alpha \rangle$, in the same mode we find that now $\Delta \hat H^2  = \hbar^2 (\sigma^2 |\alpha|^4 + \omega^2 |\alpha|^2)$. The first term on the RHS is due to the fluctuations in frequency whilst the second term is due to photon number fluctuations. The usual time-energy uncertainty relation is no longer saturated.

Our second example goes beyond this, breaking the restriction that $f_g$ has negligible support on negative frequencies. Consider the situation in which we start in the vacuum state but allow the temporal variance of the mode to be reduced till it becomes shorter than the optical oscillation period. This is known as the sub-cycle regime and the time energy uncertainty relation in the heuristic form $\Delta E \Delta t \sim \hbar$ is commonly invoked to argue that virtual particles should spontaneously appear (see for example \cite{SRI96}). Under these conditions $f_g$ needs to be modified to \cite{PhysRevD.105.056023}:
\begin{equation}
f_g = {\frac{1}{(2 \pi \sigma^2)^{1/4}}} \text{sign}(\omega) \sqrt{\frac{|\omega|}{\omega_o}} e^{-i t_o(\omega-\omega_o) - \frac{(\omega-\omega_o)^2}{4 \sigma^2}},
\end{equation}
and will have non-zero support for negative frequencies. The annihilation operator of this optical mode can be written 
\begin{equation}
\hat a_g = \int_{0}^{\infty} d \omega f_g(\omega) \hat a_{\omega} + \int_{0}^{\infty} d \omega f_g(-\omega) \hat a^{\dagger}_{\omega}
\end{equation}
where negative frequency continuum annihilation operators are interpreted as continuum creation operators. As such we find that
\[
\langle 0| \hat a^{\dagger}_g \hat a_g |0 \rangle = \int_{0}^{\infty} d \omega |f_g(-\omega)|^2 \ne 0.
\]
We can consider the energy variance for a particular $\omega_o$ via 
\begin{equation}
\Delta \hat H^2 = \hbar^2 \omega_o^2 (\langle 0| \hat a^{\dagger}_g \hat a_g \hat a^{\dagger}_g \hat a_g |0 \rangle -\langle 0| \hat a^{\dagger}_g \hat a_g |0 \rangle^2).
\label{EV}
\end{equation}
This calculation should be a good estimate of the total energy variance if we consider the limit $\omega_o \ll \sigma$ such that the spread in energy is large compared to the energy gap $\hbar \omega_o$. In this limit we find
\begin{equation}
\Delta \hat H^2 = \hbar^2 \omega_o^2 (2 \int_{0}^{\infty} d \omega |f_g(\omega)|^2 \int_{0}^{\infty} d \omega |f_g(-\omega)|^2) = \frac{\hbar^2 \sigma^2}{\pi}.
\label{EV2}
\end{equation}
The photon distribution of the mode in this limit is in fact identical to a single mode squeezed state -- the virtual particles appear in pairs with the quadrature amplitudes exhibiting Gaussian statistics. An operational meaning can be assigned to this mode by considering a rapidly switched, ideal detector \cite{PhysRevD.105.056023}.  If the detector is switched on and off using a Gaussian switching function with variance $\Delta t^2 =1/(2 \sigma^2)$ it can strongly couple to the sub-cycle mode, ending up with the same energy variance as Eq.\ref{EV2}. This leads to the time-energy uncertainty relation \cite{SAJ26}:
\begin{equation}
\Delta \hat H^2 \Delta t^2 = \frac{\hbar^2}{2 \pi},
\label{TUC}
\end{equation}
which is consistent with the heuristic relation. Although, this result seems extreme, experimental methods for probing this regime have been demonstrated \cite{doi:10.1126/science.aac9788}.

Another extension into the relativistic domain that has been considered is the evaluation of quantum limits on the estimation of spacetime curvature parameters \cite{DOW17}. Whilst the general approach is complex and requires a number of approximations, some simple but illuminating examples can be summarized. Suppose we wish to find the minimum uncertainty bound when attempting to estimate a small deviation from flat spacetime, $\eta_{\mu \nu}$~\footnote{Flat spacetime is characterized by the metric tensor $\eta_{\mu \nu}$ whose only non-zero elements (in Cartesian coordinates) are diagonal, with the time component, $\eta_{0 0}=-1$ and the other three diagonal spatial components equal to unity.}. The metric is perturbed away from the flat metric such that $g_{\mu \nu} (S) = \eta_{\mu \nu} + S \delta^{\mu_0}_{\mu} \delta^{\nu_0}_{\nu}$ for some fixed $\mu_0$ and $\nu_0$ and $S$ is the parameter we wish to estimate. Using the properties of the QFI and the Cram\'er-Rao bound discussed in Sec.~\ref{subsec:multiparameter}, making some simplifying assumptions about the properties of $S$ and assuming the probe has a strong coherent amplitude (relative to the quantum fluctuations) the following uncertainty relation can be derived:
\begin{equation}
 \langle \delta S^2 \rangle \langle \Delta (\int dv \hat T^{\mu_0 \nu_0})^2 \rangle \geqslant \frac{\hbar}{4}.
\label{qgcramer}
\end{equation} 
where the $\hat T^{\mu \nu}$ operator represents the energy density of the probe field being used to estimate the metric parameter and the integration is over the 4-volume element $dv$. 
Interestingly, if the perturbation is in the time component of the metric, $g_{0 0}$, then this will lead to a change in the proper time , $\tau$, of a probe ``clock" in the perturbation region relative to a reference clock in the surrounding flat spacetime. Under these conditions Eq.\eqref{qgcramer} reduces to
\begin{equation}
\Delta \tau^2 \Delta \hat H^2  \geqslant \frac{\hbar^2}{4},
\label{TUCg}
\end{equation}
in line with the usual time energy uncertainty relation. Similarly, if the perturbation is in the $x$ spatial component of the metric, $g_{1 1}$, then this will lead to a change in the proper length , $X$, of a probe ``ruler" in the perturbation region and Eq.\eqref{qgcramer} reduces to
\begin{equation}
 \Delta X^2 \Delta \hat P_X^2 \geqslant \frac{\hbar^2}{4},
\label{LUCg}
\end{equation}
which is a parametrized version of the usual position momentum uncertainty relation with $\Delta \hat P_X^2$ the momentum uncertainty in the $x$ direction.

It is important to emphasize that this approach to deriving uncertainty relations implicitly assumes that the $\langle \hat T^{\mu \nu} \rangle$ terms associated with probe are much smaller than the stress-energy terms producing the perturbation, otherwise there could be back action from the probe onto the metric. What happens if back-action cannot be neglected? Currently there is no generally consistent approach for treating such a situation, however several lines of argument lead to similar conclusions \cite{doi:10.1142/S0217751X95000085,Milburn_2006, PhysRevLett.103.231301}. Let us consider position momentum uncertainty. If the uncertainty in position is made extremely small the uncertainty in momentum will become very large. In such a situation we might expect the large momentum uncertainty would produce a Lorentz contraction to the proper length we are attempting to measure. If we generically assume $\Delta X \to \Delta X \sqrt{1 - \beta \Delta \hat P_X^2}$ where $\beta$ is an unknown constant, then the usual position momentum uncertainty relation can be written 
\begin{equation}
 \Delta X^2 \Delta \hat P_X^2 \geqslant \frac{\hbar^2}{4}(1 + \beta \Delta \hat P_X^2),
\label{LUCQG}
\end{equation}
where we have assumed $\beta \Delta \hat P_X^2 \ll 1$. It is easily shown that the consequence of this modification is to introduce a minimum measurable length which can be expressed as $\Delta x_{min} = L_P M_P c \sqrt{\beta}$, where $L_P$ and $M_P$ are the Planck length and Planck mass respectively. Indeed, quantum gravity derivations of Eq.~\eqref{LUCQG} often start from the idea of a minimum length. Whilst the correction to the uncertainty principle is very small there are suggestions for how it might be detected \cite{pikovski2012probing}.


\subsection{Squeezed States}
\label{section:squeezedlight}

Finally, we turn to squeezed states, which provide a versatile tool for testing the uncertainty principle, and exploiting the many possible applications of it. The limit on how precisely we can estimate non-commuting quantities, set by Heisenberg's uncertainty principle (Eq.~\eqref{eq:VUR_Kennard}), is saturated by a class of states known as coherent states. A similar class of states that also saturates this limit is the class of squeezed states. While coherent states have equal uncertainties in both conjugate variables, such as amplitude and phase quadratures, squeezed states allow the uncertainty to be distributed asymmetrically, enabling application in precision metrology, secure communications and computing. In essence, squeezed states exploit the flexibility within the uncertainty principle: while the principle itself remains a fundamental law that cannot be violated (under some reasonable assumptions, see Sec.~\ref{subsubsecSurpass}), it can be strategically leveraged to enhance measurement accuracy in specific scenarios that would otherwise be impossible. 

The study of squeezed states, with its focus on redistributing uncertainty within the limits of the Heisenberg principle, is itself an exploration of the uncertainty principle itself. In this section, we discuss this connection briefly. In Sec.~\ref{subsec:SSWhat} we give a brief introduction to squeezed states before the many applications of squeezed states are discussed in Sec.~\ref{subsec:SSapplication}.



\subsubsection{Redistributing Uncertainty in the Interest of Physics}
\label{subsec:SSWhat}

The concept of squeezed states goes back to the era of the uncertainty principle itself. Kennard~\cite{kennard1927quantenmechanik} introduced the concept of squeezed states through his investigation of Gaussian wave packets. The squeezing operator was introduced in the 1950s, by Friedrichs~\cite{friedrichs1953mathematical}, Infeld~\cite{infeldplebanski}, and Plebanski~\cite{plebanski1956wave}. However, these works went largely unnoticed until the 1970s, when squeezed states were rediscovered~\cite{stoler1970equivalence,stoler1971equivalence,yuen1976two}. The renewed attention on the usefulness of this unique class of states generated enough momentum to spark a new line of research. Notably, the work of Caves~\cite{caves1981quantum} and Walls~\cite{walls1983squeezed} on the application of squeezed states of light in interferometry became a cornerstone of this field. Although significant theoretical progress was made, experimental efforts initially faced several challenges, see Ref.~\cite{levenson1985experimentalists} for a detailed summary. Interestingly, not too long after these difficulties were pointed out, the first observation of squeezed light was reported by Slusher \textit{et al.} using four-wave mixing in sodium atomic vapor~\cite{slusher1985squeezing,slusher}. Since then, the quantum optics community has embarked upon a decades-long pursuit of higher squeezing levels and their applications across various domains in physics~\cite{schnabel2017squeezed,taylor2013biological,taylor2014subdiffraction,li2018quantum,li2020squeezed,lawrie2019quantum,michael2019squeezing,de2020quantum,lucivero2017sensitivity,pooser2020truncated,pooser2015ultrasensitive,aasi2013enhanced,tse2019quantum,casacio2021quantum,madsen2012continuous,jacobsen2018complete,zhao2020high,zhao2023enhancing,shajilal2024improving,yoshikawa2016invited,asavanant2019generation,madsen2022quantum}.

Squeezed states of light use the amplitude and phase quadratures of the electric field ($x$, and $p$ respectively as presented in Sec.~\ref{sec:Exp_PUR}), as the conjugate variables of interest. The corresponding uncertainty relation is
\begin{equation}
\Delta x \Delta p = \frac{\hbar}{2}.
\end{equation}
Coherent states have equal uncertainties in $x$ and $p$, as shown in Fig.~\ref{fig:interferometry} (f), and are characterized by circular uncertainty ellipses. This circle represents the standard quantum limit in phase space. A squeezed state, on the other hand, is characterized by an uncertainty ellipse, where the uncertainty in one quadrature is reduced below the limit of a coherent state, while the uncertainty in the conjugate quadrature is correspondingly increased~\cite{walls1983squeezed}, as shown in Fig.~\ref{fig:interferometry} (e). Squeezed states effectively reduce the noise in the variable of interest, at the price of increased noise in the conjugate variable.

In quantum optics, squeezing can be understood in terms of sideband correlations around a carrier frequency. A laser beam, for example, has a well-defined carrier frequency, but quantum fluctuations are present in the sidebands-frequencies slightly above and below the carrier. These fluctuations are often represented as noise in the quadratures of the light field, such as amplitude (in-phase) and phase (out-of-phase) components of the light. In a coherent state, the noise in the amplitude and phase quadratures is uncorrelated, and both exhibit fluctuations at the standard quantum limit~(SQL). However, in a squeezed state, quantum correlations arise between the sidebands, such that in one quadrature the noise is reduced, while in the other quadrature the noise increases~\cite{loudon1987squeezed}.

When discussing squeezed states, we often focus on the squeezing of quadrature amplitudes, particularly within the Gaussian regime. However, squeezed states are not necessarily limited to the squeezing of quadrature amplitudes, nor do they need to be Gaussian. There are optical states of light are non-Gaussian, such as squeezed cat states~\cite{sychev2017enlargement,takase2022generation,wang2022experimental,winnel2024deterministic}, and superpositions of vacuum and Fock states~\cite{wodkiewicz1987squeezing}. Furthermore, the squeezing of physical quantities other than quadrature amplitudes has been reported in the literature, including the squeezing of Stokes parameters~\cite{bowen2002polarization,heersink2005efficient,kalinin2023observation,andrianov2024polarization} and photon number squeezing~\cite{tapster1987generation,friberg1996observation,abe2000wideband,wang2020observation,ding2024observation}. We also note that for certain non-Gaussian states it is possible to achieve simultaneous squeezing in both x and p within localized regions of phase space, as exemplified by the GKP states~\cite{gottesman2001encoding}. In these states, the phase space consists of a grid of sharply peaked structures, each locally squeezed in both quadratures. While this may appear to violate the uncertainty principle, the relation remains respected because the grid extends throughout phase space, leading to large overall uncertainties in x and p. This localized squeezing is central to the use of GKP states in error correction, where small displacement errors in x and p can be corrected provided they remain well below the spacing between neighboring peaks.

Similar to squeezed light, in atomic systems, the redistribution of uncertainty in the spin component through specific interactions leads to spin squeezing~\cite{kitagawa1993squeezed}. Spin squeezing can be explained through the collective spin operators. For an ensemble of \( N \) two-level atoms (or spin-\(\frac{1}{2}\) particles), the collective spin operators are defined as:
\begin{equation}
S_x = \sum_{i=1}^{N} \frac{\sigma_x^{(i)}}{2}, \quad S_y = \sum_{i=1}^{N} \frac{\sigma_y^{(i)}}{2}, \quad S_z = \sum_{i=1}^{N} \frac{\sigma_z^{(i)}}{2}
\end{equation}
where \( \sigma_x^{(i)}, \sigma_y^{(i)}, \sigma_z^{(i)} \) are the Pauli matrices for the \( i \)-th particle. These spin components obey a Heisenberg-like uncertainty relation:
\begin{equation}
\Delta S_x \, \Delta S_y \geqslant \frac{|\langle S_z \rangle|}{2}
\end{equation}
A spin-squeezed state reduces the uncertainty in one spin component (e.g., \( S_x \)) while increasing it in the conjugate component (e.g., \( S_y \)). A widely used measure of spin squeezing was introduced by Wineland et al.~\cite{wineland1992spin}:
\begin{equation}
\xi^2 = \frac{N (\Delta S_{\bar{n}_\perp})^2}{|\langle \mathbf{S_{\bar{n}}} \rangle|^2}
\end{equation}
where \( (\Delta S_{\bar{n}\perp})^2 \) is the variance of the spin component along the direction \( \bar{n}_\perp \), which is perpendicular to the spin \( \mathbf{S_{\bar{n}}} \) in the direction \( \bar{n} \). If \( \xi^2 < 1 \), the state is considered spin-squeezed and can be used to achieve quantum-enhanced sensitivities in applications ranging from time keeping to precision measurements~\cite{meyer2001experimental,louchet2010entanglement,riedel2010atom,bornet2023scalable,franke2023quantum,strobel2014fisher,ockeloen2013quantum,schleier2010states,bohnet2014reduced,sewell2012magnetic}.

\subsubsection{Applications of Squeezed States}
\label{subsec:SSapplication}
Before concluding, we highlight several current and near-future applications of squeezed states across fundamental physics, quantum information technologies, and precision measurements.

For \textit{fundamental physics}, squeezed states of light have become important tools for probing the limits of our understanding of nature. A landmark example is their role in gravitational-wave astronomy, where injecting squeezed vacuum states into gravitational wave interferometers has improved sensitivity, extending the observable reach of these detectors and enabling the detection of ever fainter signals from the cosmos~\cite{aasi2013enhanced,ganapathy2023broadband}. Beyond this, squeezed states of light enable the exploration of intrinsic quantum phenomena. For instance, they can be used to generate optical Schr\"odinger cat states -- superpositions of coherent states with macroscopically distinguishable phases -- through photon subtraction~(or addition) or conditional measurements~\cite{ourjoumtsev2006generating,neergaard2006generation,parigi2007probing,chen2024generation,takase2021generation,endo2025high,takase2021generation}. These non-Gaussian states exhibit negativity in their Wigner function, a clear signature of non-classicality~\cite{endo2025high,chen2024generation}. 
A two-mode squeezed vacuum state (TMSV) can be created by interfering two single mode squeezed states on a 50:50 beam-splitter. They are the most entangled continuous variable (CV) state for a fixed photon number~\cite{bowen2003biased,braunstein2005quantum}, and are the standard resource in CV entanglement experiments, including quantum teleportation, quantum dense coding, quantum secret sharing and quantum error correction.~\cite{furusawa1998unconditional,li2002quantum,lassen2010quantum,aoki2009quantum,lance2004tripartite,
zhao2023enhancing,shajilal2024improving,conlon2024verifying}. Moreover, it has been shown that TMSV states can violate Bell-type inequalities when appropriately measured~\cite{cavalcanti2011large,thearle2018violation,abiuso2021measurement,brask2012robust,bjerrum2023proposal,moradi2025long,ishihara2025longa}.

Squeezed states of light are also central to CV \textit{quantum computation}, in particular measurement-based quantum computation (MBQC)~\cite{larsen2021fault}. In MBQC, quantum information is processed through local, adaptive measurements on a highly entangled resource state rather than by applying a sequence of unitary gates. A special class of entangled states known as cluster states provides the universal substrate for this approach~\cite{asavanant2019generation,larsen2019deterministic}. Once the cluster state is prepared, the computation proceeds solely through single-site measurements, eliminating the need for complex, coherent control of individual qubits and replacing it with the upfront challenge of generating large-scale, high-quality entangled states~\cite{raussendorf2001one,menicucci2006universal}. In the CV setting, conventional qubits of the cluster states are replaced by CV modes encoded either in squeezed states or in GKP states~\cite{bourassa2021blueprint}. Together, these resources enable universal quantum computation~\cite{larsen2021fault,tzitrin2021fault,walshe2025linear,ostergaard2025octo}. However, as in all quantum computing architectures, fault tolerance is essential to ensure that errors accumulate slower than they can be corrected. In CV MBQC, this requirement is typically expressed in terms of a squeezing threshold: the minimum level of squeezing needed for error-corrected, fault-tolerant quantum computation. Recent advances in error-correction protocols and cluster-state architectures have significantly lowered the required squeezing threshold -- from about 15 dB~\cite{walshe2019robust,menicucci2014fault} to 9.75 dB~\cite{aghaee2025scaling} – by employing qubit surface codes and assuming that every node of the cluster state is implemented with GKP states.

A prominent example demonstrating the computational power of squeezed states is Gaussian Boson Sampling (GBS)~\cite{hamilton2017gaussian,rahimi2015can}. In GBS, squeezed states of light are injected into a large linear optical network, and the output photon-number statistics are sampled at the detectors. For systems with many squeezed modes and sufficiently low loss, it is widely believed that simulating these output distributions on a classical computer is computationally intractable~\cite{hamilton2017gaussian,boixo2018characterizing,oh2024classical}. Indeed, two experimental demonstrations have reported evidence for quantum advantage using GBS~\cite{madsen2022quantum, zhong2020quantum}, but these claims have since been challenged by newly developed classical simulation algorithms based on tensor-network calculus~\cite{oh2024classical}.

Squeezed states of light are central to advancing \textit{quantum communications}, enabling both higher performance and optimality across multiple protocols. CV quantum key distribution (QKD) schemes exploit squeezed states to achieve secret key rates beyond those attainable with coherent states~\cite{gottesman2001secure,madsen2012continuous,gehring2015implementation,nguyen2025practical} and ultimately pave the way toward device-independent QKD~\cite{moradi2025long,ishihara2025longa}. They also provide the necessary resources for generating multipartite entangled states, enabling advanced cryptographic primitives such as quantum secret sharing~\cite{lance2004tripartite,conlon2024verifying} and conference key agreement~\cite{ishihara2025longb}. Moreover, while deterministic quantum repeaters for continuous-variable systems remain a significant challenge, proposals combining squeezed states with quantum error correction offer promising solutions~\cite{dias2017quantum}. Finally, squeezed states constitute the optimal resources for several quantum communication protocols in terms of communication capacity~\cite{pirandola2020advances}, underscoring their fundamental and practical importance.

As discussed in detail in Sec.~\ref{sec:SQLHS}, squeezed states of light have been used in \textit{quantum metrology} to achieve the Heisenberg scaling. Besides the squeezed-light-enhanced gravitational-wave detectors mentioned above, squeezed light has also been employed across a wide range of quantum sensing applications. Examples include the absolute calibration of photodiodes with high quantum efficiency~\cite{vahlbruch2016detection}, magnetometry~\cite{li2018quantum}, Raman spectroscopy~\cite{de2020quantum}, microscopy~\cite{casacio2021quantum}, and multiparameter estimation~\cite{bradshaw2018ultimate,bradshaw2017tight,steinlechner2013strong}.

Similar to squeezed states of light and their application in precision metrology, spin squeezing has become increasingly important for achieving sub-shot-noise measurements~\cite{wineland1992spin,kitagawa1993squeezed,wineland1994squeezed,cronin2009optics,riedel2010atom,Auzinsh2022SpinSqueezedMetrology,Friis2020SqueezingMetrology,colombo2022time,Mamaev2025NonGaussianSqueezing}. The connection between spin squeezing and QFI has proven particularly valuable in parameter estimation theory, providing new pathways for spin-squeezing-enhanced metrology~\cite{ma2009fisher,ma2011quantum,zhong2014quantum}. In Ramsey spectroscopy, spin-squeezed states have demonstrated remarkable capabilities in improving measurement precision beyond classical limits~\cite{wineland1992spin,ulam2001spin,birrittella2021optimal}. This enhancement has direct applications in advancing the precision of atomic clocks~\cite{braverman2018impact,schulte2020prospects,eckner2023realizing,robinson2024direct}. The development of more precise atomic clocks through spin squeezing techniques has significant implications for time-keeping standards~\cite{schulte2020prospects} and related experiments~\cite{delva2017test,safronova2018search,sanner2019optical,roberts2020search,delva2013atomic,grotti2018geodesy,mehlstaubler2018atomic}.

Beyond precision metrology and time-keeping, spin squeezing serves as a powerful tool for detecting and characterizing quantum entanglement~\cite{amico2008entanglement,guhne2009entanglement,horodecki2009quantum,korbicz2005spin,sorensen2001entanglement,toth2007optimal}, central to both quantum physics foundations and quantum information processing. The Kitagawa-Ueda squeezing parameter~\cite{kitagawa1993squeezed} has proven particularly valuable in characterizing pairwise entanglement in many-body spin-1/2 states. This parameter relates to negative pairwise correlations and concurrence, making it an effective measure of quantum correlations~\cite{wootters1998entanglement,ulam2001spin,wang2003spin}. Additionally, the Wineland squeezing parameter also serves as a definite multipartite entanglement witness in similar systems~\cite{sorensen2001many,wang2003relations,guhne2009entanglement}. 

The ability to generate and control spin-squeezed states provides new possibilities for quantum information processing and quantum simulation protocols~\cite{buluta2009quantum,buluta2011natural,pezze2021quantum,bornet2023scalable,santra2024squeezing}. As one example, recent research has explored the relationship between spin squeezing and quantum phase transitions, demonstrating that spin squeezing parameters can effectively identify critical points in quantum systems~\cite{wong2023quantum,halimeh2017dynamical,corps2023theory,corps2022dynamical,homrighausen2017anomalous}. The resilience of spin-squeezed states to various forms of decoherence has been extensively studied, showing promising results for practical applications~\cite{stockton2003characterizing,ulam2001spin,huelga1997improvement,ma2011quantum}. This robustness, combined with the ability to generate and measure spin squeezing in various experimental platforms~\cite{meyer2001experimental,esteve2008squeezing,riedel2010atom,gross2010nonlinear,kuzmich2000generation,takano2009spin,takano2010manipulation,schleier2010states,louchet2010entanglement,appel2009mesoscopic,chen2011conditional,leroux2010implementation}, has made it an attractive resource for quantum technologies.


\section{Discussions}
\label{conclusion}
The uncertainty principle remains a cornerstone of quantum mechanics, drawing a fundamental boundary between classical and quantum descriptions of nature. 
Yet, even a century after its inception, it continues to spark new questions -- both theoretical and experimental. 
Looking ahead, one major direction is exploring the foundations of physics. Observing the wave-like properties of increasingly massive particles, and designing sensitive experiments searching for gravitational effects in quantum systems, will help us determine the limits of our current theory of quantum mechanics. 
Perhaps with the appropriate experiment we may finally isolate a single correct interpretation of quantum mechanics.

Another key future direction is the search for new applications. 
While the uncertainty principle is foundational, its role as a practical resource is increasingly coming into focus. 
In optical quantum computing, for instance, states squeezed below the standard uncertainty limit enable truly quantum capabilities. 
Similarly, in quantum sensing, surpassing classical limits through squeezed states or entangled probes relies directly on uncertainty relations. 
Developing these technologies to the point where the general public benefit from them will be one of the most important goals for the coming century.

Increasing attention is being given to connecting disparate areas -- such as quantum communication and quantum sensing -- through a common lens shaped by the uncertainty principle, prompting a fundamental question: to what extent does uncertainty underpin the entire landscape of quantum information processing, and might it offer a unifying framework linking phenomena as varied as entanglement, nonlocality, and computational advantage?
Advancing our understanding of these questions may sharpen the boundary between the classical and quantum domains, revealing what is -- and is not -- possible in the quantum era, and illuminating the path toward genuine quantum advantage.

Despite a century of development, uncertainty relations continue to raise fundamental and practical questions. 
A central challenge is to develop a unified framework that connects the many existing formulations while clarifying the physical limitation captured by each. 
Equally important is the search for tight and computable bounds for observables with nontrivial internal structure, including symmetry-based algebraic, and geometric structures. 
Even at the level of preparation uncertainty, important questions remain about how to enrich the landscape beyond current development and how to formulate bounds that are both sharp and physically informative. 
The more operational problem of measurement uncertainty is even less settled: 
different physical platforms come with different noise models, control limitations and measurement architectures, raising the question of which notion of error and disturbance is most appropriate in each setting, and how strongly this choice depends on the underlying system. 
Addressing these issues would sharpen our understanding of the fundamental limitations imposed by different physical platforms and help identify the experimentally achievable precision limits, namely the minimal uncertainty compatible with the available resources and noise. 

Further open directions include understanding uncertainty as a quantum resource; 
clarifying its relation to complementarity, contextuality, nonlocality, scrambling and thermodynamic irreversibility; 
and extending uncertainty principles from static states to dynamical processes, higher-order quantum transformations and even scenarios with indefinite causal order. 
From a practical perspective, it is also crucial to formulate uncertainty relations under finite statistics, realistic noise, detector imperfections and device assumptions. 
The next stage of the field is therefore not only to derive stronger inequalities, but to identify uncertainty relations that are optimal, operationally meaningful, computationally tractable and directly useful for quantum technologies.

In the century ahead, we may see both a deeper understanding of the uncertainty principle and a broader reach of its influence. 
Its continuing role in shaping emerging technologies and its potential to unify core ideas in quantum theory suggest that Heisenberg's legacy is far from complete. 
We also anticipate that advances in areas such as artificial intelligence will aid the exploration of the principle itself -- helping to uncover new formulations and identify their optimal bounds, especially in regimes beyond our current reach. 
The next hundred years promise not only a fuller mapping of its limits, but perhaps also the discovery of new principles that extend or refine it -- bringing quantum ideas ever closer to practical applications and societal benefit. 
We hope the reader shares our excitement for what lies ahead.


\appendix

\section*{Appendix}


\section{Entropies}
\label{appendix:Entropies}

In this appendix, we introduce several entropic tools relevant to quantum theory, particularly in the contexts of quantum foundations and quantum information theory. We begin with differential entropy (also known as continuous entropy~\cite{Differential_Entropy}), which extends the notion of entropy to continuous probability distributions. 
This measure serves as a foundation for expressing the well-known position-momentum trade-off in terms of entropic quantities. Let $X$ be a random variable with a probability density function $f$, supported on the set $\mX$. The differential entropy, denoted as $h(X)$ or $h(f)$, is defined as
\begin{align}\label{eq:diff_ent}
    h(X):=-\int_{\mX}f(x)\log f(x)dx.
\end{align}
Unless stated otherwise, all logarithms in this review are taken to base $2$.

The second quantity is the well-known Shannon entropy~\cite{Shannon1948}. In information theory, a discrete random variable $X$ takes values $x$ from a finite set $\mX$ and is governed by a probability distribution $p: \mX \to [0,1]$. 
The information content~\cite{doi:https://doi.org/10.1002/047174882X.ch2}, also called the surprisal or self-information, of an event $X=x$ is defined as $-\log p_x$, where $p_x$ is the probability of the event occurring. As suggested by the term ``surprisal,'' the information content increases when the event is less likely, i.e., when $p_x$ is smaller. In particular, deterministic events, which occur with certainty, have no information content, as observing such an event provides no new knowledge. Conversely, the information content of highly unlikely events can become extremely large. This relationship between probability and information content serves as the foundation for the concept of Shannon entropy $H(X)$, which is defined as
\begin{align}
\label{eq:Shannon_Entropy}
    H(X):=-\sum_xp_x\log p_x.
\end{align}
Shannon's source coding theorem establishes that the Shannon entropy of a set of independent, identically distributed (i.i.d.) random variables determines the minimum number of bits needed for efficient compression in the asymptotic limit~\cite{MacKay2002}. Compressing below this threshold inevitably causes information loss. 
Schumacher compression offers a quantum analogue to this classical concept, substituting Shannon entropy $H$ with von Neumann entropy $S$~\cite{PhysRevA.51.2738}. For a quantum state $\rho$, its von Neumann entropy is 
\begin{align}
\label{eq:von_Neumann_Entropy}
    S(\rho):=-\Tr[\rho\log\rho].
\end{align}
Going forward, for notational simplicity, we will not distinguish between the Shannon entropy and the von Neumann entropy (both denoted $H$); the nature of the underlying system -- classical or quantum -- will be evident from the context. 

R\'enyi entropy $H_{\alpha}$ generalizes  Shannon entropy by introducing a parameter $\alpha$ that adjusts sensitivity to different probability distributions, offering a more flexible and nuanced measure of information~\cite{renyi1961measures}. By interpolating between key entropy measures -- recovering Shannon entropy $H$ as $\alpha\to1$, collision entropy $H_{\text{coll}}$ at $\alpha=2$, and min-entropy $H_{\min}$ as $\alpha\to\infty$ -- it provides a unified framework for diverse applications in cryptography, statistical inference, coding theory, and quantum information. Its ability to emphasize rare or dominant events makes it particularly powerful for analyzing uncertainty, resource quantification, and distinguishability in complex systems. The R\'enyi entropy of order $\alpha$ for a random variable $X$, defined for $0<\alpha<\infty$ with  $\alpha\neq1$, is given by
\begin{align}
\label{eq:Renyi_Entropy}
    H_{\alpha}(X):=\frac{1}{1-\alpha}\log \sum_x p_x^{\alpha}.
\end{align}
By introducing the $\alpha$-norm of the probability distribution $\vec{p}:= (p_x)$ as $\|\vec{p}\|_{\alpha}$, 
\begin{align}
\label{eq:alpha_Norm}
    \|\vec{p}\|_{\alpha}:=\left(\sum_x p_x^{\alpha}\right)^{1/\alpha},
\end{align}
the R\'enyi entropy of $X$ can be reformulated as
\begin{align}
\label{eq:Renyi_Entropy_alpha_Norm}
    H_{\alpha}(X)=\frac{\alpha}{1-\alpha}\log \|\vec{p}\|_{\alpha}.
\end{align}
A key property of R\'enyi entropy is that it is monotonically decreasing in $\alpha$; specifically, for $\alpha\leqslant\beta$, we have
\begin{align}\label{eq:Renyi_monotone}
    H_{\alpha}(X)\geqslant H_{\beta}(X).
\end{align}
In the quantum domain, the R\'enyi entropy of order $\alpha\in(0,1)\cup(1,\infty)$ for a positive semi-definite $\rho\geqslant0$ with $\Tr[\rho]\neq0$ is defined as a one-parameter generalization of the von Neumann entropy
\begin{align}\label{eq:Renyi_Entropy_Quantum}
    H_{\alpha}(\rho):=\frac{1}{1-\alpha}\log\frac{\Tr[\rho^{\alpha}]}{\Tr[\rho]}.
\end{align}

Mirroring the extension of entropy to the quantum setting, divergences can likewise be generalized. The standard R\'enyi divergence $D_{\alpha}$, between two positive semi-definite operators $\rho$ and $\tau$, with $\rho\neq0$, is defined as
\begin{align}\label{eq:Renyi_Divergence_Standard}
    D_{\alpha}(\rho\|\sigma):=\frac{1}{\alpha-1}\log\frac{\Tr[\rho^{\alpha}\sigma^{1-\alpha}]}{\Tr[\rho]}.
\end{align}
For a more axiomatic treatment of this generalization and its relation to the Kullback–Leibler divergence~\cite{1320776d-9e76-337e-a755-73010b6e4b64}, we refer the interested reader to Ref.~\cite{PETZ198657,muller2013quantum,Wilde2014}. A distinct and operationally more relevant variant -- widely used in quantum information theory, quantum resource theories, and in particular the formulation of entropic time-energy uncertainty relations (see Sec.~\ref{sec:QCTO}) -- is the sandwiched R\'enyi divergence, defined by~\cite{muller2013quantum,Wilde2014}
\begin{align}\label{eq:Renyi_Divergence_Sandwiched}
    \widetilde{D}_{\alpha}(\rho\|\sigma)
    :=
    \frac{1}{\alpha-1}\log\frac{
    \Tr[(\sigma^{\frac{1-\alpha}{2\alpha}}\rho\sigma^{\frac{1-\alpha}{2\alpha}})^{\alpha}]
    }
    {\Tr[\rho]}.
\end{align}

Building upon these R\'enyi divergences, we now examine bipartite quantum systems and their associated conditional entropies. Let $\mS_{AB}$ denote the set of sub-normalized quantum states on the composite Hilbert system $\mH_A \otimes \mH_B$, consisting of positive semi-definite operators with trace less than or equal to one. For a quantum state $\rho_{AB} \in \mS_{AB}$, the conditional von Neumann entropy of $\rho_{AB}$ given $B$ is defined as
\begin{align}\label{eq:Conditional_von_Neumann_Entropy}
    H(A|B)_{\rho}
    :=
    &
    H(\rho_{AB})-H(\rho_B)\\
    =&
    -\inf_{\sigma_{B}\in\mS_B}D_1(\rho_{AB}\|\1_{A}\otimes\sigma_B)
\end{align}
Here, $\rho_B$ denotes the reduced state of $\rho_{AB}$ on subsystem $B$, obtained by tracing out subsystem $A$. The identity operator on the corresponding Hilbert space is written as $\1$. The general conditional entropies are described as
\begin{align}\label{eq:alpha_Conditional_Entropies}
    H_{\alpha}(A|B)_{\rho}
    :=
    &\sup_{\sigma_{B}\in\mS_B}-D_{\alpha}(\rho_{AB}\|\1_{A}\otimes\sigma_B),
\end{align}
where the optimization is over sub-normalized states $\sigma_B$ on subsystem $B$. An analogous expression holds for $\widetilde{H}_{\alpha}(A|B)_{\rho}$, obtained by replacing $D_{\alpha}$ with the sandwiched R\'enyi divergence $\widetilde{D}_{\alpha}$ in Eq.~\eqref{eq:Renyi_Divergence_Sandwiched}. In particular, the conditional min- and max-entropies arise as limiting cases of this family and take the following forms
\begin{align}
    H_{\min}(A|B)_{\rho}
    :=&\sup_{\sigma_{B}\in\mS_B}-D_{\max}(\rho_{AB}\|\1_{A}\otimes\sigma_B),\label{eq:Entropy_Condi_Min}\\
    H_{\max}(A|B)_{\rho}
    :=&\sup_{\sigma_{B}\in\mS_B}-D_{\frac{1}{2}}(\rho_{AB}\|\1_{A}\otimes\sigma_B).\label{eq:Entropy_Condi_Max}
\end{align}
Note that here $D_{\max}=\lim_{\alpha\to\infty}\widetilde{D}_{\alpha}$.

To explore the memory-assisted uncertainty principle (see Sec.~\ref{subsec:QU_with_CM} and~\ref{subsec:QU_with_QM}), we will further describe smooth entropies~\cite{doi:10.1142/S0219749908003256}. Given two sub-normalized quantum states $\rho$ and $\sigma$, their purified distance is defined as
\begin{align}
    P(\rho,\sigma):=\sqrt{1-\bar{F}(\rho,\sigma)^2},
\end{align}
where $\bar{F}(\rho,\sigma)$ denotes the generalized fidelity between $\rho$ and $\sigma$, given by 
\begin{align}
    \bar{F}(\rho,\sigma):=
    \left\|
    \sqrt{\rho\oplus(1-\Tr[\rho])}
    \sqrt{\sigma\oplus(1-\Tr[\sigma])}
    \right\|_1.
\end{align}
Based on this metric, we can define smooth versions of the min- and max-entropies by optimizing over all sub-normalized quantum states within a small purified distance from the original state
\begin{align}
    H^{\varepsilon}_{\min}(A|B)_{\rho}&:=\sup_{\sigma\in\mB^{\varepsilon}(\rho)}
    H_{\min}(A|B)_{\sigma},\label{eq:Entropy_Smooth_Min}\\
    H^{\varepsilon}_{\max}(A)_{\rho}&:=\inf_{\sigma\in\mB^{\varepsilon}(\rho)}
    H_{\max}(A)_{\sigma}.\label{eq:Entropy_Smooth_Max}
\end{align}
The $\varepsilon$-ball around $\rho$ is formally defined as
\begin{align}
    \mB^{\varepsilon}(\rho):=\{\sigma\,|\, \sigma\geqslant0, \Tr[\sigma]\leqslant1, P(\rho,\sigma)\leqslant\varepsilon\}.
\end{align}
Given a bipartite quantum state $\rho_{AB}$ and any purification $\psi_{ABC}$ of it, the smooth max-entropy and smooth min-entropy are dual~\cite{5550419,5208530}, in the sense that their values are related by
\begin{align}\label{eq:Entropy_Min_Max}
    H^{\varepsilon}_{\max}(A|B)_{\rho}:=
    -H^{\varepsilon}_{\min}(A|C)_{\psi}.
\end{align}
For a more in-depth discussion of smooth entropies, we refer the reader to Refs.~\cite{doi:10.1142/S0219749908003256,5208530,5319753,5550419,tomamichel2015quantum}.


\section{Majorization Theory}
\label{appendix:Majorization}

Majorization is a fundamental mathematical concept that originated in Schur's analysis of Hadamard's determinant inequality~\cite{schur22klassevon} and was later formalized through the influential work of Hardy, Littlewood, and P\'olya in unifying a broad class of inequalities under convex functions~\cite{hardy1929some,hardy1952inequalities}. Over time, it has found widespread applications across diverse fields, including matrix analysis, statistics, signal processing, communication theory, economics, and, more recently, quantum information science~\cite{PhysRevLett.83.436,PhysRevLett.86.5184,nielsen2001majorization,PhysRevLett.91.057902,PhysRevA.87.042307,doi:10.1073/pnas.1411728112,VanHerstraeten2023continuous}. In particular, majorization plays a central role in entanglement theory, exemplified by Nielsen's theorem~\cite{PhysRevLett.83.436} characterizing pure bipartite state transformations under local operations and classical communication (LOCC)~\cite{Chitambar2014}. In the following, we briefly review the concept of majorization and demonstrate its relevance in understanding quantum uncertainty.

Let $\mathbb{P}^{d,\downarrow}$ denote the ordered probability simplex, which consists of all $d$-dimensional probability vectors with components arranged in non-increasing order. Within this set, weak majorization automatically becomes majorization, as all vectors share the same total sum of one. For any vector $\vec{x}\in\mathbb{P}^{d,\downarrow}$, the following chain of majorization relations can always be constructed
\begin{align}\label{eq:uncertainty_certainty}
    (\frac{1}{d}, \frac{1}{d}, \ldots, \frac{1}{d})
    \prec
    \vec{x}
    \prec
    (1,0,\ldots,0).
\end{align}

The necessary and sufficient condition for the majorization relation to hold is based on the concept of doubly stochastic matrix. For a $d$-dimensional system, the set of all such matrices is denoted by $D_d$. Specifically, for any vectors $\vec{x}, \vec{y} \in \mathbb{R}^{d}$, we have $\vec{x}\prec\vec{y}$ if and only if $\vec{x}$ lies in the convex hull of all permutations of $\vec{y}$. In other words, there exists a doubly stochastic matrix $D$ such that~\cite{arnold2012majorization}
\begin{align}\label{eq:x_Dy}
    \vec{x}=D\vec{y}.
\end{align}
A doubly stochastic matrix is a square matrix with non-negative entries, where the sum of the elements in each row and each column equals one. This condition is established by Birkhoff–von Neumann theorem~\cite{birkhoff1946tres}, which states that any doubly stochastic matrix can be expressed as a convex combination of permutation matrices. Functions that reverse the majorization order are known as Schur-concave functions. Specifically, a function $f$ defined on the $d$-dimensional probability simplex is Schur-concave if $f(\vec{x})\geqslant f(\vec{y})$ whenever $\vec{x}\prec\vec{y}$. R\'enyi entropies (see Eq.~\eqref{eq:Renyi_Entropy}) provide a prominent example of Schur-concave functions.

\begin{figure}[t]
    \centering   
    \includegraphics[width=0.42\textwidth]{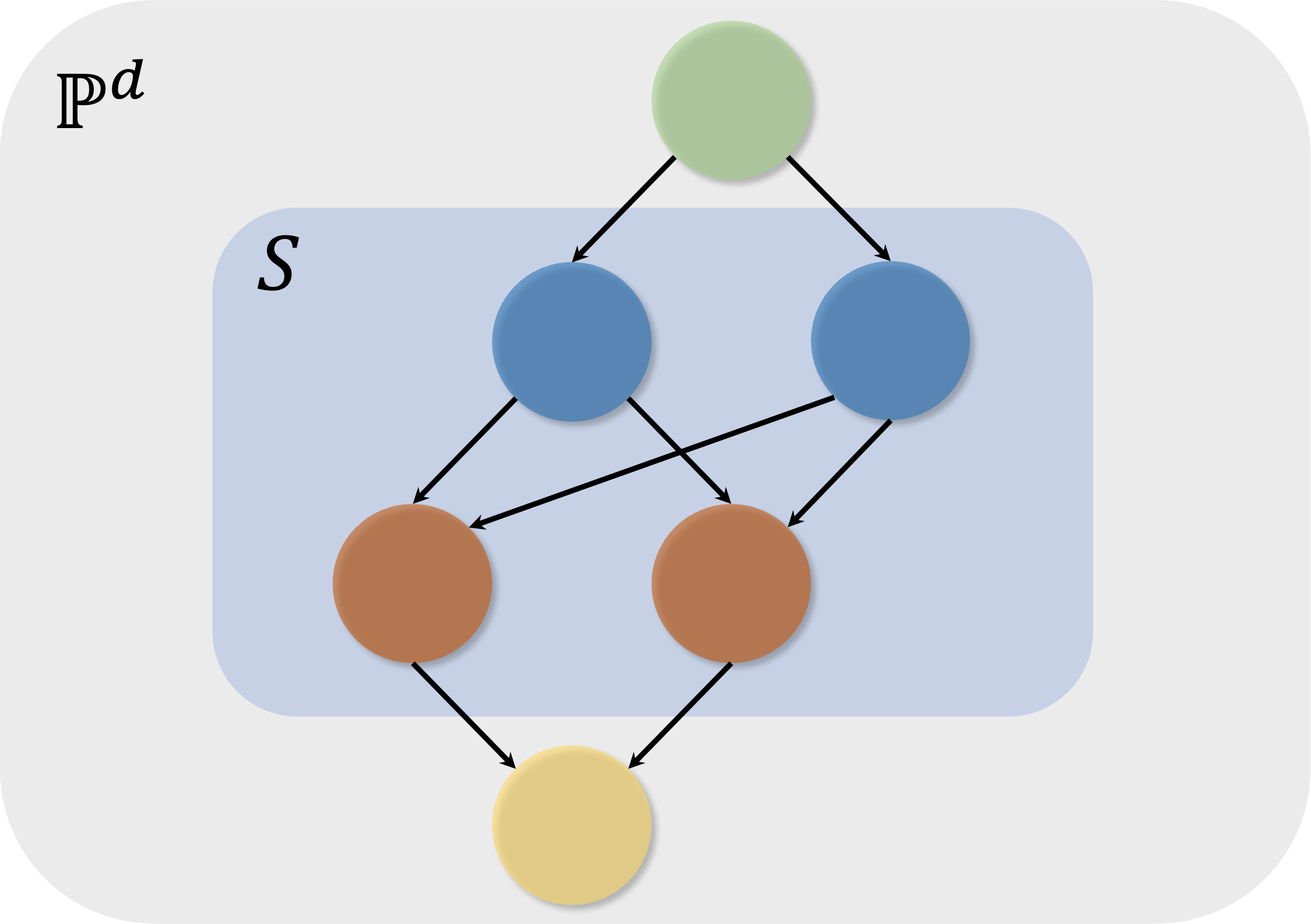}
    \caption{(Color online) \textbf{Majorization Lattice}.
    we use a lattice to represent probability vectors, with $S$ denoting the set of joint uncertainties expressed as probability vectors, such as $\vec{p}\otimes\vec{q}$ (see Eq.~\eqref{eq:Maj_DP}) and $\lambda\vec{p}\oplus(1-\lambda)\vec{q}$ (see Eq.~\eqref{eq:Maj_DS}). The arrows indicate majorization. For instance, an arrow from the green lattice to the blue lattice signifies that the probability vector associated with the green lattice majorizes the probability vector associated with the blue lattice. For a given set of joint uncertainties $S$, the GLB and LUB generally fall outside the set  $S$, although they still reside within the probability simplex.
    }
    \label{fig:Majorization_Lattice}
\end{figure}

A fundamental property of majorization is that the probability simplex, when endowed with the majorization, forms not only a lattice but a complete lattice, as illustrated in Fig.~\ref{fig:Majorization_Lattice}. This means that for any subset $S \subseteq\mathbb{P}^{d,\downarrow}$, there exist both a unique greatest lower bound (GLB) $\wedge S$ and a unique least upper bound (LUB) $\vee S$ with respect to majorization. In particular, if a vector $\vec{y}$ majorizes all elements in $S$, and all elements in $S$ majorize a vector $\vec{x}$ -- that is, $\vec{x}\prec \vec{z}\prec \vec{y}$ for all $\vec{z}\in S$ -- then it follows that
\begin{align}
    \vec{x}\prec\wedge S
    \quad\text{and}\quad
    \vee S\prec\vec{y}.
\end{align}

In the analysis of uncertainty, particularly through the framework of majorization, a central objective is the explicit construction of the least upper bound $\vee S$ of a set $S$ under the majorization order. This construction involves two steps. The first step is to compute a sequence of intermediate values $\varpi_{i}$ and $\omega_i$, defined by
\begin{align}\label{eq:varpi}
    \varpi_{i}:=\max_{\vec{z}\in S}\sum_{j=1}^{i}z_j
    \quad\text{and}\quad
    \omega_i:=\varpi_{i}-\varpi_{i-1}.
\end{align}
Here, $z_j$ denotes the $j$-th component of the vector $\vec{z}$, which is assumed to be sorted in non-increasing order. By convention, we define $\varpi_{0}=0$. Using the values 
$\omega_i$, we construct a vector 
\begin{align}\label{eq:Maj_omega}
    \vec{\omega}_S:= (\omega_{1}, \omega_{2}, \ldots, 0),
\end{align}
that serves as an upper bound for the set $S$, in the sense that 
\begin{align}
    S\prec\vec{\omega}_S.
\end{align}
However, in general, $\vec{\omega}_S$ is neither in non-increasing order nor optimal. As an illustrative example, consider the set $S=\{(0.6, 0.15, 0.15, 0.1)^{\T}, (0.5, 0.25, 0.2, 0.05)^{\T}\}$. In this case, the derived upper bound is $\vec{\omega}_S= (0.6, 0.15, 0.2, 0.05)^{\T}$, yet it is not the tightest possible bound. The LUB $\vee S$ in terms of majorization would be $(0.6, 0.175, 0.175, 0.05)^{\T}$~\cite{PhysRevResearch.3.023077,PhysRevLett.130.240201,Yuan2023}. 

To address this issue -- forming the second step of our construction -- we define an additional procedure known as the flatness process~\cite{992785}, denoted by $\mF$. Let $\vec{z}\in\mathbb{P}^{d,\downarrow}$ be a $d$-dimensional probability vector. Define $j$ to be the smallest index in $\{2, \ldots,d\}$ such that $z_j>z_{j-1}$ and let $k$ be the largest index in $\{1, \ldots,j-1\}$ satisfying $z_{k-1}\geqslant a:= (x_k+\cdots +x_j)/(j-k+1)$. We then define a transformation $\mT$ by
\begin{align}
\mT (\vec{z}) := \left(z_{1}^{\prime}, \ldots, z_{d}^{\prime}\right)
\end{align}
with
\begin{align}
z_{l}^{\prime} = 
     \begin{cases}
       a & \text{for}\quad l = k, \ldots, j, \\
       z_{l} & \text{otherwise.} \\ 
     \end{cases}
\end{align}
The flatness process $\mF$ is defined by repeatedly applying the transformation $\mT$ until the vector is fully arranged in non-increasing order.


\section{Quantum Lottery Games}
\label{appendix:Lottery_Games}

\begin{figure}[t]
    \centering   
    \includegraphics[width=0.48\textwidth]{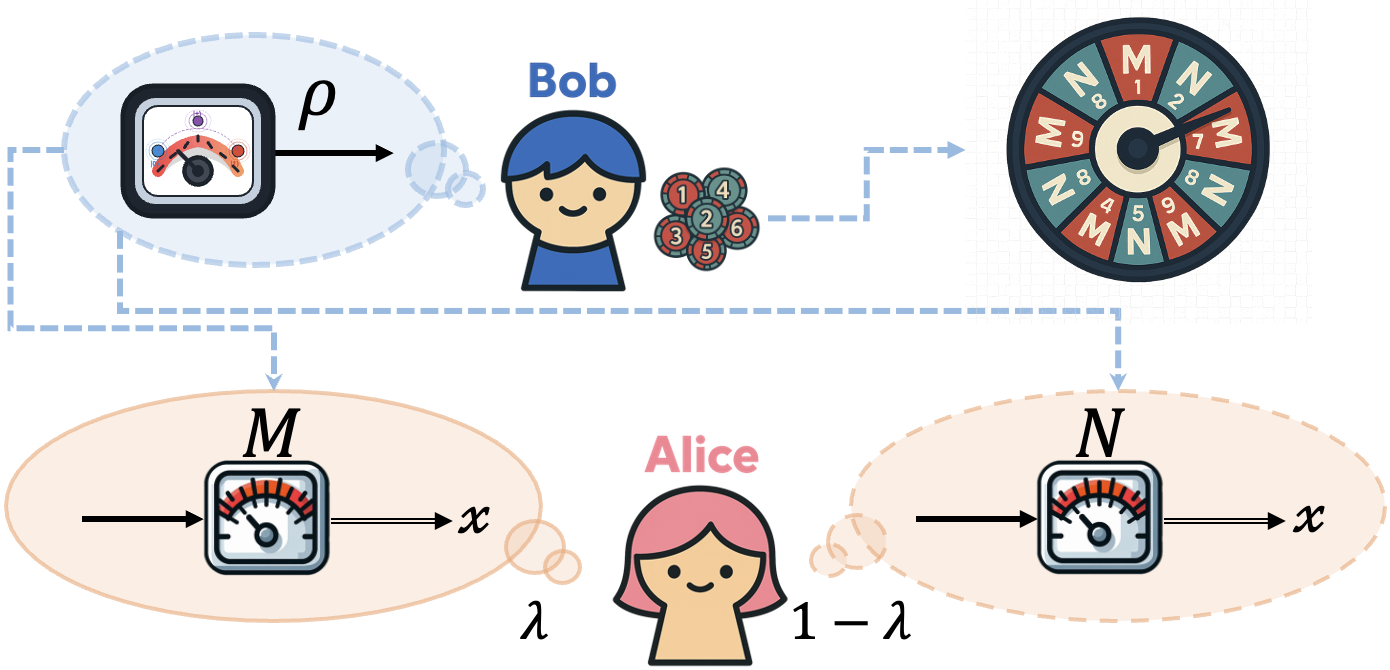}
    \caption{(Color online) \textbf{Lottery Games}. First, Alice sends Bob a classical specification of her measurements $M$ and $N$. With this information, Bob prepares and submits a state that is optimally tailored to the measurements. Alice then performs a single measurement, choosing $M$ with probability $\lambda$ or $N$ with probability $1-\lambda$, obtains outcome, and announces her choice. Finally, Bob is granted $k$ wagers on a ``quantum roulette,'' each wager assigned to a distinct outcome, and he wins the game if at least one wager coincides with Alice's result.
   }
    \label{fig:Lottery_Games}
\end{figure}

Analogous to the guessing game framework (see Fig.~\ref{fig:Guessing_Game}) for entropic uncertainty relations discussed in Sec.~\ref{subsec:Guessing_Games}, one can also cast majorization‐based uncertainty relations as an interactive protocol -- here termed the quantum lottery game. In this formulation, Alice and Bob proceed through the following steps: Alice begins by conveying to Bob a classical specification of her measurement ensemble; Bob, armed with this information, prepares and submits a state optimized against that ensemble. Alice then selects one measurement -- choosing $M$ with probability $\lambda$ and $N$ with probability $1-\lambda$ -- to obtain outcome, and discloses her choice. Bob is allocated $k$ wagers, which he places on a ``quantum roulette'' (see Fig.~\ref{fig:Lottery_Games}), each wager is assigned to a specific outcome of one of the possible measurements. He wins if at least one wager matches Alice's outcome. The function $\varpi_{\oplus,k}(\lambda)$ introduced in Eq.~\eqref{eq:Maj_DP_DS_1} characterizes the maximal winning probability with which Bob can secure a win when allotted exactly $k$ wagers as a function of the measurement mixing parameter $\lambda$, thereby providing an operational, betting‐odds characterization of majorization uncertainty relations in Eq.~\eqref{eq:Maj_DS}.


\section{Quantum Metrology}




\subsection{Quantum Fisher Information Uncertainty Relation}
\label{apenQFIuncerP}
In this appendix the QFI uncertainty relation, Eq.~\eqref{eq:SLDCRB:UP} in the main text, is derived. As in the main text, it is assumed that the parameter of interest is encoded through some Hamiltonian $H$, as $\ket{\psi_\theta}=\mathrm{e}^{-\mathrm{i}H\theta}\ket{\psi_0}$. The derivative of this state with respect to $\theta$ is then
\begin{equation}
\label{eq:apen:deriv}
    \ket{\psi'_\theta}=-\mathrm{i}H\ket{\psi_\theta}.
\end{equation}

To derive the corresponding SLD operator we shall use the fact that for pure states $\rho=\rho^2$. Then, the derivative can be written as
\begin{equation}
    \frac{\partial\rho}{\partial\theta}=\frac{\partial\rho^2}{\partial\theta}=\rho\frac{\partial\rho}{\partial\theta}+\frac{\partial\rho}{\partial\theta}\rho.
\end{equation}
By comparison with Eq.~\eqref{eq:SLDopBG}, the SLD operator can be determined as
\begin{equation}
\begin{split}
    L&=2\frac{\partial\rho}{\partial\theta}=2(\ket{\psi_\theta}\bra{\psi'_\theta}+\ket{\psi'_\theta}\bra{\psi_\theta})\\
&=2(\ket{\psi_\theta}\bra{\psi_\theta}(\mathrm{i}H)-\mathrm{i}H\ket{\psi_\theta}\bra{\psi_\theta}),
\end{split}
\end{equation}
where in the second line we use Eq.~\eqref{eq:apen:deriv}. Substituting this expression into Eq.~\eqref{eq:SLDQFIBG}, the QFI can be determined as
\begin{equation}
\begin{split}
        J_\text{S}&=\text{Tr}[\rho L_\theta L_\theta]\\
        &=4[\bra{\psi_\theta}H^2\ket{\psi_\theta}-|\bra{\psi_\theta}H\ket{\psi_\theta}|^2]\\
        &=4(\Delta H^2).
    \end{split}
\end{equation}
Using the SLDCRB (Eq.~\eqref{eq:sldbound}), we arrive at a form of uncertainty principle, Eq.~\eqref{eq:SLDCRB:UP} in the main text.  
\begin{equation}
    \delta\theta^2\geqslant\frac{1}{4M(\Delta H)^2}.
\end{equation}


\subsection{The Standard Quantum Limit}
\label{apenQFISQl}
Here we discuss the SQL for general protocols, before analysing some specific examples considered in the main text. For optical interferometry, when $\ket{\psi_1}\ket{\psi_2}=\ket{N}\ket{0}$ in configuration Fig.~\ref{fig:interferometry} (b), the resulting state is separable and the SQL is derived in Sec. V.a of Ref.~\cite{demkowicz2015quantum}. 

For qubits, we allow the use of $N$ qubits in the configuration shown in Fig.~\ref{fig:RosettaStone} (a). In this setting, Eq.~\eqref{eq:SLDCRB:UP}, can be used to derive the SQL. For separable probes evolving under an additive generator
\(H=\sum_{j=1}^{N}h^{(j)}\), the variance is additive,
\begin{equation}
(\Delta H)^2=N(\Delta h)^2\;.
\end{equation}
Substitution into the QFI uncertainty relation, Eq.~\eqref{eq:SLDCRB:UP}, 
therefore gives
\begin{equation}
    \delta\theta \sim \frac{1}{\sqrt{MN}}\;,
\end{equation}
which is the standard quantum limit.

\subsubsection{Coherent States}
\label{apen:cohQFI}
For the configuration shown in Fig.~\ref{fig:interferometry} (a) with \mbox{$\ket{\psi_1}=\ket{\alpha}$} we now derive the QFI using Eq.~(10) of Ref.~\cite{pinel2013quantum}. Note that for a coherent state, both the purity and covariance matrix are unchanged under a rotation, meaning the QFI can be expressed as
\begin{equation}
    J_\text{S}=\Delta x_\theta \sigma^{-1}\Delta x_\theta,
\end{equation}
where $x_\theta=(\langle x\rangle,\langle p\rangle)$, $\Delta x_\theta=\partial x_\theta/\partial\theta$ and $\sigma$ is the covariance matrix, the $2\times2$ identity matrix for a coherent state. Evaluating this using the terms from Sec.~\ref{subsubsecSQL} gives $J_\text{S}=4\alpha^2$, in agreement with Eq.~\eqref{eq:SQLcoherent1}.


\subsubsection{Single Photons}
\label{apensubsubsinglephoton}
Here we derive the SQL for single photon states, Eq.~\eqref{Eq:SQLsinglePhotons}. We start by noting that the variance in our value of $N_-$ given $N\times M$ repetitions of the experiment is equal to
\begin{equation}
(\delta N_-)^2=NMp_-(1-p_-).
\end{equation}
Then the variance in $4N_-/NM$ is given by 
\begin{equation}
   (\delta \frac{4N_-}{NM})^2 =\frac{16p_-(1-p_-)}{NM}.
\end{equation}
Finally, the variance in the square root of this quantity, i.e. the variance in our estimate of $\theta$ is given by
\begin{equation}
    (\delta\tilde{\theta})^2=\frac{1-p_-}{NM}.
\end{equation}


\subsection{Heisenberg Scaling}
\label{apenHscale2}
From Eq.~\eqref{eq:SLDCRB:UP} it is evident that states with larger $\Delta H^2$ can obtain more sensitive estimates of $\theta$. It is known that~\cite{giovannetti2006metrology,roy2008exponentially,boixo2007generalized}
\begin{equation}
\label{equation:HamiltonianEigs}
    \Delta H\leqslant\frac{1}{2}(\lambda_\text{max}-\lambda_\text{min}),
\end{equation}
where $\lambda_\text{max}$ and $\lambda_\text{max}$ denote the maximum and minimum eigenvalues of $H$ respectively. Let us first consider the scenario in Fig.~\ref{fig:RosettaStone} (a), with $H=\sum_{j=1}^{N}h^{(j)}$. From the previous section, we know that for a separable state $\Delta H^2=N\bra{\psi}h^{2}\ket{\psi}-N\bra{\psi}h\ket{\psi}^2$ implying that $\Delta H^2\leqslant \sqrt{N}(\lambda_\text{max,s}-\lambda_\text{min,s})$, where we use $\lambda_\text{max,s}$ and $\lambda_\text{min,s}$ to denote the maximum and minimum eigenvalues of $h$ which acts on a single mode system.

On the other hand, for an entangled state $\ket{\psi}$, by creating an entangled superposition of the eigenvectors corresponding to $\lambda_\text{max,s}$ and $\lambda_\text{min,s}$ for the single mode state, a state can be created with $\Delta H^2\leqslant N(\lambda_\text{max,s}-\lambda_\text{min,s})/2$. This $\sqrt{N}$ improvement over the separable case is the Heisenberg scaling.

\subsubsection{Heisenberg Scaling in Ramsey Interferometry}
We now present a simple example to illustrate this -- Ramsey interferometry with a $N$ qubit state, estimating a rotation about the $z$-axis, $\ket{\psi(\theta)}=\text{e}^{\mathrm{i}h\theta}\ket{\psi}=\text{e}^{\sum_{j=1}^{N}\mathrm{i}\sigma_z^{(j)}\theta/2}\ket{\psi}$, where $\sigma_z^{(j)}$ denotes $\sigma_z$ acting on the $j$th qubit. We write $H=\sum_{j=1}^{N}\sigma_z^{(j)}/2=\sigma_z\otimes \1_{2^{N-1}}+\1_2\otimes\sigma_z\otimes \1_{2^{N-2}}+\cdots$. For this $H$, the maximum and minimum eigenvalues are $\pm N/2$, giving $\delta\theta\geqslant1/N$ as expected. See also Ref.~\cite{pezze2009entanglement} for a discussion on entanglement and the QFI.


\subsubsection{Heisenberg Scaling for \texorpdfstring{$N00N$}{N00N} States}
\label{apenHscale1}

Here we use the observable $A_N$ acting on a phase shifted $N00N$ state (Eq.~\eqref{eq:phaseshiftedN00N}) to derive the Heisenberg scaling. For this we use the following standard formula for propagating uncertainty:
\begin{equation}
\Delta\tilde{\theta}=\frac{\Delta A_N}{\abs{\partial\langle A_N\rangle/\partial\theta}}=\frac{1}{N}.
\end{equation}
For this observable, it is easy to compute 
\begin{equation}
\langle A_N\rangle=\text{cos}(N\theta),
\end{equation}
and
\begin{equation}
( \Delta A_N)^2=\text{sin}^2(N\theta).
\end{equation}
Substituting these terms in reveals the Heisenberg scaling as promised.


\subsection{Useful Properties of the Quantum Fisher Information}
\label{QFIpropapen}
Here we list some useful properties of the QFI, reproduced from Ref.~\cite{demkowicz2009quantum}.
\begin{itemize}
\item The QFI is linear in the direct sum. For estimating a parameter $\theta$, we let $\rho(\theta)$ and $\sigma(\theta)$ be two density matrices supported on orthogonal subspaces, $S_{\rho(\theta)}\perp S_{\sigma(\theta)}$, which remain orthogonal for infinitesimal changes in $\theta$, then
\begin{equation}
J_\text{S}[p\rho(\theta)\oplus (1-p)\sigma(\theta)]=pJ_\text{S}[\rho(\theta)]+(1-p)J_\text{S}[\sigma(\theta)].
\end{equation}
\item The QFI is convex
\begin{equation}
J_\text{S}[p\rho(\theta)+ (1-p)\sigma(\theta)]\leqslant pJ_\text{S}[\rho(\theta)]+(1-p)J_\text{S}[\sigma(\theta)].
\end{equation}
\item For pure states, $\rho(\theta)=\ket{\psi(\theta)}\bra{\psi(\theta)}$, the QFI is given by
\begin{equation}
J_\text{S}=4[\bra{\psi'(\theta)}\ket{\psi'(\theta)}-\abs{\bra{\psi'(\theta)}\ket{\psi(\theta)}}^2],
\end{equation}
where $\ket{\psi'(\theta)}=\partial\ket{\psi(\theta)}/\partial\theta$.
\end{itemize}

We add two more useful points to this list:
\begin{itemize}
\item The SLD operators can be obtained using the following equation
\begin{equation}
\label{eq:apen:LDef}
    L_i=2\sum_{jk}\frac{\ket{\psi_j}\bra{\psi_k}}{\lambda_j+\lambda_k}\bra{\psi_j}\partial_i\rho\ket{\psi_k},
\end{equation}
where $\rho=\sum_j\lambda_j\ket{\psi_j}\bra{\psi_j}$ and $\partial_i\rho$ denotes the partial derivative of $\rho$ with respect to parameter $i$~\cite{pinel2013quantum}. 

\item For single parameter estimation, the QFI can be written as
\begin{equation}
    J_\text{S}=-2\frac{\partial^2 F(\rho(\theta),\rho(\theta+\epsilon))}{\partial \epsilon^2}\bigg\rvert_{\epsilon=0},
\end{equation}
where $F(\rho_1,\rho_2)$ denotes the fidelity between $\rho_1$ and $\rho_2$~\cite{pinel2013quantum}.
\end{itemize}

\section*{Acknowledgments}

Lorc\'{a}n O. Conlon, Biveen Shajilal, Syed M. Assad, and Ping Koy Lam were supported by A*STAR C230917010, Emerging Technology, A*STAR C230917004, Quantum Sensing, A*STAR C23091703, and Q.InC Strategic Research and Translational Thrust. Lorc\'{a}n O. Conlon and Ping Koy Lam were supported by the National Research Foundation, Singapore through the National Quantum Office, hosted in A*STAR, under its Centre for Quantum Technologies Funding Initiative (S24Q2d0009). Lorc\'{a}n O. Conlon was supported by the Templeton Foundation, grant 63121. Jie Zhao and Tim C. Ralph were supported by the Australian Research Council Centre of Excellence for Quantum Computation and Communication Technology (Project No. CE170100012). Ulrik L. Andersen was supported by Danish National Research Foundation (bigQ, DNRF0142), EU ERC project ClusterQ (101055224, ERC-2021-ADG), EU Horizon Europe (QSNP, no. 101114043), NNF project CBQS (NNF 24SA0088433), and Innovation Foundation Denmark (CyberQ). Yunlong Xiao acknowledges support from A*STAR under its Central Research Funds and Career Development Fund (C243512002).


\bibliography{References}


\end{document}